\documentclass[fleqn,usenatbib]{mnras}

\usepackage{txfonts}
\usepackage[T1]{fontenc}
\DeclareRobustCommand{\VAN}[3]{#2}
\let\VANthebibliography\thebibliography
\def\thebibliography{\DeclareRobustCommand{\VAN}[3]{##3}\VANthebibliography}

\usepackage{graphicx}	
\usepackage{longtable}
\usepackage{soul}
\usepackage{ulem}
\usepackage{xcolor}
\usepackage{multirow}
\usepackage{textcomp}
\usepackage{caption}
\usepackage{subcaption}
\usepackage{caption}
\usepackage{array}
\usepackage{hyperref}
\usepackage{natbib}

\title[Spectral age distribution of RLAGN]{Spectral age distribution for radio-loud active galaxies in the XMM-LSS field}

\author[S. Pinjarkar et al.]{Siddhant Pinjarkar$^{1}$\thanks{E-mail: s.pinjarkar2@herts.ac.uk},
Martin J. Hardcastle$^{1}$,
Jeremy J. Harwood$^{1}$,
Dharam V. Lal$^{2}$,
Peter W. Hatfield$^{3}$,\newauthor
Matt J. Jarvis$^{3,4}$,
Zara Randriamanakoto$^{5,6}$,
and Imogen H. Whittam$^{3,4}$.
\\
$^{1}$Centre for Astrophysics Research, Department of Physics, Astronomy and Mathematics, University of Hertfordshire, College Lane, Hatfield AL10 9AB, UK\\
$^{2}$National Centre for Radio Astrophysics - Tata Institute of Fundamental Research, Post Box 3, Ganeshkhind P.O., Pune 411007,
India\\
$^{3}$Astrophysics, University of Oxford, Denys Wilkinson Building, Keble Road, Oxford, OX1 3RH, UK\\
$^{4}$Department of Physics and Astronomy, University of the Western Cape, Robert Sobukwe Road, Bellville 7535, South Africa\\
$^{5}$South African Astronomical Observatory, P.O. Box 9, Observatory 7935, South Africa\\
$^{6}$Department of Physics, University of Antananarivo, P.O. Box 906, Antananarivo, Madagascar\\}

\date{Accepted XXX. Received YYY; in original form ZZZ}

\pubyear{2023}

\begin{document}
\label{firstpage}
\pagerange{\pageref{firstpage}--\pageref{lastpage}}
\maketitle

\begin{abstract}
Jets of energetic particles, as seen in FR type-I and FR type-II sources, ejected from the center of Radio-Loud AGN affect the sources surrounding intracluster medium/intergalactic medium. Placing constraints on the age of such sources is important in order to measure the jet powers and determine the effects on feedback. To evaluate the age of these sources using spectral age models, we require high-resolution multi-wavelength data. The new sensitive and high-resolution MIGHTEE survey of the XMM-LSS field along with data from the Low Frequency Array (LOFAR) and the Giant Metrewave Radio Telescope (GMRT) provide data taken at different frequencies with similar resolution, which enables us to determine the spectral age distribution for radio loud AGN in the survey field. In this study we present a sample of 28 radio galaxies with their best fitting spectral age distribution analyzed using the Jaffe-Perola (JP) model on a pixel-by-pixel basis. Fits are generally good and objects in our sample show maximum ages within the range of 2.8 Myr to 115 Myr with a median of 8.71 Myr. High-resolution maps over a range of frequencies are required to observe detailed age distributions for small sources and high-sensitivity maps will be needed in order to observe fainter extended emission. We do not observe any correlation between the total physical size of the sources and their age and we speculate both dynamical models and the approach to spectral age analysis may need some modification to account for our observations.
\end{abstract}



\section{Introduction}
\subsection{\normalfont{\textit{Radio-loud active galactic nuclei}}}
\label{RLAGN}
Active galactic nuclei (AGN) are driven by the accretion of matter onto the central supermassive black hole of a galaxy. In this paper, we use the term radio-loud AGN (RLAGN) to refer to objects which have strong radio emission related to the active nucleus, in general exceeding the radio emission due to star formation in their host galaxy. Traditionally RLAGNs have been classified morphologically into two different types depending on the jet morphology, namely, Fanaroff-Riley type I and type II (FRI and FRII), named after the Faranoff and Riley morphological distinction for central brightened and edge brightened sources (\citealt{FanaroffandRiley1974}). The jets that terminate in a hotspot and remain relativistic throughout are categorised as edge-brightened FRII radio galaxies, whereas the jets that are relativistic initially and decelerate through kpc scales are categorised as centre-brightened FRI radio galaxies. Hence, the main extended structures that are seen in the RLAGN are: the jet, the hotspot, and the lobes, seen as FRII's and jets which decelerate to form plumes of lobes, seen as FRI's (e.g., \citealt{Bridelandperley1984}, \citealt{Harwoodetal2015}, \citealt{Blandfordetal2019}, \citealt{HardcastleandCroston2020}). These structures can be pc to Mpc in size (e.g., \citealt{RDBlandfordandMJRees1974},~\citealt{Urryetal1995},~\citealt{Raffertyetal2006}).\par  
Energy is dissipated in the jets and hotspots of these sources, which accelerate electrons to relativistic speeds and give rise to intense radio emission. Hotspots advance through the external medium and leave behind the material that forms the lobes (see~\citealt{BegelmanBlandfordandRees1984} review). The expansion of the lobes does work on the external environment which can heat the gas around it and affect the gas cooling rates; such effects are known as "AGN feedback" processes (e.g., \citealt{Crotonetal2006, Boweretal2006}). These processes in turn transfer energy onto the intracluster medium (ICM) or intergalactic medium (IGM) (e.g., \citealt{Mcnamaraetal2012}), causing the environment around the jets to heat. Hence, it is important to understand the power of AGN mechanism which requires us to focus our attention towards the AGN energetics and the time they spend providing feedback. Furthermore, a constrained plasma age measurement can also give us insights into the dynamics of such powerful radio galaxies.\par 
 
\subsection{\normalfont{\textit{Spectral ageing models}}}
\label{SAM}
Synchrotron radiation is responsible for the radio emission that is observed in jets and lobes as was first realized in 1956 by examining the polarization and spectrum of the emission (\citealt{Baade1956}, see \citealt{RybickiandLightman1979} for details on radiation mechanism). The energy losses for synchrotron radiation producing regions under fixed magnetic field strengths scale directly with frequency, and so they can be expressed in terms of a frequency-dependent volume emissivity (\citealt{Longair2010}). However to calculate this magnetic field strengths must be known. It is possible to estimate magnetic field strengths by assuming the minimum energy conditions i.e. the field strength that gives the lowest total energy requirement for the system \citep{Burbidge1956}. The minimum energy conditions are close to the equipartition of energy, where the total particle energy density is equal to the total magnetic energy density (e.g, \citealt{Longair2010}, \citealt{HardcastleandCroston2020}).\par 
We need to observe a range of radio frequencies in order to find the shape of the spectrum (e.g., \citealt{JPleahyetal1987}). As the time scale ($\tau$) of the energy loss for the synchrotron process varies inversely with electron energy ($\tau$ $\sim$ 1/E), electrons at higher energies radiate their energy more rapidly as compared to electrons at low energies. This means that an increasingly curved spectrum should be observed over time (e.g., \citealt{Kardashev1962}, \citealt{JaffeandPerola1973}, \citealt{Harwoodetal2015}).\par   
There are three widely discussed models that assume the generation of single injection electron energy distribution at the jet termination point. The models are named after their originators: \cite{Kardashev1962, Pacholczyk1970} (KP model), \cite{JaffeandPerola1973} (JP model), and \cite{Tribble1993} (Tribble model). Assuming that the electron population can be described by a power law at the point of acceleration, the electron distribution at acceleration can be given by the normalization factor ($N_0$) and the power law index ($\delta$) of the initially injected electron energy distribution \citep{Pacholczyk1970}; \textbf{$N(E) = N_{0}E^{-\delta}$}, where $N(E)$ is number of electrons at a given energy. Along the lobes, electrons lose energy through synchrotron losses as they propagate, which allows us to evaluate the age distribution for the AGN structures \citep{Harwoodetal2013}. For electron energy distribution, defined above, when subjected to synchrotron and inverse-Compton losses, the intensity at given frequency is given by:
\begin{equation} \label{eq1}
	\\ I(\nu) = P_{c}N_{0}B\int_{0}^{\pi/2}\sin^{2}\theta d\theta\int_{0}^{E_{T}^{-1}}E^{-\delta}\times(1-E_{T}E)^{\delta-2}F(x)dE
\end{equation}\\
where $F(x)$ is the kernel for monoenergetic synchrotron emission and $P_{c}$ are constants \citep{Pacholczyk1970}. $E_{T}$ are model losses where the intensity is dependent on the pitch angle $\theta$, Energy E, and constant magnetic field B. The injection index ($\alpha_{inj}$) defines the initial power law spectrum of the electrons where they are accelerated at the hotspots and is related to $\delta$ by $\alpha_{inj}= ({\delta-1})/{2}$. Here and throughout the paper we define spectral index ($\alpha$) in the sense that flux $\propto \nu^{-\alpha}$. For time $t$, since acceleration, for an electron under radiative losses, the pitch angle is assumed to be constant for the KP model (\citealt{Kardashev1962, Pacholczyk1970}), whereas the pitch angle in the JP model is assumed to become isotropic over short time-scales (\citealt{JaffeandPerola1973}). The losses and lifetime relation for the KP and the JP model is given by equations \ref{KP} and \ref{JP} respectively:
\begin{equation} \label{KP}
	\\ E_{T}{\propto} B^{2}({\sin}^{2}{\theta})t 
\end{equation}\\
\begin{equation} \label{JP}
	\\ E_{T}{\propto} B^{2}{\langle}{\sin}^{2}{\theta}{\rangle}t 
\end{equation}\\
A third model, the Tribble model (\citealt{Tribble1993}), evaluates the electron population using the assumption that the magnetic field strength is spatially variable (see~\citealt{Harwoodetal2013, Harwoodetal2015} for detailed model studies). \cite{Harwoodetal2013, Harwoodetal2015} state that in some circumstances the KP model provides a better description of the observed spectra than the JP model although the JP model is more plausible physically because pitch angle scattering is expected to take place efficiently in realistic magnetic field configurations. Fits using the Tribble model tend to have similar goodness of fit to the KP model while maintaining the physical motivation of the JP model, but its application is computationally expensive. \par 
Observations of the power law index injected by the particle acceleration for hotspots initially showed it to be between 0.5 and 0.6 (\citealt{Meisenheimeretal1989, Carillietal1991b}) where shock theory limits the value at 0.5 (\citealt{Bell1978}). \cite{Konar2013} studied dynamics of Double-Double radio galaxies by measuring the injection index and jet power. They report injection index values between 0.5 to 0.85 and show that the injection index values are strongly dependent on jet power. A broader coverage that includes lower frequencies has shown injection index values between 0.7 and 0.8, suggesting additional acceleration mechanism and/or absorption processes (\citealt{Harwoodetal2013, Mahatmaetal2019}).\par 
Another lifetime estimation method that is used to calculate the ages of the AGN is called dynamical modeling. Such a model uses the radio lobe expansion speeds and the instantaneous source size to determine the dynamical ages (e.g., \citealt{Machalskietal2006, Harwoodetal2017}). Evaluating ages using the two methods, dynamical ageing, and spectral ageing, a discrepancy between the two has been observed (\citealt{Mahatmaetal2019}). The incorrect use of equipartition magnetic field estimates for the lobes can partly account for the difference in age values for the two methods (\citealt{Mahatmaetal2019}). \cite{Harwoodetal2016} and \cite{Turneretal2018} have also suggested electron mixing, the mixing of electron populations over an extended region as a contributing factor. Furthermore, narrow-bandwidth observations (even in sensitive surveys) might not account for the oldest radiating particles which can affect the spectral shape; correction for this requires us to include much more data at low radio frequencies which is often not available. Studies carried out until now (e.g., \citealt{Harwoodetal2013, Harwoodetal2015, Mahatmaetal2019}) have pointed out that there are only a limited number of sources available to form a representative sample of RLAGN population for a portion of the sky. In addition there is also a lack of high sensitivity surveys with higher resolution at higher frequencies.\par  
\subsection{\normalfont{\textit{Questions to answer}}}
\label{OQA}
The aim of the study is to examine the age distribution of the radio galaxy population using spectral age analysis. This requires a large sample of radio sources from sensitive, well-resolved surveys. This study uses a sample systematically extracted from a multi-frequency sky survey, which has not been done before. This study can also be used as an overview of what to expect if we want to perform the spectral age analysis for a larger sample and the different problems that may be encountered (discussed in detail below). In our analysis of the spectral ages, we use data from four surveys at different frequencies, the early science release of the  MeerKAT International GHz Tiered Extragalactic Exploration (MIGHTEE) survey of the XMM-LSS field \citep{Heywood2022}, a survey of XMM-LSS with the Low Frequency Array \citep{HaleandJarvisetal2018}, a survey of XMM-LSS using the Giant Meter-wave Radio Telescope (GMRT) \citep{SmolcicSlausandNovaketal2018} and the early science SuperMIGHTEE survey in the XMM-LSS at band 3 (300-500 MHz), also with the GMRT, (Lal, Taylor, et al., submitted). To evaluate the spectral age of the sources, we use the Broadband Radio Astronomy Tools (\textsc{BRATS}\footnote[1]{http://www.askanastronomer.co.uk/brats/}) software package \citep{Harwoodetal2013, Harwoodetal2015}. The package evaluates the spectral ages of the radio galaxies on a pixel-by-pixel basis.\par 
Within this paper we therefore aim to answer the following questions using the spectral age analysis of our sample:
\begin{enumerate}[i]
	\item What is the average age, the oldest age and the maximum age distribution observed for resolved sources?
	\item What are the different observable morphologies and how many sources fall into each class?
	\item What is the relationship between the age distribution maps and the source morphology?
	\item What can be a good source selection criterion for a given resolution to perform spectral age analysis?
	\item Is there a correlation observed between the source size and the spectral age and what does it say about the dynamics of the radio sources?    
\end{enumerate}
Section 2 describes the data processing steps applied for the spectral age analysis of the targets. In Section 3 and 4 we discuss the results obtained from our analysis. The conclusions derived from the analysis are given in Section 5. In this study we use a cosmology in which $H_0$ = 70 km s$^{-1}$ Mpc$^{-1}$, $\Omega_{m}$ = 0.3 and $\Omega_{\Lambda}$ = 0.7.
\section{Data Reduction and Analysis}
\subsection{\normalfont{\textit{Data Extraction and Organization}}}
\label{DEO}
The MeerKAT International GHz Tiered Extragalactic Exploration (MIGHTEE) (\citealt{Jarvisetal2016}) survey is providing radio continuum, spectral line, and polarization information for four (COSMOS, XMM-LSS, ECDFS, and ELAIS-S1) well studied extra-galactic deep fields, using observations with the South African MeerKAT telescope. The MeerKAT is equipped to observe in three bands, namely UHF (544 – 1088 MHz), L-band (856 – 1712 MHz), and S-band (1750 – 3500 MHz), where the dense core region of dishes (three-quarter collecting area) spans over 1 km in diameter and spreads out to provide a maximum baseline of 8 km. The MIGHTEE survey will cover $\approx$20 deg$^{2}$ over the four extragalactic deep fields at a central frequency of  $\approx$1284 MHz with $\approx$ 1000 h of observations with the L-band receivers. The early science data release provides an area of 3.5 deg$^{2}$ in XMM-LSS (a three-pointing mosaic), with the thermal noise of 1.5 {$\mu$}Jy/beam (the image is also limited by classical confusion, so the measured noise in the centre of the image is around 4.5 {$\mu$}Jy/beam) and a resolution of 8.2 arcsec, where the effective frequency changes across the map. Over 20,000 radio components in the XMM-LSS field were extracted to form a catalog (\citealt{Heywood2022} for more information about the MIGHTEE survey and details on data processing steps). The Low Frequency Array (LOFAR) has made observations of the XMM-LSS field at 120-168 MHz. The observations in the field reach a central rms of 280 {$\mu$}Jy/beam at 144 MHz and provide a resolution of 7.5{$\times$}8.5 arcsec (\citealt{HaleandJarvisetal2018}). The GMRT survey is a 610 MHz radio continuum survey covering a 25 deg$^{2}$ area in the XMM-LSS field and towards the XXL-N field. The rms achieved in the XMM-LSS field is around 200{$\mu$}Jy/beam and the resolution of the mosaic is around 6.5 arcsec (\citealt{SmolcicSlausandNovaketal2018}). In addition, we use maps from the superMIGHTEE project where the observations target the MIGHTEE XMM-LSS early science region. The region used in this study is covered by a mosaic of 4 pointings at band-3 with a total solid angle of 6.22 deg$^{2}$. The Band-3 radio frequency covers the range 300 to 500 MHz, out of which we use the narrow band data at frequencies 320 MHz, 370 MHz, 420 MHz, and 460 MHz, the resolution of which is 10 arcsec with a flux calibration error of 5\% (Lal, Taylor, et al., submitted). LOFAR, the GMRT, and MeerKAT all have shortest baselines that are short enough to sample all of our target sources adequately, also shown in Table \ref{tableuvrange}; the three telescopes at our observing frequencies are sensitive to all structure on scales less than $\sim$ 20 arcmin. \par

\begin{table*}
	\centering
	\caption{Baseline and uv coverage information for the MeerKAT, the LOFAR and the GMRT telescopes.} 
	\label{tableuvrange}
	\begin{tabular}{l r r r r r}
		\hline
		Telescopes & Frequency (MHz) & Shortest baseline (m) & Longest baseline (m) & Minimum uv distance ($\lambda$) & Maximum uv distance ($\lambda$) \\
		\hline\hline
		MeerKAT & 1280 & 20 & 8000 & 85 & 34000\\
		LOFAR & 144 & 320 & 90000 & 150 & 43000 \\
		GMRT & 610 & 60 & 25000 & 120 & 51000 \\
		GMRT & 390 & 60 & 25000 & 80 & 32000 \\
		GMRT & 460 & 60 & 25000 & 90 & 38000\\
		\hline
	\end{tabular}
\end{table*}
Due to MeerKAT's high sensitivity, the data obtained from the survey makes it our obvious choice for mining the radio sources. The MeerKAT data fits well as a reference for sources from other survey frequencies as it is known that spectral curvature is easily observable at GHz frequencies for moderately aged radio sources. Our first task was to look for coordinates of the radio galaxies in the catalogue generated using \textsc{PyBDSF} (Python Blob Detector and Source Finder, \citealt{PyBDSF}), which contained around 20,000 radio sources, to identify radio galaxies in the image. As the spectral age analysis requires us to look for sources at multiple frequencies, we required sources that are also present in the LOFAR and GMRT surveys. The MIGHTEE survey dataset was searched for extended sources, which we expected to have an elliptical shape due to the elongation of the lobes along the jet axis. Hence all cataloged radio sources which had deconvolved major axis values greater than 10 arcsec in the MIGHTEE survey were identified as potential extended sources. The double lobes of some powerful AGN can be misinterpreted as two different sources and so a visual inspection of the MeerKAT mosaic was conducted to look for any additional sources that could be included in the list of extended sources. We identified 12 sources using visual inspection. \par
\subsection{\normalfont{\textit{Creation of the sample}}}
\label{CS}
A total of 120 extended sources were identified from the catalog and the visual inspection. We next looked for counterparts of MeerKAT sources in the LOFAR and GMRT images. Any point source, or a source that was surrounded by artifacts, in any of the surveys, was removed from the list of extended sources. When the list of extended source coordinates was matched with the LOFAR and the GMRT surveys, using a match radius of 1 arcsec, we found that these comparatively low-sensitivity surveys could not detect the structure that was seen in the highly sensitive MeerKAT images. Hence, any source that did not show the structural features of an extended source in all the surveys was removed from the list, which reduced the sample to 41 radio sources. The position of the sources was listed and FITS cutouts were created, centered at the radio position. For the MeerKAT data, we updated the FITS headers of the cutouts to account for the spatially varying effective frequency, which allows the correct frequency to be used in the spectral age analysis.\par 
We visually inspected these 41 sources and used DS9~\citep{DS9} to find their counterparts in the SDSSr (Sloan Digital Sky Survey), WISE 3.4, and WISE 4.6 (Wide-field Infrared Survey at 3.4 $\mu$m and 4.6 $\mu$m). Once we found a counterpart at or around the radio coordinate, we then used NASA/IPAC Extragalactic Database (\textsc{NED}\footnote[2]{https://ned.ipac.caltech.edu/}) near position search to look for the counterparts and their recorded redshift. Counterparts that gave spectroscopic redshift information were updated in our sample list; if there was no spectroscopic redshift information present, then we used the photometric redshifts reported by \cite{Hatfieldetal2022}. There were some regions absent from the photometric redshift study by \cite{Hatfieldetal2022} which did not report redshift values for some of our sources, such as target 14, and target 22 (see Table \ref{tableinfo}). For such sources, we searched for counterparts in NED to obtain a photometric redshift, described above and added them to the sample list. Hence, apart from the redshifts reported by \cite{Hatfieldetal2022}, we additionally obtained redshift values using optical counterparts present in the SDSS, the Cosmic Assembly Near-infrared Deep Extragalactic Legacy Survey-Ultra Deep field (CANDELS-UDS), the Spitzer Wide-area InfraRed Extragalactic Survey (SWIRE) photometric redshift catalog, and the Subaru/XMM–Newton Deep Field (SXDF) survey (\citealt{SDSS, CANDELS_UDS, theSWIREsurvey, SXDF}). Three sources with optical counterpart but with no available redshift information were excluded from the sample. We finally obtained a sample of 28 targets (see Table \ref{tableinfo} for target details).\par 
\subsection{\normalfont{\textit{Data Processing}}}
\label{DR}
As the \textsc{BRATS} software package uses a pixel-by-pixel analysis to determine the spectral age which allows us to view spectral features as a function of position, it is important to make sure that the radio maps are aligned accurately in pixel space across all frequencies (\citealt{Harwoodetal2013}). Using LOFAR's elliptical beam with the circular beam of MeerKAT and GMRT would mean that the radiation received per beam area would be different across frequencies. Hence, the radio maps were smoothed and regridded to give equal beam size for each source using the \textit{imsmooth} and \textit{imregrid} commands from Common Astronomy Software Applications (CASA)~\citep{CASA}. The resolution obtained after the smoothing and regridding was a circular PSF of FWHM 10 arcsec with a pixel size of 1 arcsec. As a check to see if the smoothing and regridding has worked properly, \textsc{DS9} was used to verify that the total flux of each source before and after the processing was approximately the same at any given frequency. The typical difference between the total flux values before and after smoothing and regridding  was less than 1\% for sources at any given frequency. After matching for sky coordinates, to reduce the effects of misalignment we account for frequency-dependent phase shifts and obtain accurate alignment of the radio maps. We used the Gaussian-fitting method which uses a point source around the target to fit a Gaussian at all frequencies and choose an appropriate reference pixel to align the images (refer to \citealt{Harwoodetal2013} for a detailed discussion on alignment). As a check for alignment, a Gaussian was again fitted to the same point source in the resulting images and a maximum difference of 0.1 pixels between two maps for any source was set as the threshold to indicate misalignment. No misalignment was observed for the sources.    
\subsection{\normalfont{\textit{Parameter Determination}}}
\label{PD}
Using \textsc{BRATS's} wide range of spectral age model fitting tools, we can easily determine the on-source properties, evaluate and visualize distributions and run statistical tests on the radio maps. A detailed description of the software package is available in the \textsc{BRATS} cookbook\footnote[3]{http://www.askanastronomer.co.uk/brats/downloads/bratscookbook.pdf}. By using the \textit{load} command in \textsc{BRATS} we loaded the radio maps at each frequency. In order to ensure any background sources were excluded from the model fitting, region files that loosely encompass the target sources were created using DS9. In cases where a bright core was observed (which are not expected to be properly described by models of spectral ageing) an exclusion region was added to ensure that only lobe emission was considered. A background region file was also defined in order to determine the off-source rms thermal noise. Using the rms values, an initial source detection was performed by assuming a 5{$\sigma$} cut-off. However, we also wanted to account for the uncertainties arising due to modeling of the extended emission during imaging. As described by~\cite{Harwoodetal2013}, we assume an on-source noise multiplier of 3. Flux calibration uncertainties were assumed to be 10{\%}; this gives a good match to flux scale uncertainties for the datasets used~\citep{Heywood2022, HaleandJarvisetal2018} and we used consistent flux scale uncertainties for all datasets to avoid bias towards a particular frequency range in the fitting. Using the {\textit{setregions}} command in BRATS we defined the number of pixels present in the target (see \textsc{BRATS} cookbook for details on the pixel detection techniques,~\citealt{Harwoodetal2013}).\par 

\begin{table*}
	\centering
	\caption{Properties of the sources in our sample, the radio source name where each name should be prefixed with MGTC to indicate discovery in the MIGHTEE continuum survey \citep{Delhaizeetal2020}, the host galaxy coordinates for the respective radio source, the redshift where P represents photometric redshift and S represents spectroscopic redshift value, the total flux from LOFAR survey, the radio luminosity measured at 144 MHz (LOFAR), the magnetic field strength evaluated using the flux density from the LOFAR survey assuming field at 0.4 times equipartition, source total physical size and the size of the lobes. The total physical size of the radio source was measured using largest extent of the source as seen in the MeerKAT map, given in kpc (column 8) and the size of the two lobes (comma seperated) under analysis, measured using MeerKAT maps, is given in arcsec. For sources with single measurement, only one lobe is seen in the MeerKAT maps. The morphology of the sources is given in the superscript of 'Radio source IAU name' column, where FR type-I and type-II are represented as I and II, HT is head-tail source and F is for any unidentified flagged source. The errors on the flux density and radio luminosity are dominated by LOFAR flux calibration uncertainties and so are of order 10\%.} 
	\label{tableinfo}
	\begin{tabular}{c c c c m{1.2cm} m{1.5cm} m{1.5cm} m{0.8cm} m{1.0cm}}
		\hline
		Target & MGTC name & Host galaxy (RA, DEC) & Redshift (z) & Total Flux from LOFAR (Jy)& Radio luminosity at 144 MHz ($10^{25}$ W/Hz) & Magnetic Field Strength ($\times 10^{-10}$ T) & Total source size (kpc) & Lobes size (arcsec)\\
		\hline\hline
		\\
		1 & $J021531.81-044050.9^{HT}$ & 02h15m31.25s -04$^{\circ}$40'58.9" & $0.37^{1}$(P) & 0.108 & 4.77 & 1.55 & 253 & 49\\ [1.25ex]
		
		2 & $J021500.04-045346.3^{II}$ & 02h15m00.16s -04$^{\circ}$53'47.6" & $0.89^{2}$(S) & 0.845 & 279 & 3.98 & 533 & (34, 33)\\[1.25ex]
		
		3 & $J021724.39-051255.5^{II}$ & 02h17m24.40s -05$^{\circ}$12'51.8" & $0.92^{3}$(S) & 0.037 & 13.30 & 1.43 & 1045 & (58, 58)\\[1.25ex]
		
		4 & $J021953.22-051826.9^{II}$ & 02h19m53.02s -05$^{\circ}$18'23.9" & $0.07^{2}$(S) & 0.156 & 0.12 & 2.84 & 80 & (24, 32)\\[1.25ex]
		
		5 & $J021956.07-052803.3^{I}$ & 02h19m56.08s -05$^{\circ}$28'08.2" & $0.28^{2}$(S) & 0.179 & 4.01 & 1.96 & 439 & (62, 62)\\[1.25ex]
		
		6 & $J022050.78-051013.4^{I}$ & 02h20m50.62s -05$^{\circ}$10'18.5" & $0.95^{1}$(P) & 0.032 & 12.60 & 2.24 & 360 & (18, 22)\\[1.25ex]
		
		7 & $J022334.41-045838.5^{II}$ & 02h23m34.30s -04$^{\circ}$58'39.5" & $0.17^{2}$(S) & 0.063 & 0.53 & 1.96 & 186 & (30, 32)\\[1.25ex]
		
		8 & $J022349.44-041221.6^{II}$ & 02h23m49.47s -04$^{\circ}$12'20.5" & $1.74^{1}$(P) & 0.213 & 335 & 5.23 & 448 & (25, 26)\\ [1.25ex] 
		
		9 & $J022325.17-042724.3^{II}$ & 02h23m25.30s -04$^{\circ}$27'24.4" & $0.61^{2}$(S) & 0.100 & 14.20 & 2.83 & 396 & (29, 28)\\ [1.25ex] 
		
		10 & $J022414.01-052823.6^{II}$ & 02h24m13.94s -05$^{\circ}$28'19.3" & $0.77^{2}$(S) & 0.893 & 211 & 3.77 & 696 & (42, 42)\\ [1.25ex] 
		
		11 & $J022511.19-045431.7^{II}$ & 02h25m10.98s -04$^{\circ}$54'33.2" & $0.23^{2}$(S) & 0.055 & 0.82 & 2.14 & 206 & (26, 22)\\ [1.25ex] 
		
		12 & $J022428.18-044952.5^{II}$ & 02h24m28.00s -04$^{\circ}$49'53.3" & $0.49^{1}$(P) & 0.052 & 4.40 & 2.29 & 312 & (24, 26)\\ [1.25ex] 
		
		13 & $J021635.17-044658.6^{II}$ & 02h16m35.18s -04$^{\circ}$46'58.2" & $1.02^{1}$(P) & 0.090 & 41.50 & 3.53 & 342 & (17, 18)\\ [1.25ex] 
		
		14 & $J021827.16-045439.2^{I}$ & 02h18m27.16s -04$^{\circ}$54'41.6" & $0.23^{1}$(P) & 0.633 & 9.80 & 3.10 & 216 & (30, 25)\\ [1.25ex] 
		
		15 & $J021926.45-051536.0^{II}$ & 02h19m26.48s -05$^{\circ}$15'34.6" & $1.36^{1}$(P) & 0.019 & 16.70 & 3.08 & 308 & (15, 16)\\ [1.25ex] 
		
		16 & $J021943.19-043113.3^{II}$ & 02h19m43.26s -04$^{\circ}$31'12.8" & $0.71^{1}$(P) & 0.012 & 2.28 & 1.80 & 333 & (20, 22)\\ [1.25ex] 
		
		17 & $J022038.83-043722.7^{I}$ & 02h20m38.76s -04$^{\circ}$37'22.6" & $1.19^{1}$(P) & 0.065 & 42.60 & 3.38 & 311 & (19, 19)\\ [1.25ex] 
		
		18 & $J022135.20-044855.7^{F}$ & 02h21m35.10s -04$^{\circ}$48'54.5" & $0.80^{1}$(P) & 0.009 & 2.26 & 1.91 & 311 & (15, 17)\\ [1.25ex] 
		
		19 & $J022230.57-044706.2^{II}$ & 02h22m30.43s -04$^{\circ}$47'05.2" & $1.83^{1}$(P) & 0.122 & 217 & 5.20 & 346 & (20, 20)\\ [1.25ex] 
		
		20 & $J022254.71-041358.2^{F}$ & 02h22m54.61s -04$^{\circ}$13'59.2" & $1.53^{1}$(P) & 0.047 & 55.60 & 3.64 & 396 & (20, 25)\\ [1.25ex] 
		
		21 & $J022256.56-042449.9^{II}$ & 02h22m56.50s -04$^{\circ}$24'49.9" & $1.18^{1}$(P) & 0.154 & 99.60 & 4.21 & 319 & (14, 21)\\ [1.25ex] 
		
		22 & $J022457.43-051656.0^{II}$ & 02h24m57.34s -05$^{\circ}$16'55.7" & $1.40^{4}$(P) & 0.199 & 190 & 5.00 & 385 & (21, 22)\\ [1.25ex] 
		
		23 & $J022410.08-044607.5^{II}$ & 02h24m09.93s -04$^{\circ}$46'07.5" & $1.45^{1}$(P) & 0.135 & 140 & 5.15 & 352 & (17, 17)\\ [1.25ex] 
		
		24 & $J021600.96-043238.5^{II}$ & 02h16m01.01s -04$^{\circ}$32'40.8" & $0.99^{1}$(P) & 0.015 & 6.51 & 1.11 & 1171 & (56, 75)\\ [1.25ex] 
		
		25 & $J021845.17-041438.9^{F}$ & 02h18m45.35s -04$^{\circ}$14'30.2" & $0.79^{1}$(P) & 0.042 & 10.70 & 1.78 & 831 & (39, 54)\\ [1.25ex] 
		
		26 & $J021634.43-045507.6^{II}$ & 02h16m34.96s -04$^{\circ}$55'06.4" & $0.91^{1}$(P) & 0.033 & 11.60 & 1.49 & 1391 & (46, 49)\\ [1.25ex] 
		
		27 & $J021658.68-044917.3^{II}$ & 02h16m59.06s -04$^{\circ}$49'20.8" & $1.32^{5}$(S) & 0.371 & 311 & 3.10 & 1269 & (66, 63)\\ [1.25ex] 
		
		28 & $J021944.61-044845.9^{I}$ & 02h19m44.64s -04$^{\circ}$48'45.1" & $0.93^{1}$(P) & 0.028 & 10.30 & 3.03 & 310 & (18, 19)\\ [1.25ex] 
		\hline
	\end{tabular} \\
	\footnotesize{$^1$\cite{Hatfieldetal2022},$^2$\cite{SDSS}, $^3$\cite{CANDELS_UDS}, $^4$\cite{theSWIREsurvey}, $^5$\cite{SXDF}}
\end{table*}

Other parameter dependencies involved the determination of redshifts and magnetic field strength. In order to determine the magnetic field strengths of the radio lobes, we used the \textsc{PySYNCH} code (\citealt{Hardcastleetal1998}), where we defined an elliptical region around the lobes with major and minor axis values and fitted a power law spectrum. With the help of redshifts (see section \ref{CS}), lobe flux, and a reference frequency (preferably at low frequency as it is least affected by spectral ageing, we use the LOFAR survey) we were able to determine the fixed magnetic field strengths for the set of target lobes. We also assumed the minimum and maximum Lorentz factor for the particle distribution to be 1 and 100000 respectively. We have assumed no protons and uniform filling factor. The estimated values for the field strengths are calculated at equipartition, although an X-ray study of edge-brightened radio sources (\citealt{Crostonetal2005}) attempted to evaluate the magnetic field strengths of lobes and the possible lobe particle population, and found that the magnetic field strength lies within 35\% of the equipartition value. This estimate was confirmed by~\cite{InesonandCroston2017} with a larger representative sample, where they observed the magnetic field strengths to have a median ratio of 0.4 of the equipartition value. Hence we assumed the lobe field strength values to be lower than equipartition by a factor of 0.4. These magnetic field estimates are affected by the statistical uncertainties on the input parameters but also by systematic uncertainties on our input assumptions and so we do not attempt to estimate magnetic field uncertainties or propagate them through our analysis. \par 
We also assumed the value of the injection index as 0.5 for the sources as earlier work eg.~\cite{Carillietal1991b} reported injection index values to be between 0.5 and 0.7. However other works have found steeper injection indices. \cite{Harwoodetal2013} found injection index values greater than 0.8 for two FR type-II sources, 3C 436 and 3C 300. They suggested several reasons for such behavior, such as poor model fit at low frequencies, inclusion of emission from strongly interacting jets, weak shock termination in FR type-II, and single injection particle distribution assumption. Furthermore, studies by \cite{Harwoodetal2015} and \cite{Harwoodetal2017} found similar steeper values of injection index, even when they had included more low-frequency data in their analysis. A recent investigation by~\cite{Mahatmaetal2019}, a study that attempts to solve the spectral age and dynamical age discrepancy problem by observing two powerful cluster center radio sources using high-resolution JVLA and deep XMM-Newton and Chandra observations, reports the injection index to be around 0.6. In the debate around the true injection index values, it should be noted that the studies conducted until now lack a representative radio source population that can help effectively constrain injection index value using high sensitivity and high resolution radio maps. Until such a population is available, it should be safe to assume a value of 0.5 for the injection index because this is the lowest possible value and has been widely used in the literature until now. The aim of this study is to examine the dispersion in the spectral age distribution of the targets. If the injection index was intrinsically steeper, this would shift the observed age distribution to lower values. The parameter information is summarized in Table \ref{tableinfo}.    
\subsection{\normalfont{\textit{Spectral Analysis}}}
\label{SA}
We perform the bulk of the spectral age fitting using the JP model due to its physical plausibility, less computational work, and the ability to provide us with the upper limits for the oldest recorded age (Section \ref{SAM}). However, for a few well resolved sources we also investigate the use of the KP and the Tribble models. For a maximum and minimum age range, flux density values were determined and a {$\chi^{2}$} test was performed as a check of the goodness-of-fit of the model, given by: 
\begin{equation} \label{eq9}
\\{\chi^{2}}=\sum_{\nu=1}^{N}\left(\frac{S_{i,\nu}-S_{model,\nu}}{{\Delta}S_{i,\nu}}\right)^{2}
\end{equation}   
where, at $N$ given frequencies $\nu$, $S_{i,\nu}$ is the observed flux density in region i, $S_{model,\nu}$ is the model flux, and ${\Delta}S_{i,\nu}$ is the total observed region uncertainty that depends on fractional flux calibration error (see \citealt{Harwoodetal2013} and the BRATS cookbook for details). The BRATS software package performs a grid search to look for the spectral age that best fits the model. By performing a broad search over a range of defined maximum and minimum ages, we can evaluate a best-fitting age over the grid. For more accurate age evaluation, the software automatically repeats the search for the age interval that produced the best fit in the previous cycle, until the desired accuracy is reached. Hence, under the selected target region we obtain a pixel by pixel age estimate which ultimately provides an age distribution for radio lobes. For our sample, we have calculated the source magnetic field strengths at 0.4 of equipartition (as discussed in section \ref{PD}) and have run the BRATS package command \textit{fitjpmodel}, \textit{fitkpmodel}, and \textit{fitjptribble} to perform the JP, the KP and the Tribble model fitting respectively.

\section{Results}
\label{R}
The sample consists of 5 FR type-I, 17 FR type-II, one head-tail source, and 5 sources with anomalous structures. For sources where a strong central core was present, the core flux was excluded from the age analysis as it is not described by the models of spectral ageing. The analysis evaluates model values by varying the parameter values for injection index, magnetic field strength, and the age intervals along with their respective redshifts.\par 
Before executing spectral age fitting, as an initial check, we generate the spectral index maps for the targets using the BRATS \textit{specindex} command. We are looking for discrepant spectral index values in order to identify any misalignment, identify regions where there are anomalies or artifacts around the source, and review the data quality before performing any computationally intensive spectral age fitting. The majority of the maps show the flattest spectral index value in the lobes for the locations which are, according to morphology, the hot-spots of the sources, and then the spectral index value steepens along the ejection axis towards the core, this is typical spectral behavior for FR type-II sources. We can also see that some targets show the flattest spectral index values near the core and continue towards both edges of the lobes gradually increasing the spectral index value. This behavior is typical for FR type-I sources. We expect to see similar patterns for our sample in terms of age distribution, depending on the type of FR morphology we observe and simultaneously confirming our observation about the structure of the source (see Appendix \ref{A1} for further discussion). Looking at the spectral index maps, we do not observe any anomalous data hence the radio maps are aligned properly and the quality of data is good enough to perform spectral analysis.\par
\begin{figure*}
	\begin{subfigure}{\columnwidth}
		\includegraphics[width=3.3in]{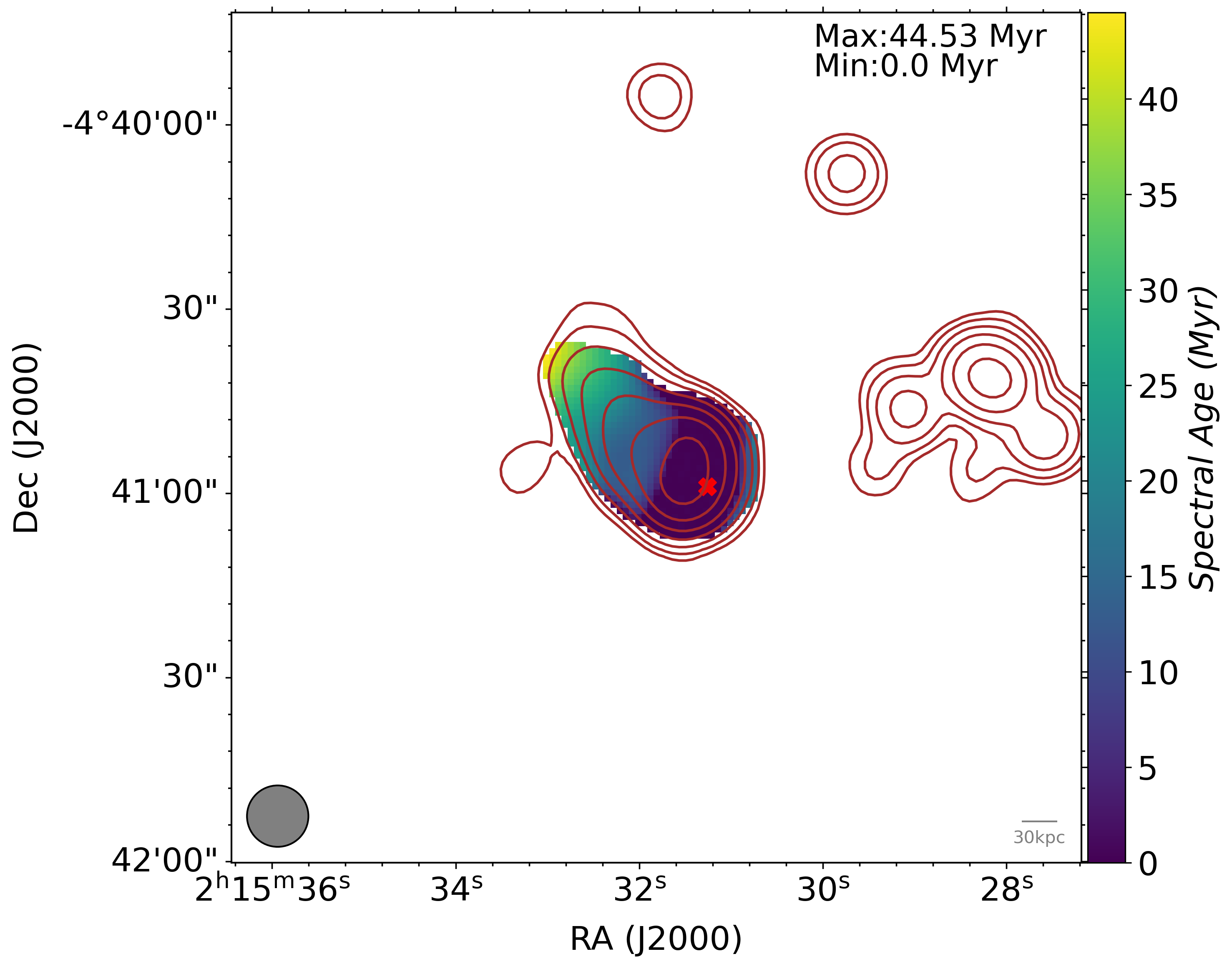}
		\caption{Target 1 (J021531.81-044050.9)}
		\label{t1a}
	\end{subfigure}
	\begin{subfigure}{\columnwidth}
		\includegraphics[width=3.3in]{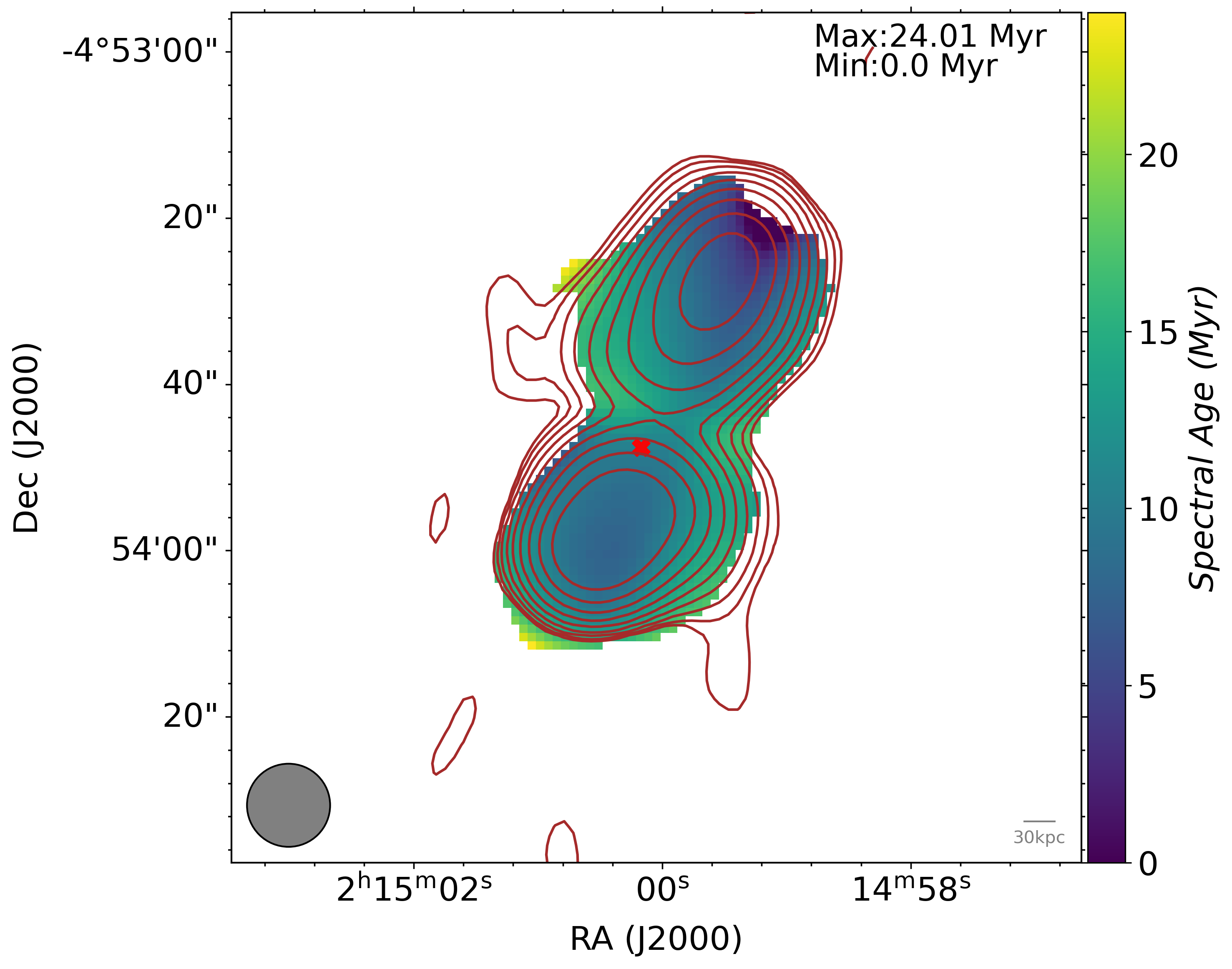}
		\caption{Target 2 (J021500.04-045346.3)}
		\label{t1b}
	\end{subfigure}	
	\begin{subfigure}{\columnwidth}
		\includegraphics[width=3.3in]{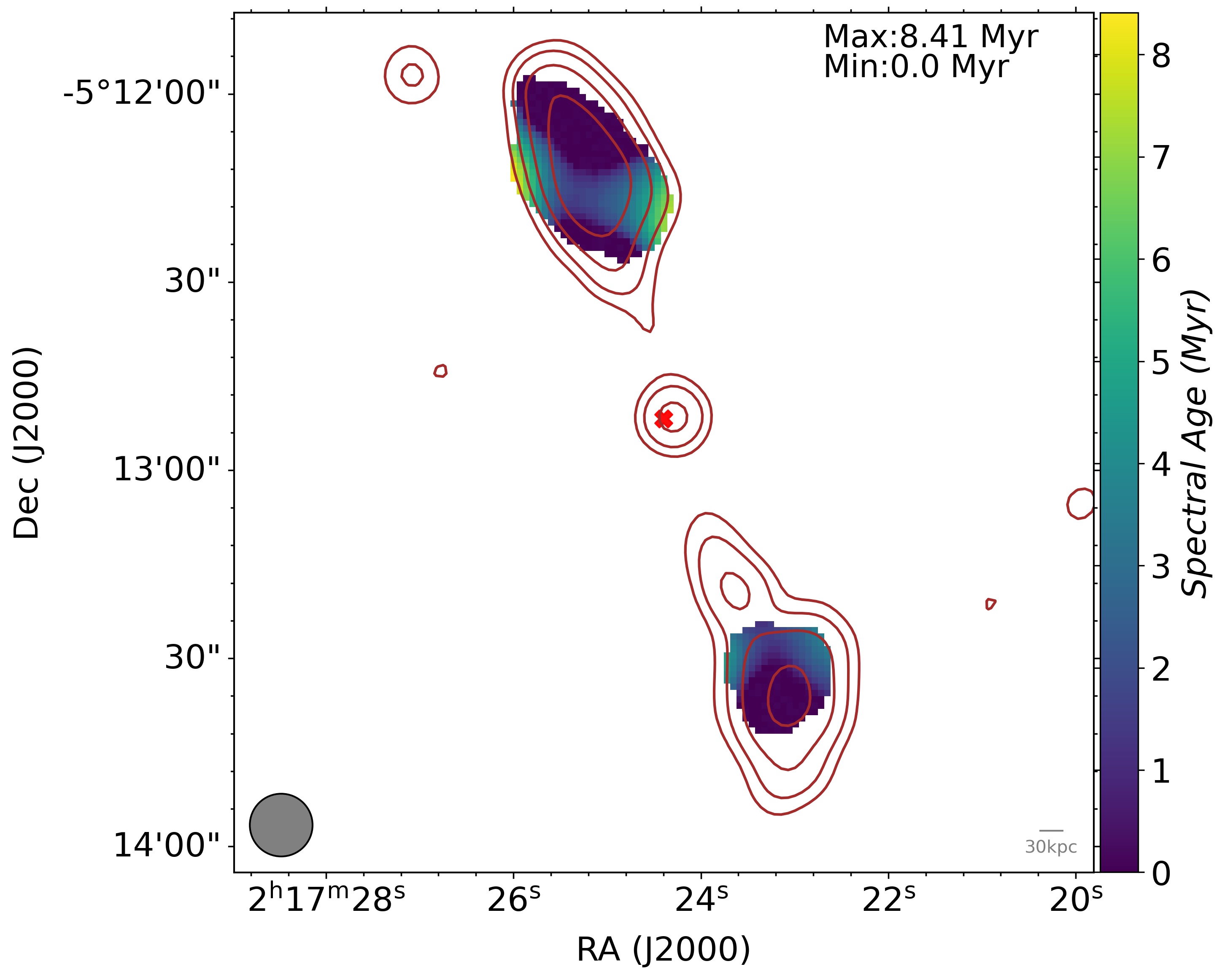}
		\caption{Target 3 (J021724.39-051255.5)}
		\label{t1c}
	\end{subfigure}
	\begin{subfigure}{\columnwidth}
		\includegraphics[width=3.3in]{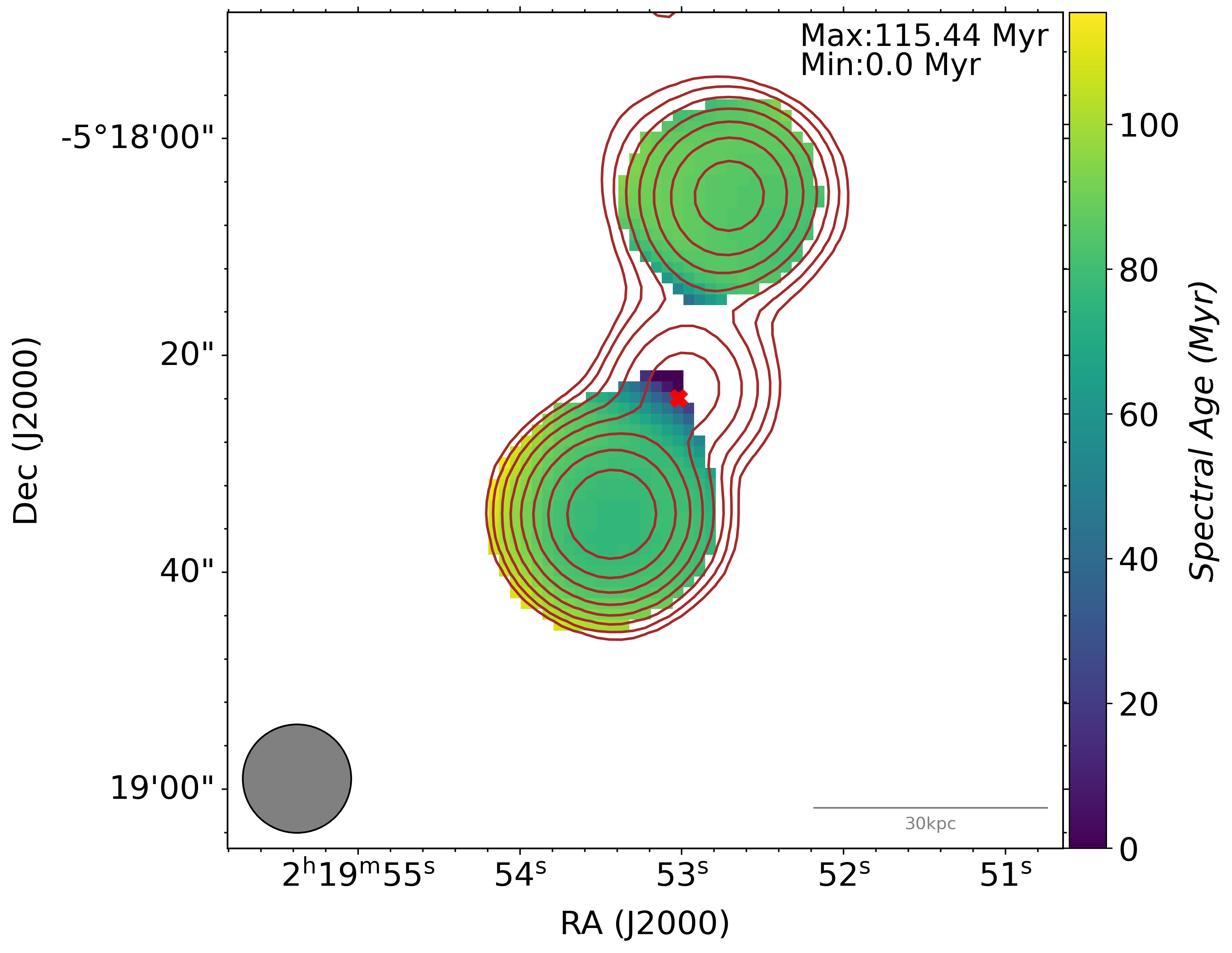}
		\caption{Target 4 (J021953.22-051826.9)}
		\label{t1d}
	\end{subfigure}		
	\caption{Spectral age maps of target 1 to target 4 with contours overlaid from the MeerKAT 1.2 GHz survey at 8.2 arcsec resolution and host galaxy position marked with a cross. The grey solid circle represents circular PSF beam of size 10 arcsec.}
	\label{t1}
\end{figure*}

\begin{figure*}
	\begin{subfigure}{\columnwidth}
		\includegraphics[width=3.3in]{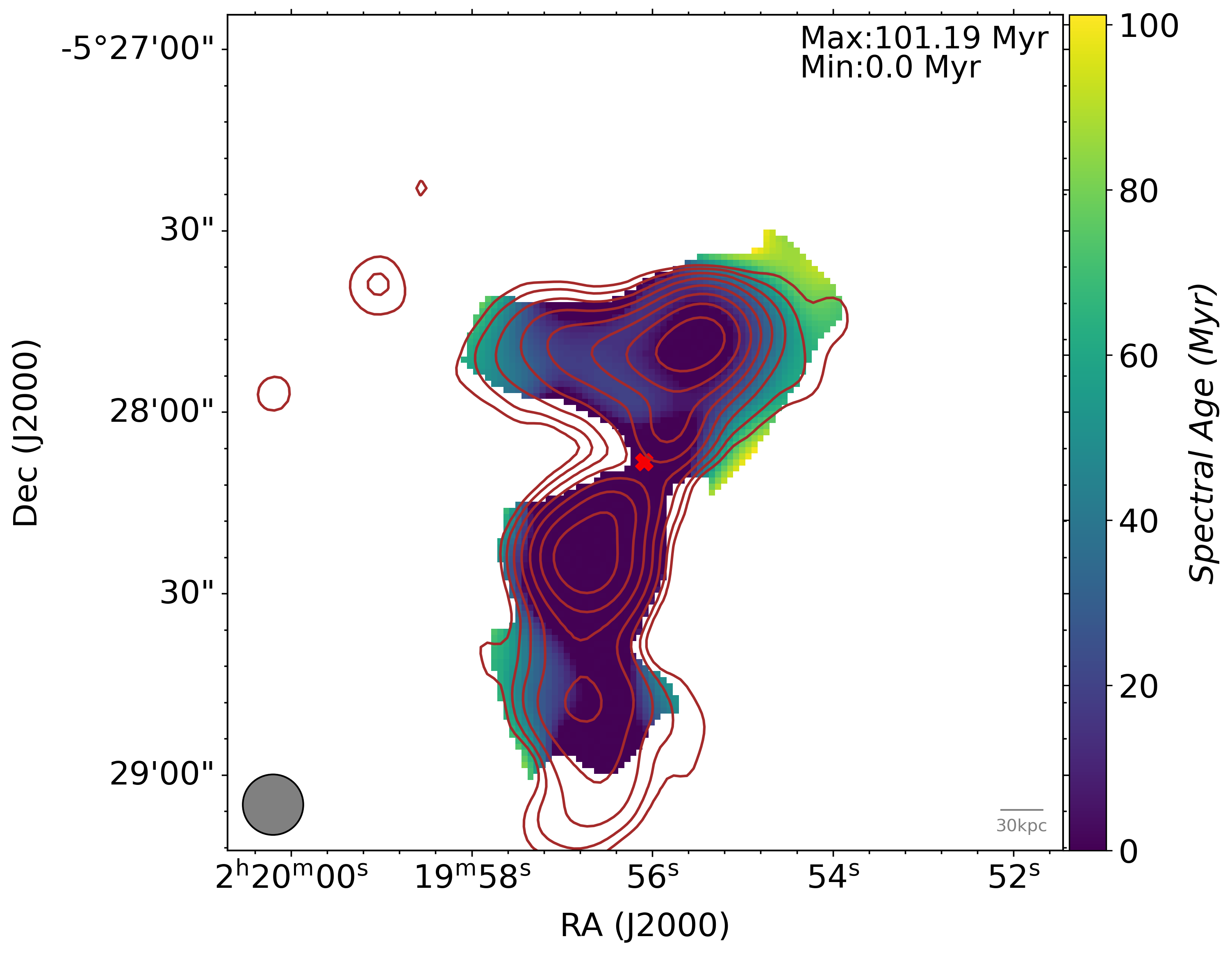}
		\caption{Target 5 (J021956.07-052803.3)}
		\label{t2a}
	\end{subfigure}
	\begin{subfigure}{\columnwidth}
		\includegraphics[width=3.3in]{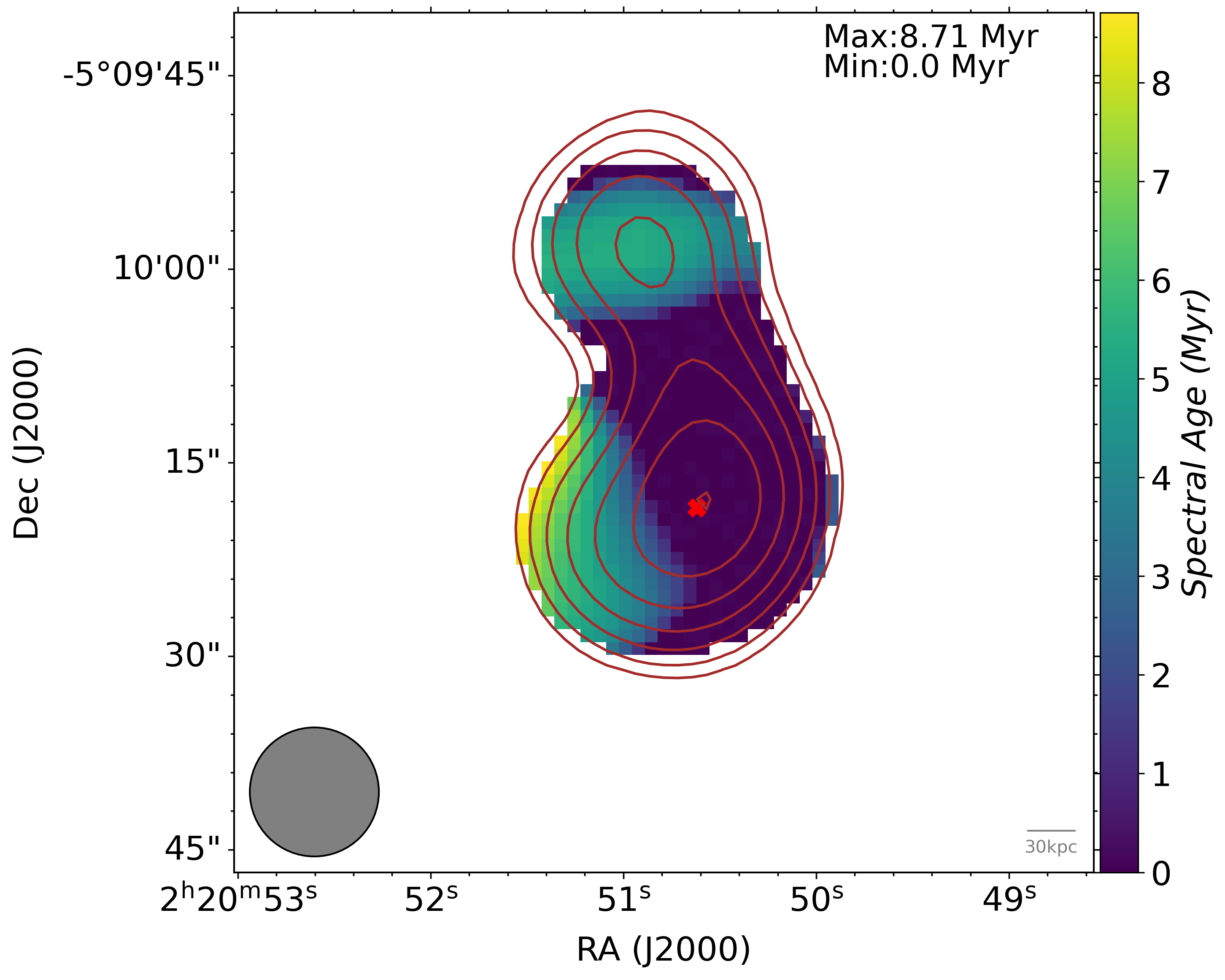}
		\caption{Target 6 (J022050.78-051013.4)}
		\label{t2b}
	\end{subfigure}	
	\begin{subfigure}{\columnwidth}
		\includegraphics[width=3.3in]{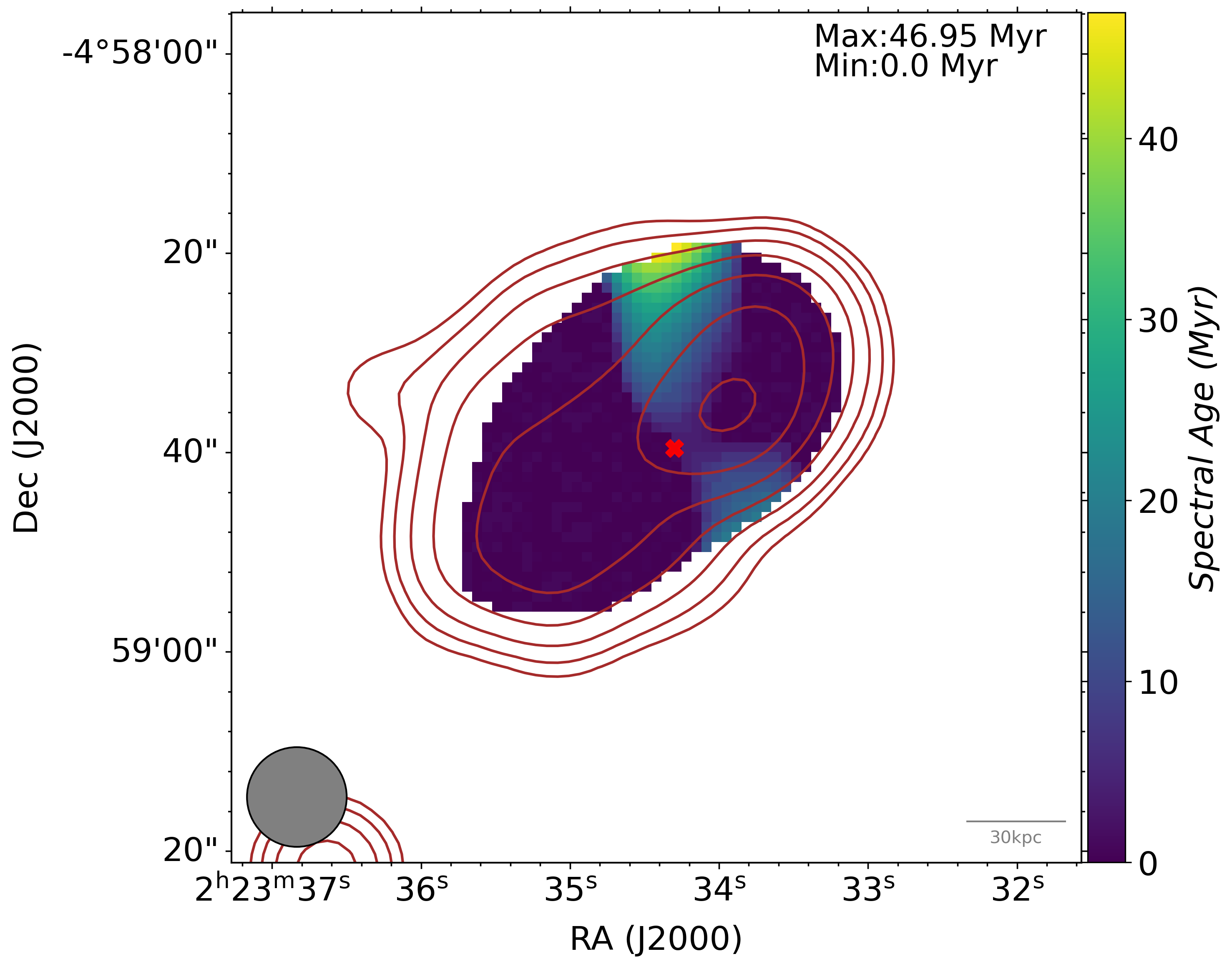}
		\caption{Target 7 (J022334.41-045838.5)}
		\label{t2c}
	\end{subfigure}
	\begin{subfigure}{\columnwidth}
		\includegraphics[width=3.3in]{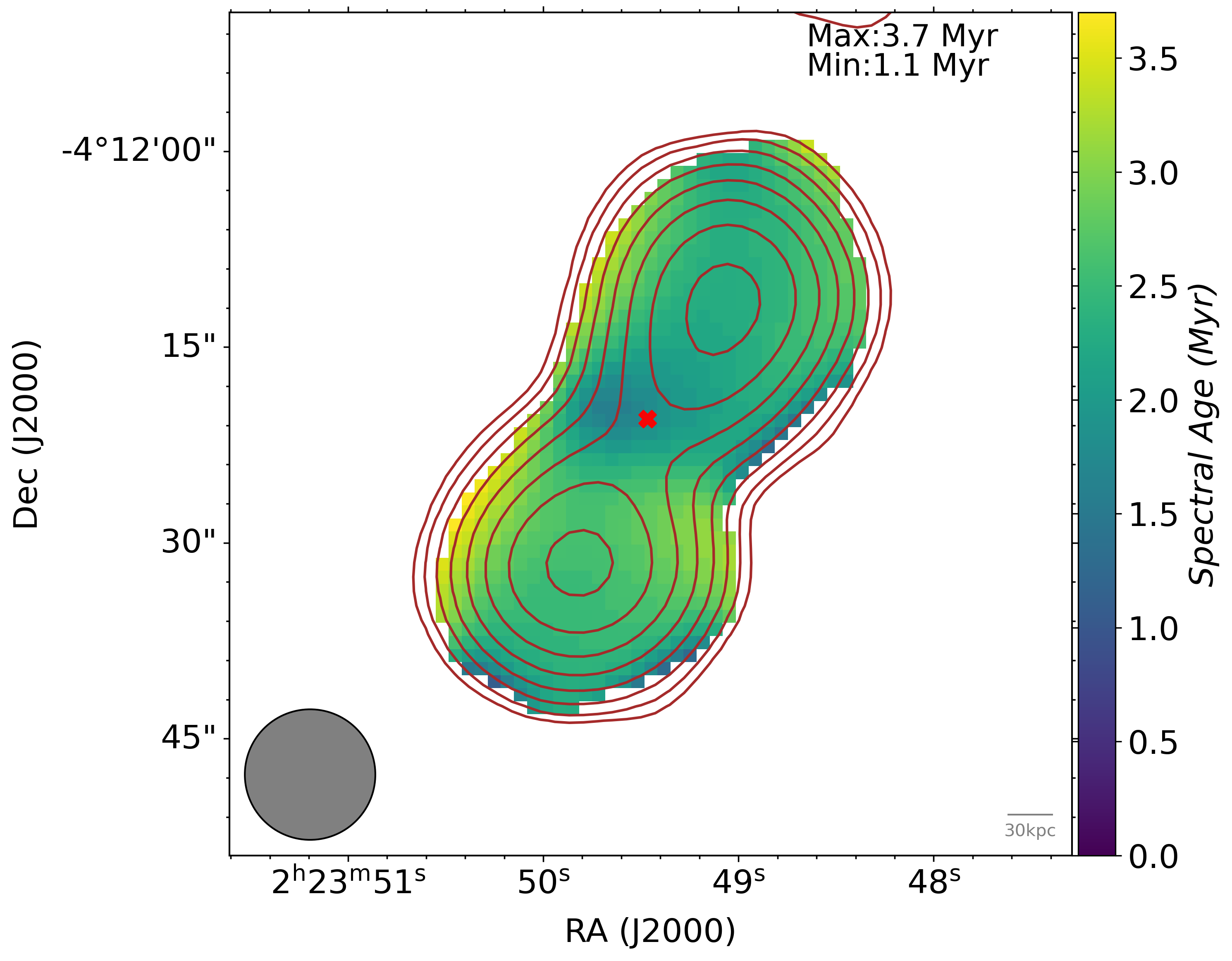}
		\caption{Target 8 (J022349.44-041221.6)}
		\label{t2d}
	\end{subfigure}		
	\caption{Spectral age maps of target 5 to target 8 with contours overlaid from the MeerKAT 1.2 GHz survey at 8.2 arcsec resolution and host galaxy position marked with a cross. The grey solid circle represents circular PSF beam of size 10 arcsec.}
	\label{t2}
\end{figure*}

\begin{figure*}
	\begin{subfigure}{\columnwidth}
		\includegraphics[width=3.3in]{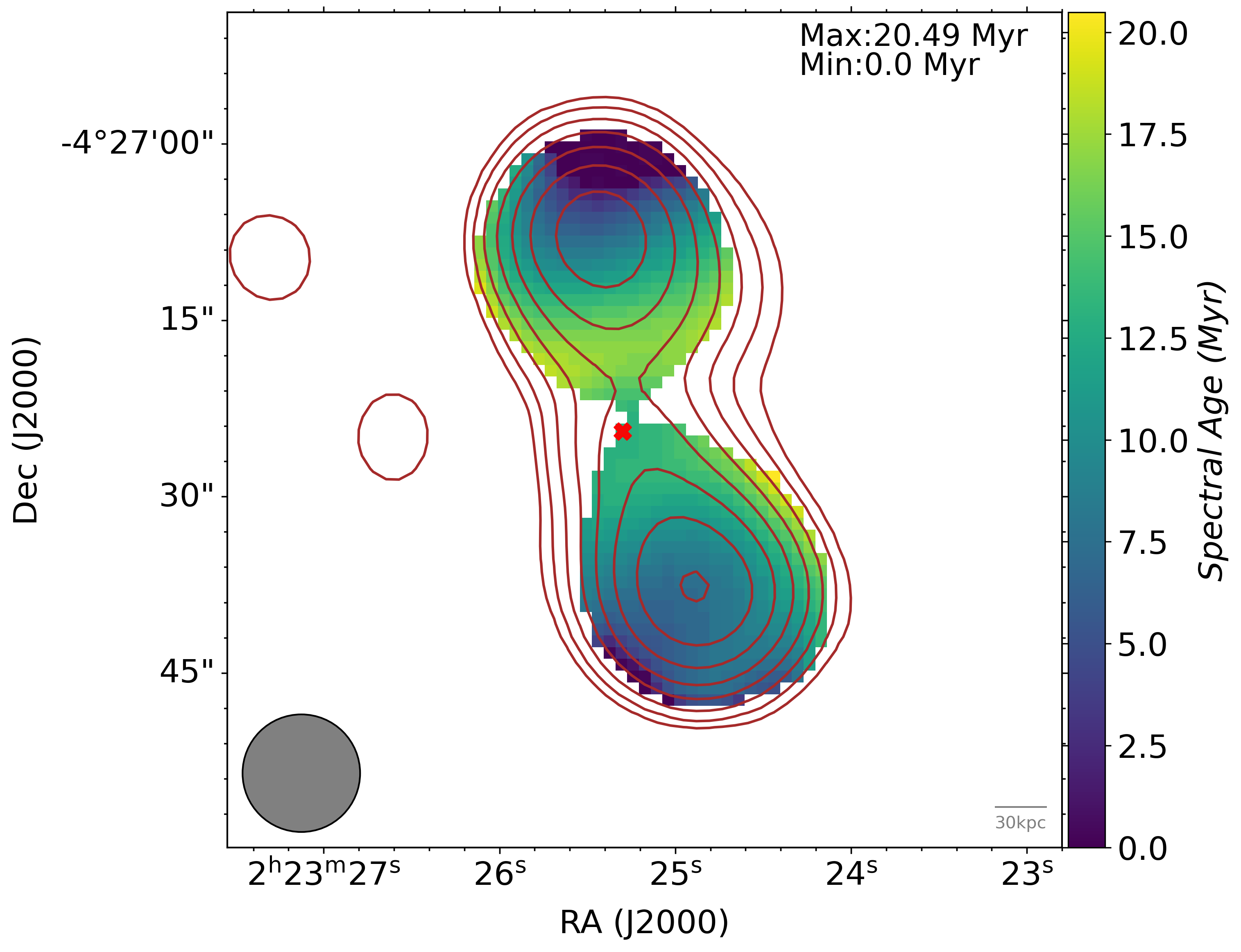}
		\caption{Target 9 (J022325.17-042724.3)}
		\label{t3a}
	\end{subfigure}
	\begin{subfigure}{\columnwidth}
		\includegraphics[width=3.3in]{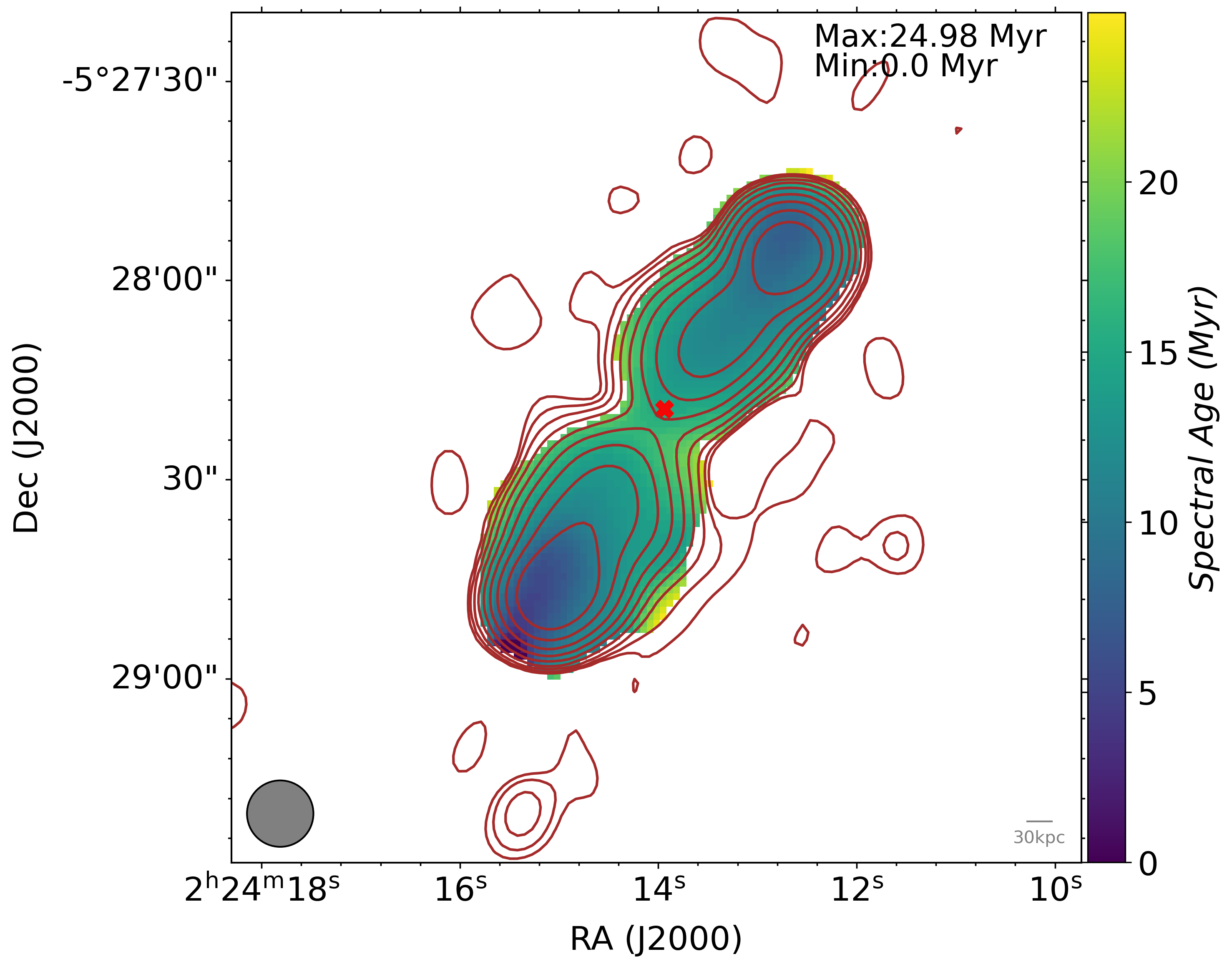}
		\caption{Target 10 (J022414.01-052823.6)}
		\label{t3b}
	\end{subfigure}	
	\begin{subfigure}{\columnwidth}
		\includegraphics[width=3.3in]{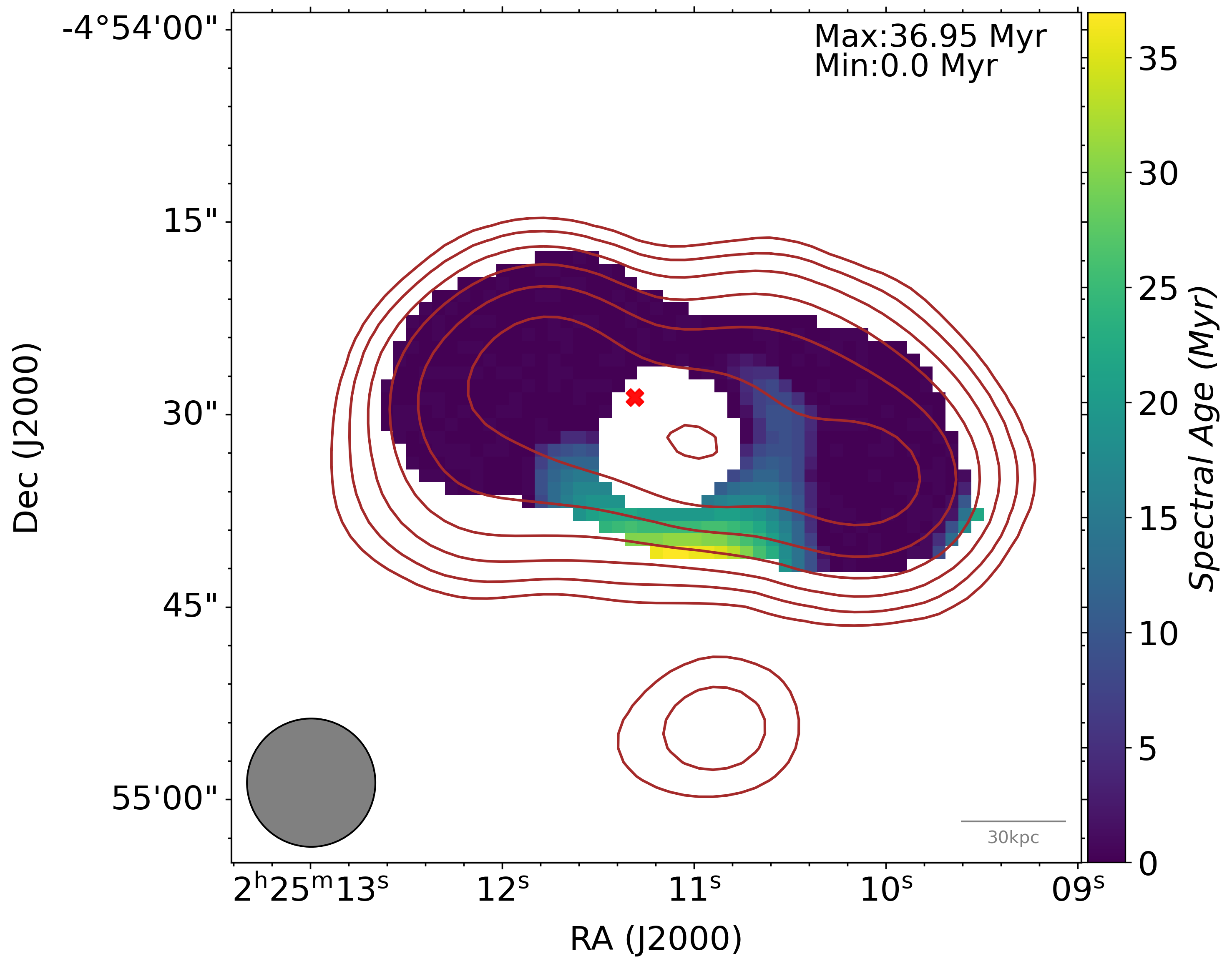}
		\caption{Target 11 (J022511.19-045431.7)}
		\label{t3c}
	\end{subfigure}
	\begin{subfigure}{\columnwidth}
		\includegraphics[width=3.3in]{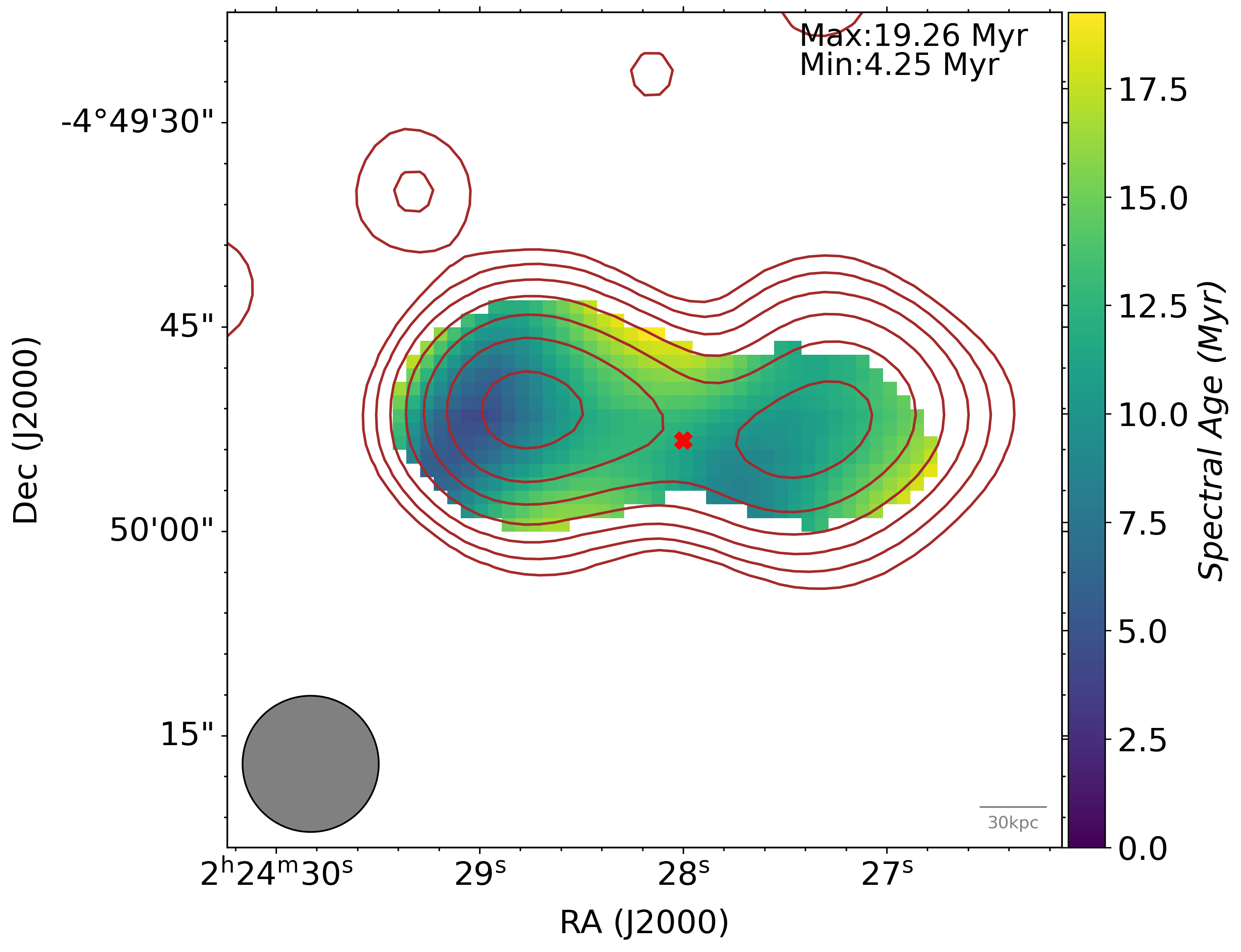}
		\caption{Target 12 (J022428.18-044952.5)}
		\label{t3d}
	\end{subfigure}		
	\caption{Spectral age maps of target 9 to target 12 with contours overlaid from the MeerKAT 1.2 GHz survey at 8.2 arcsec resolution and host galaxy position marked with a cross. The grey solid circle represents circular PSF beam of size 10 arcsec.}
	\label{t3}
\end{figure*}

\begin{figure*}
	\begin{subfigure}{\columnwidth}
		\includegraphics[width=3.3in]{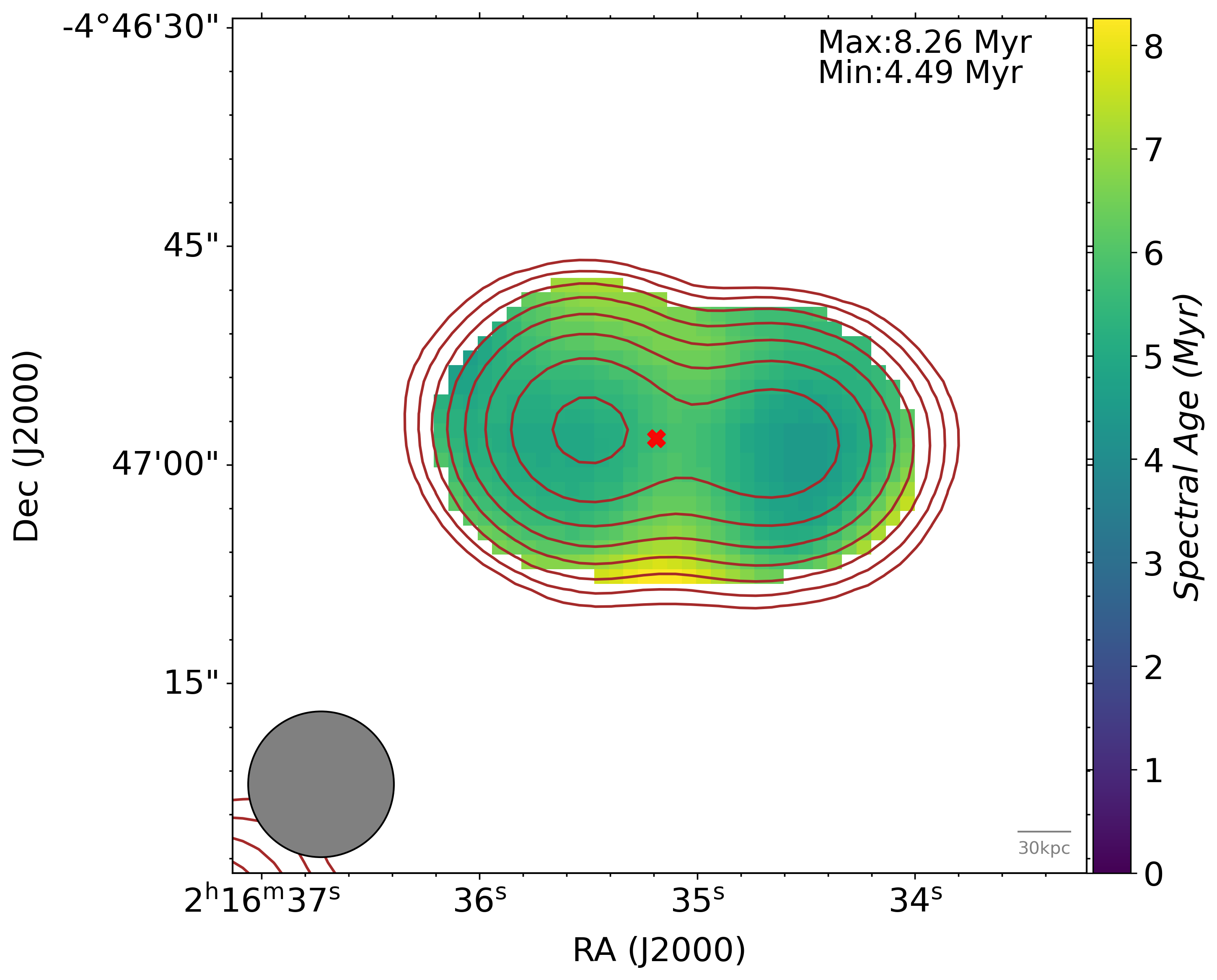}
		\caption{Target 13 (J021635.17-044658.6)}
		\label{t4a}
	\end{subfigure}
	\begin{subfigure}{\columnwidth}
		\includegraphics[width=3.3in]{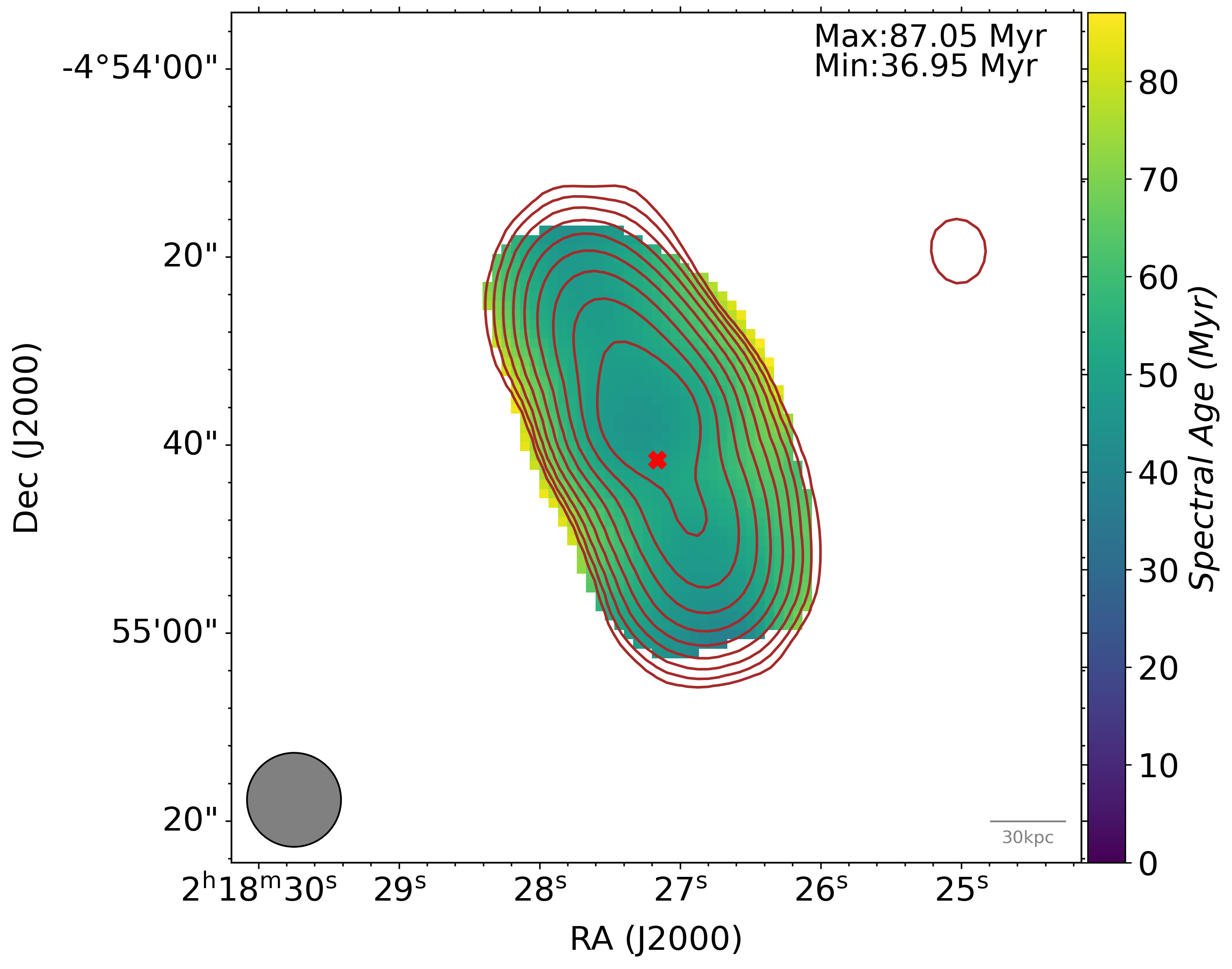}
		\caption{Target 14 (J021827.16-045439.2)}
		\label{t4b}
	\end{subfigure}	
	\begin{subfigure}{\columnwidth}
		\includegraphics[width=3.3in]{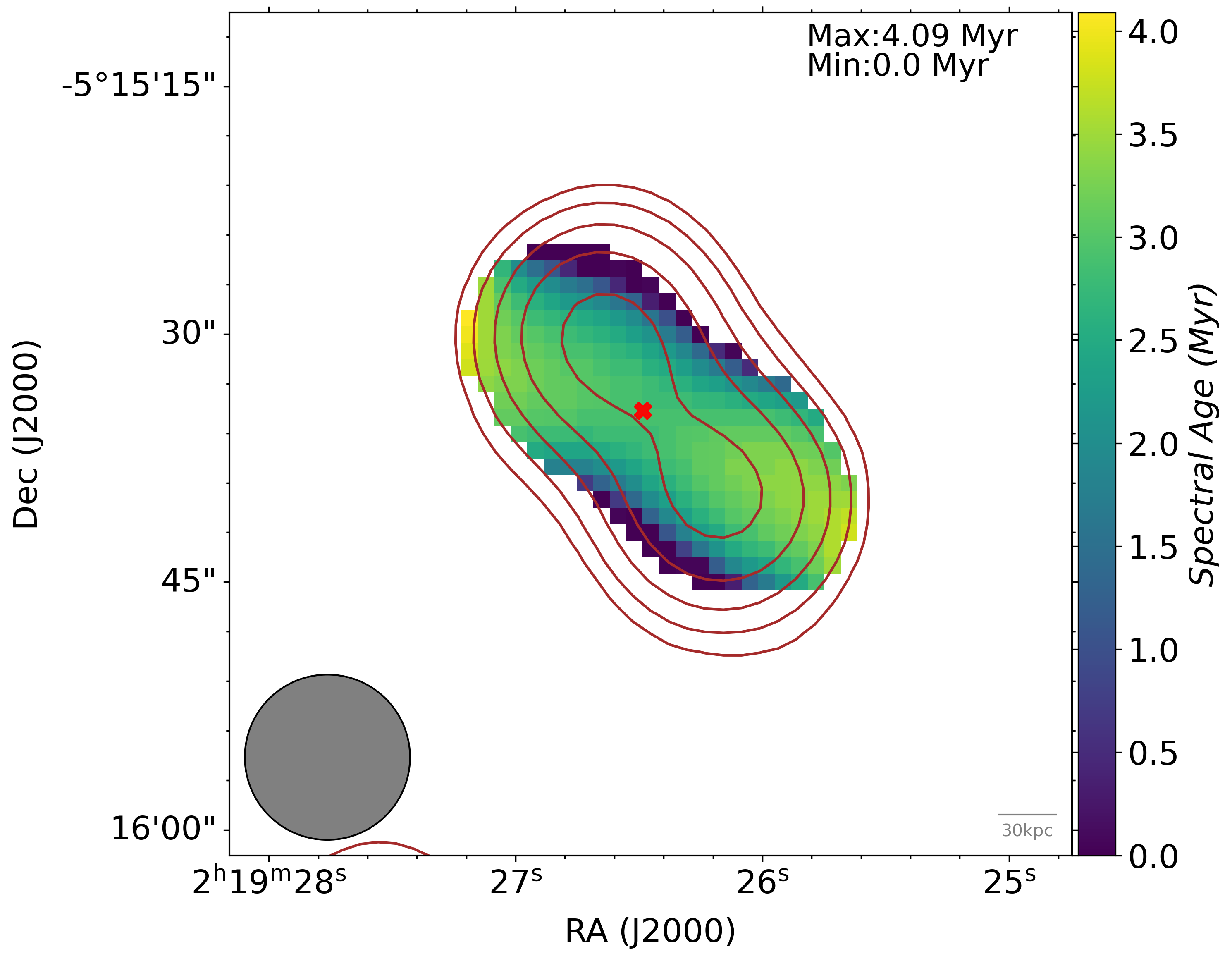}
		\caption{Target 15 (J021926.45-051536.0)}
		\label{t4c}
	\end{subfigure}
	\begin{subfigure}{\columnwidth}
		\includegraphics[width=3.3in]{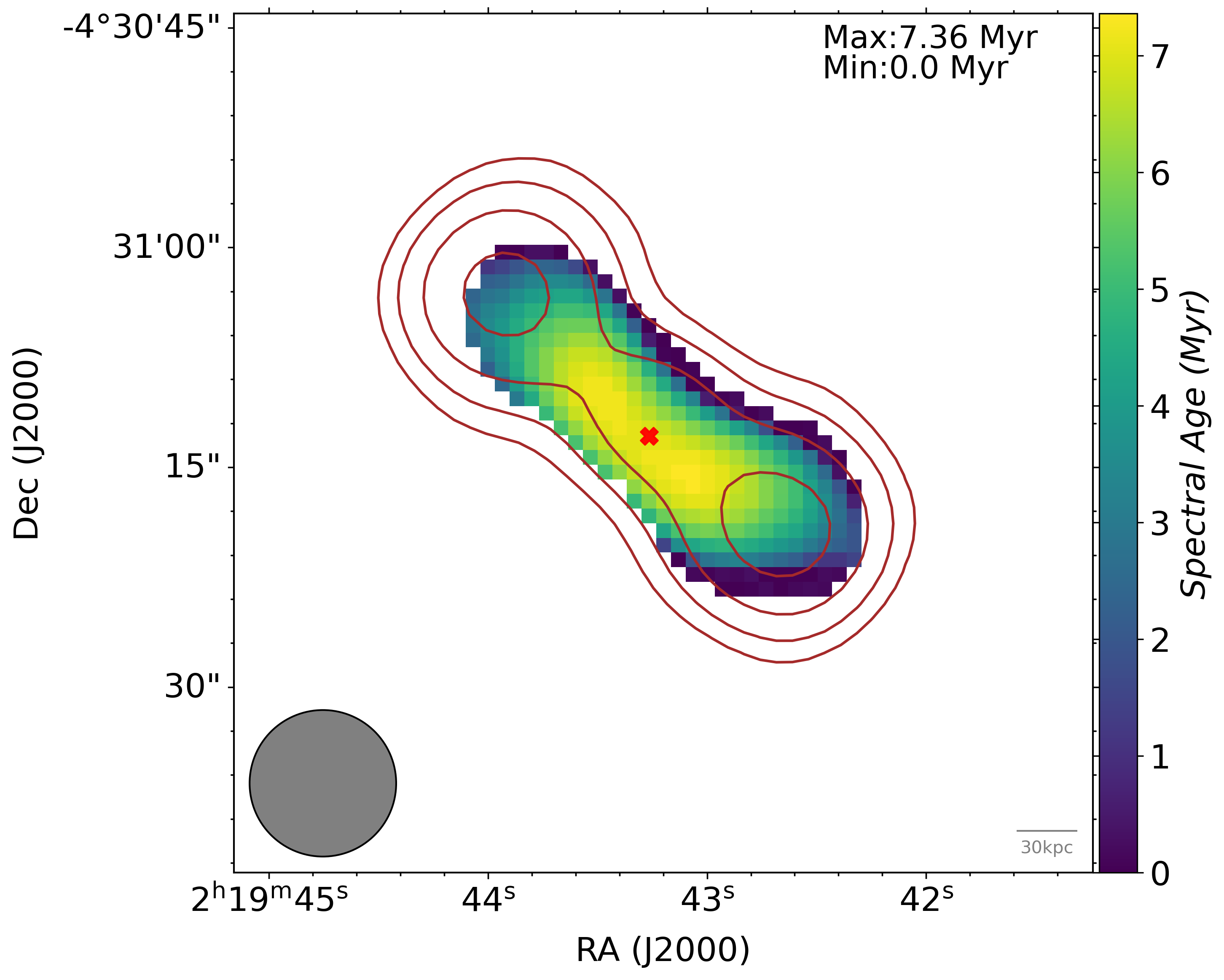}
		\caption{Target 16 (J021943.19-043113.3)}
		\label{t4d}
	\end{subfigure}		
	\caption{Spectral age maps of target 13 to target 16 with contours overlaid from the MeerKAT 1.2 GHz survey at 8.2 arcsec resolution and host galaxy position marked with a cross. The grey solid circle represents circular PSF beam of size 10 arcsec.}
	\label{t4}
\end{figure*}

\begin{figure*}
	\begin{subfigure}{\columnwidth}
		\includegraphics[width=3.3in]{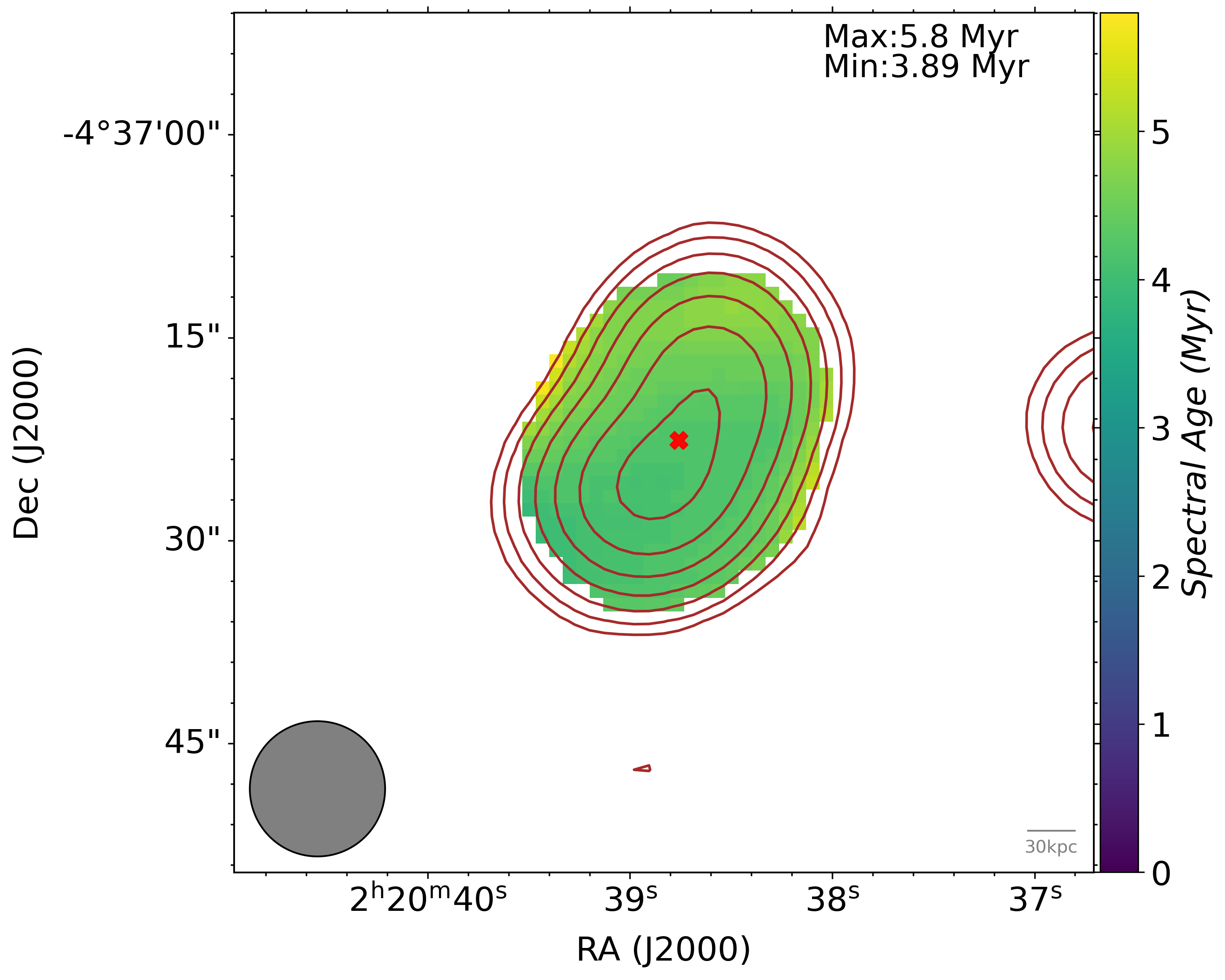}
		\caption{Target 17 (J022038.83-043722.7)}
		\label{t5a}
	\end{subfigure}
	\begin{subfigure}{\columnwidth}
		\includegraphics[width=3.3in]{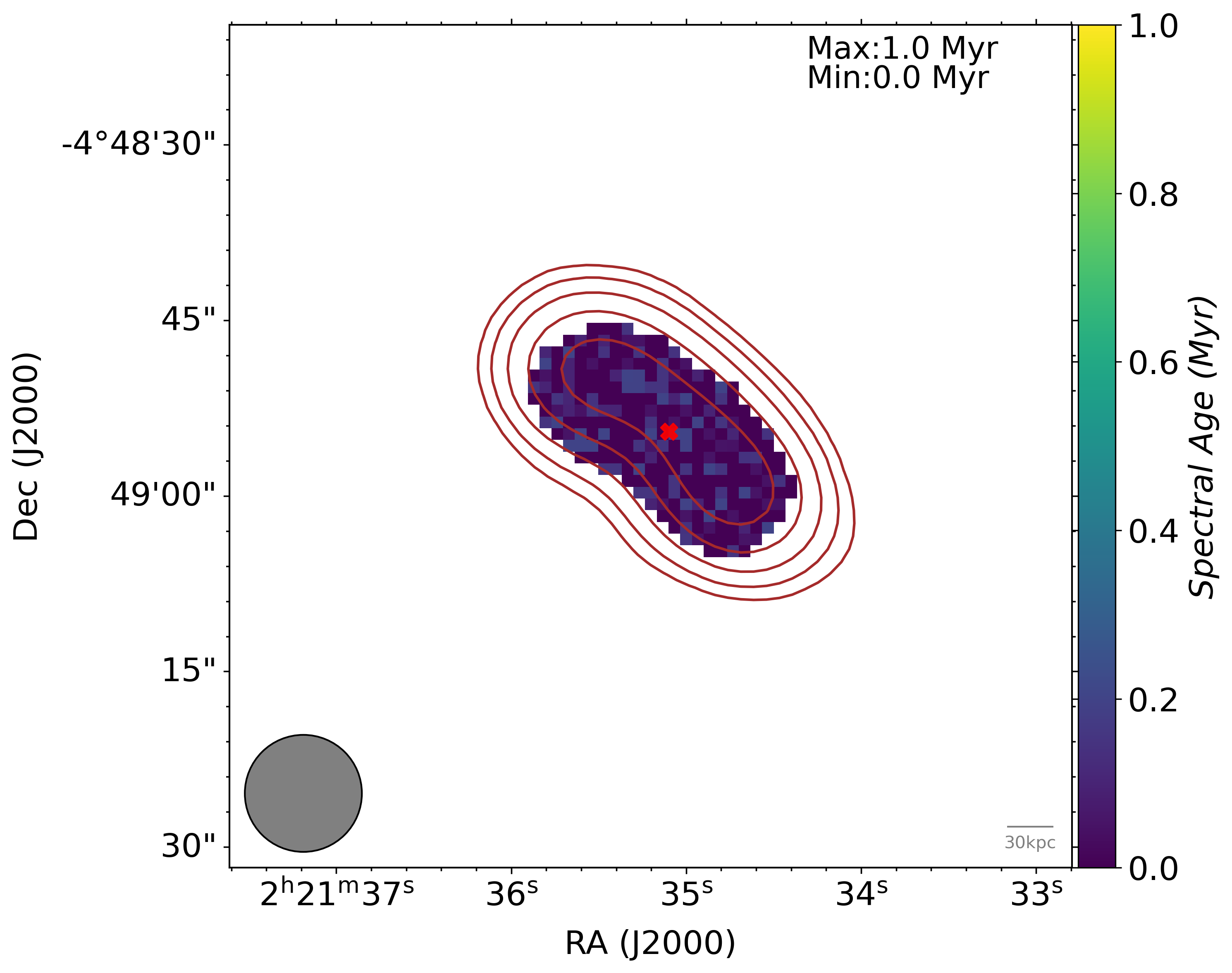}
		\caption{Target 18 (J022135.20-044855.7)}
		\label{t5b}
	\end{subfigure}	
	\begin{subfigure}{\columnwidth}
		\includegraphics[width=3.3in]{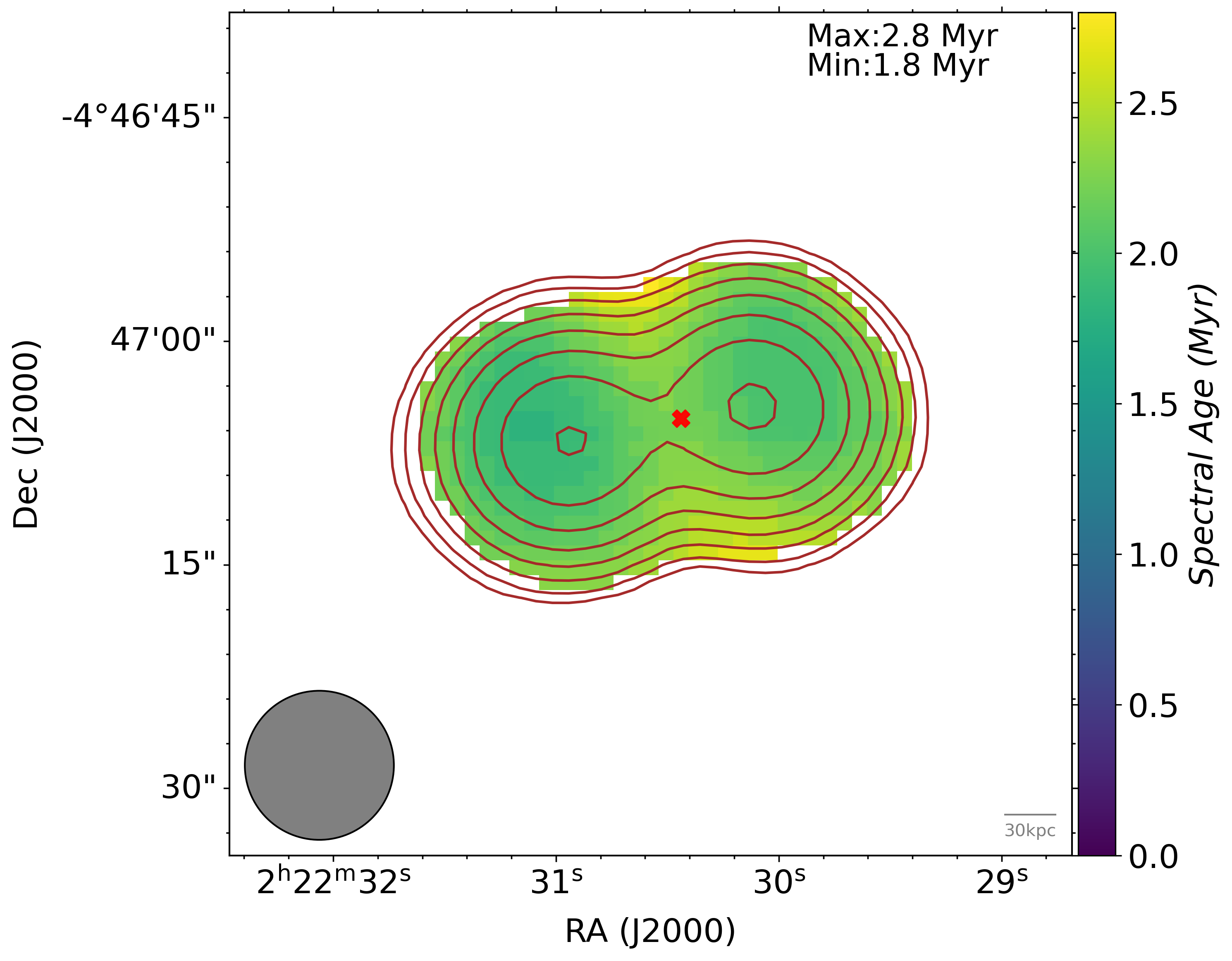}
		\caption{Target 19 (J022230.57-044706.2)}
		\label{t5c}
	\end{subfigure}
	\begin{subfigure}{\columnwidth}
		\includegraphics[width=3.3in]{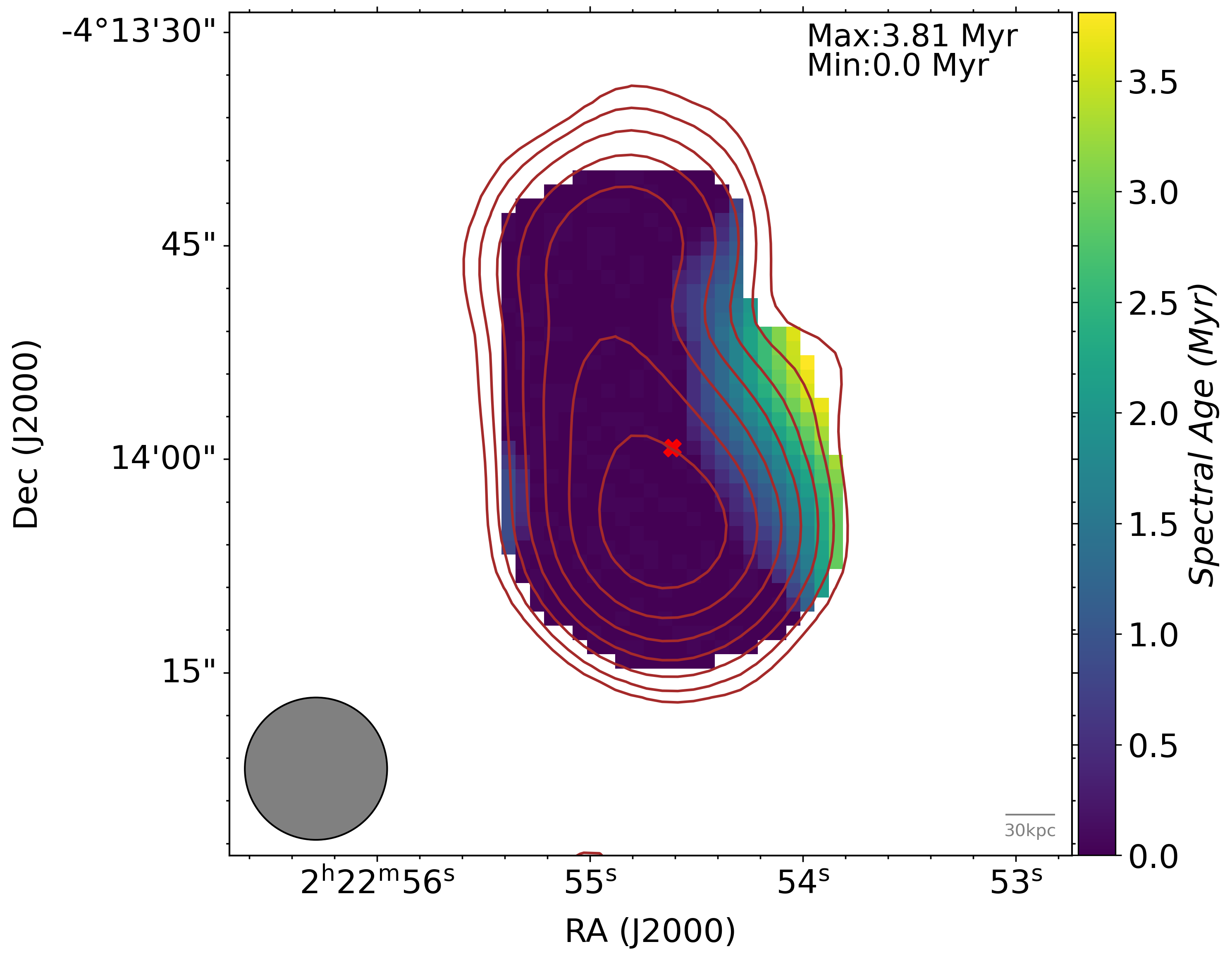}
		\caption{Target 20 (J022254.71-041358.2)}
		\label{t5d}
	\end{subfigure}		
	\caption{Spectral age maps of target 17 to target 20 with contours overlaid from the MeerKAT 1.2 GHz survey at 8.2 arcsec resolution and host galaxy position marked with a cross. The grey solid circle represents circular PSF beam of size 10 arcsec.}
	\label{t5}
\end{figure*}

\begin{figure*}
	\begin{subfigure}{\columnwidth}
		\includegraphics[width=3.3in]{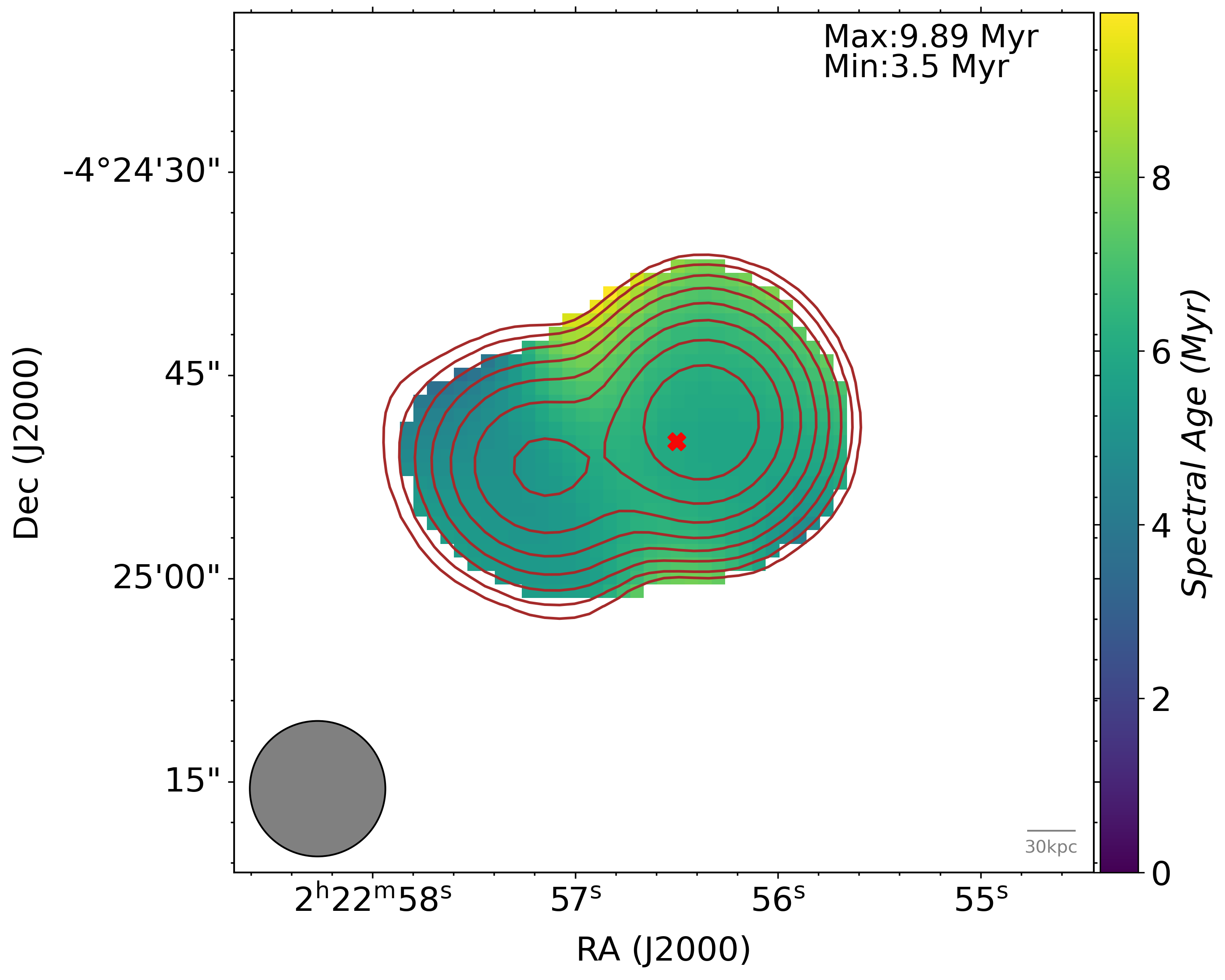}
		\caption{Target 21 (J022256.56-042449.9)}
		\label{t6a}
	\end{subfigure}
	\begin{subfigure}{\columnwidth}
		\includegraphics[width=3.3in]{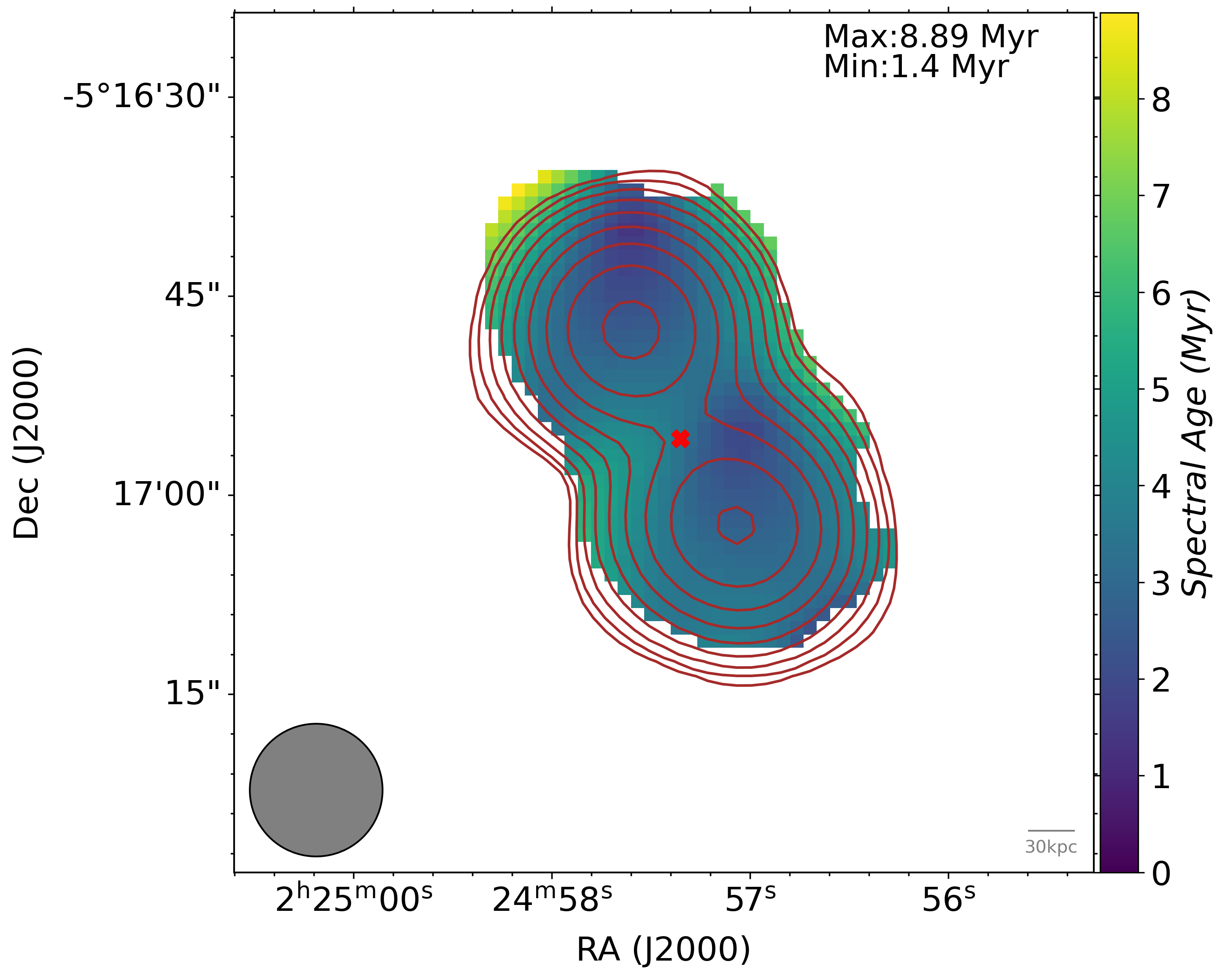}
		\caption{Target 22 (J022457.43-051656.0)}
		\label{t6b}
	\end{subfigure}	
	\begin{subfigure}{\columnwidth}
		\includegraphics[width=3.3in]{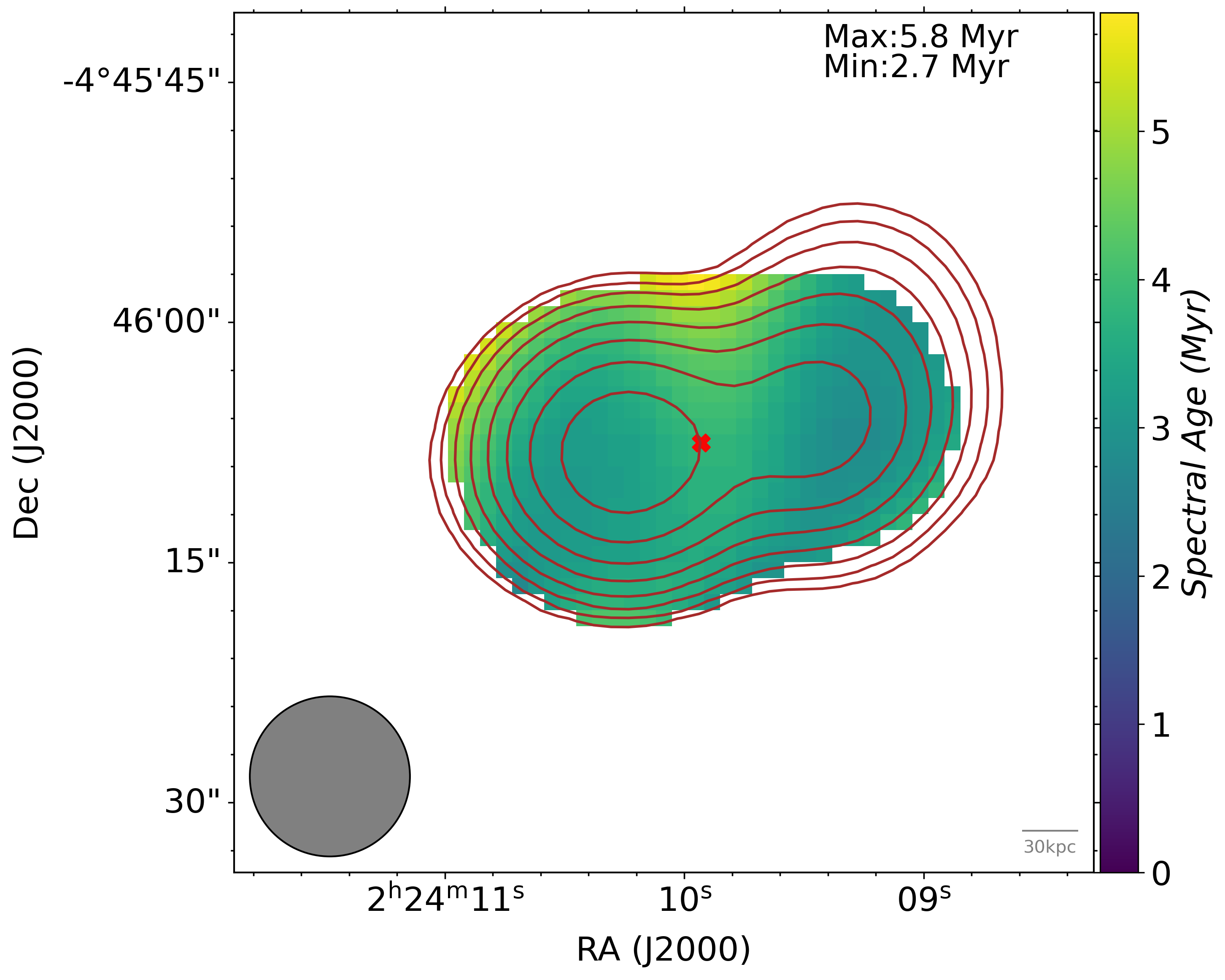}
		\caption{Target 23 (J022410.08-044607.5)}
		\label{t6c}
	\end{subfigure}
	\begin{subfigure}{\columnwidth}
		\includegraphics[width=3.3in]{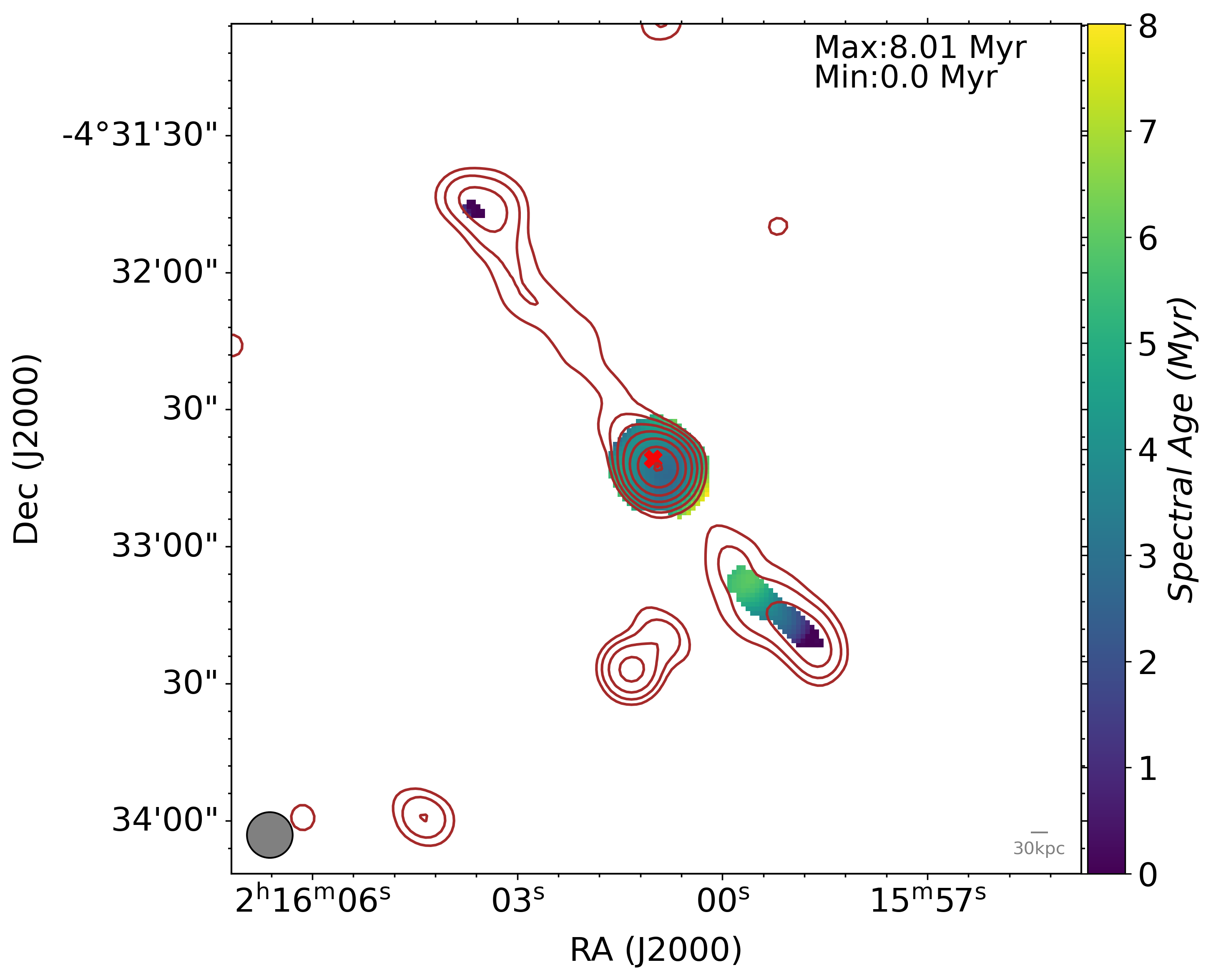}
		\caption{Target 24 (J021600.96-043238.5)}
		\label{t6d}
	\end{subfigure}		
	\caption{Spectral age maps of target 21 to target 24 with contours overlaid from the MeerKAT 1.2 GHz survey at 8.2 arcsec resolution and host galaxy position marked with a cross. The grey solid circle represents circular PSF beam of size 10 arcsec.}
	\label{t6}
\end{figure*}

\begin{figure*}
	\begin{subfigure}{\columnwidth}
		\includegraphics[width=3.3in]{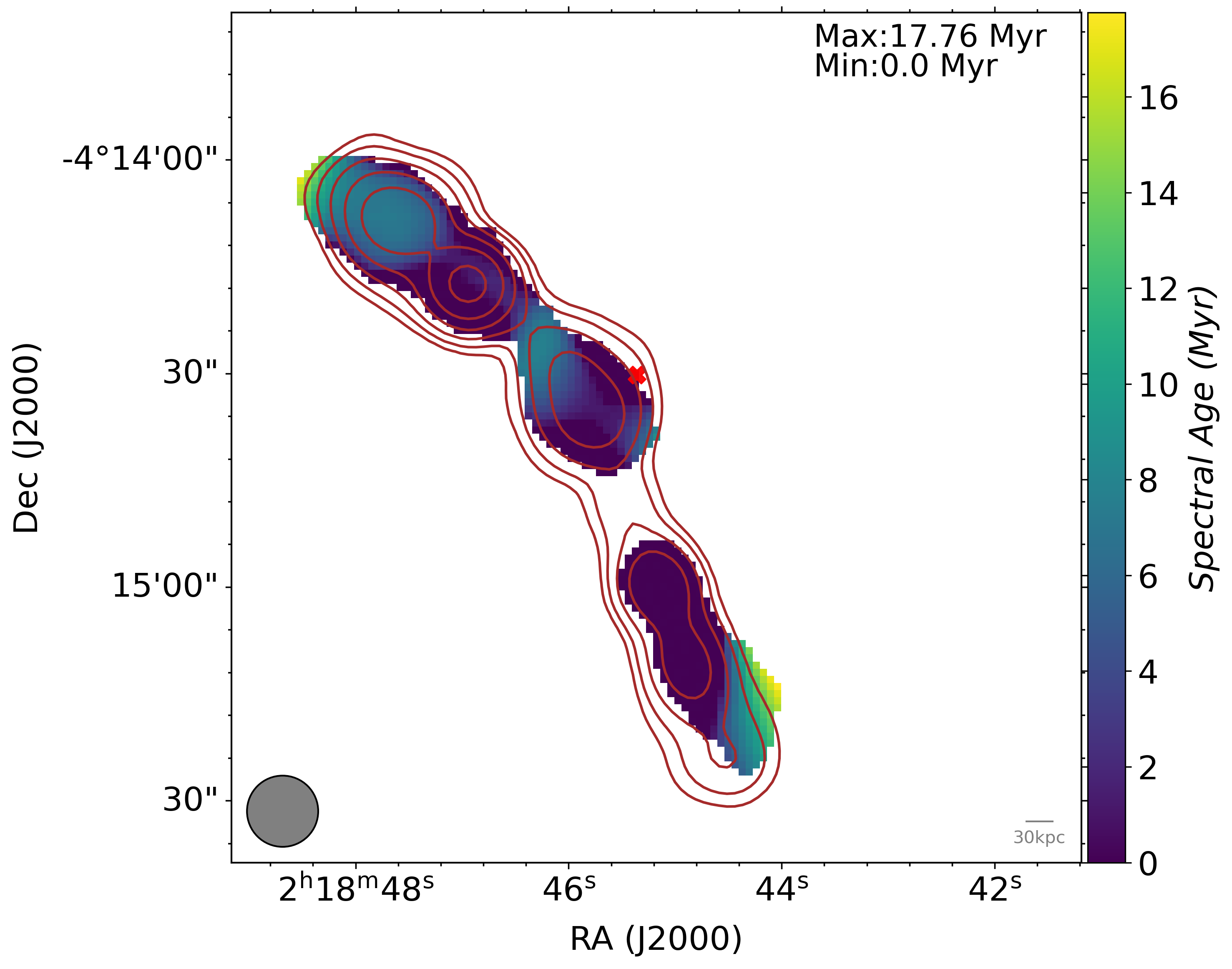}
		\caption{Target 25 (J021845.17-041438.9)}
		\label{t7a}
	\end{subfigure}
	\begin{subfigure}{\columnwidth}
		\includegraphics[width=3.3in]{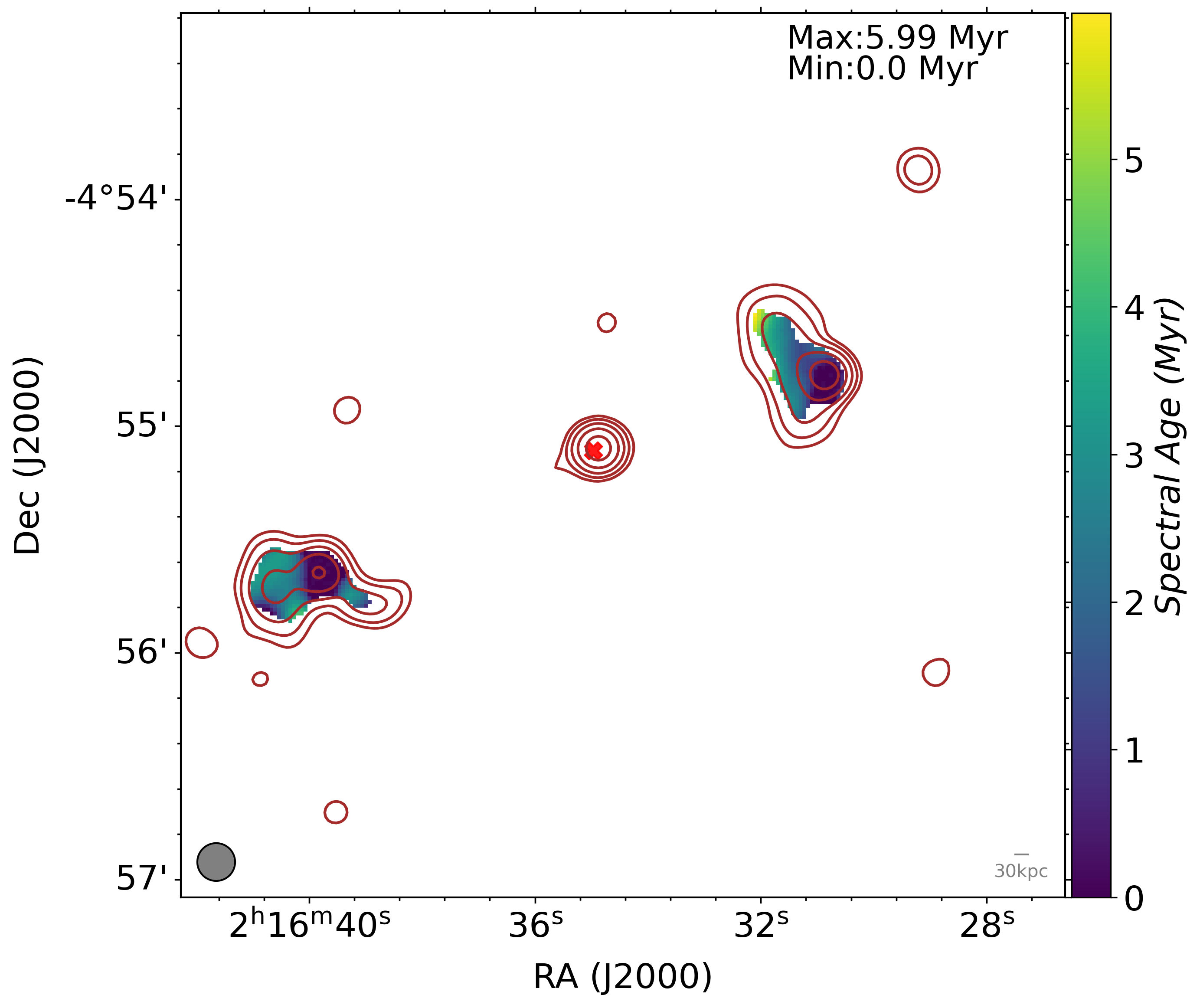}
		\caption{Target 26 (J021634.43-045507.6)}
		\label{t7b}
	\end{subfigure}	
	\begin{subfigure}{\columnwidth}
		\includegraphics[width=3.3in]{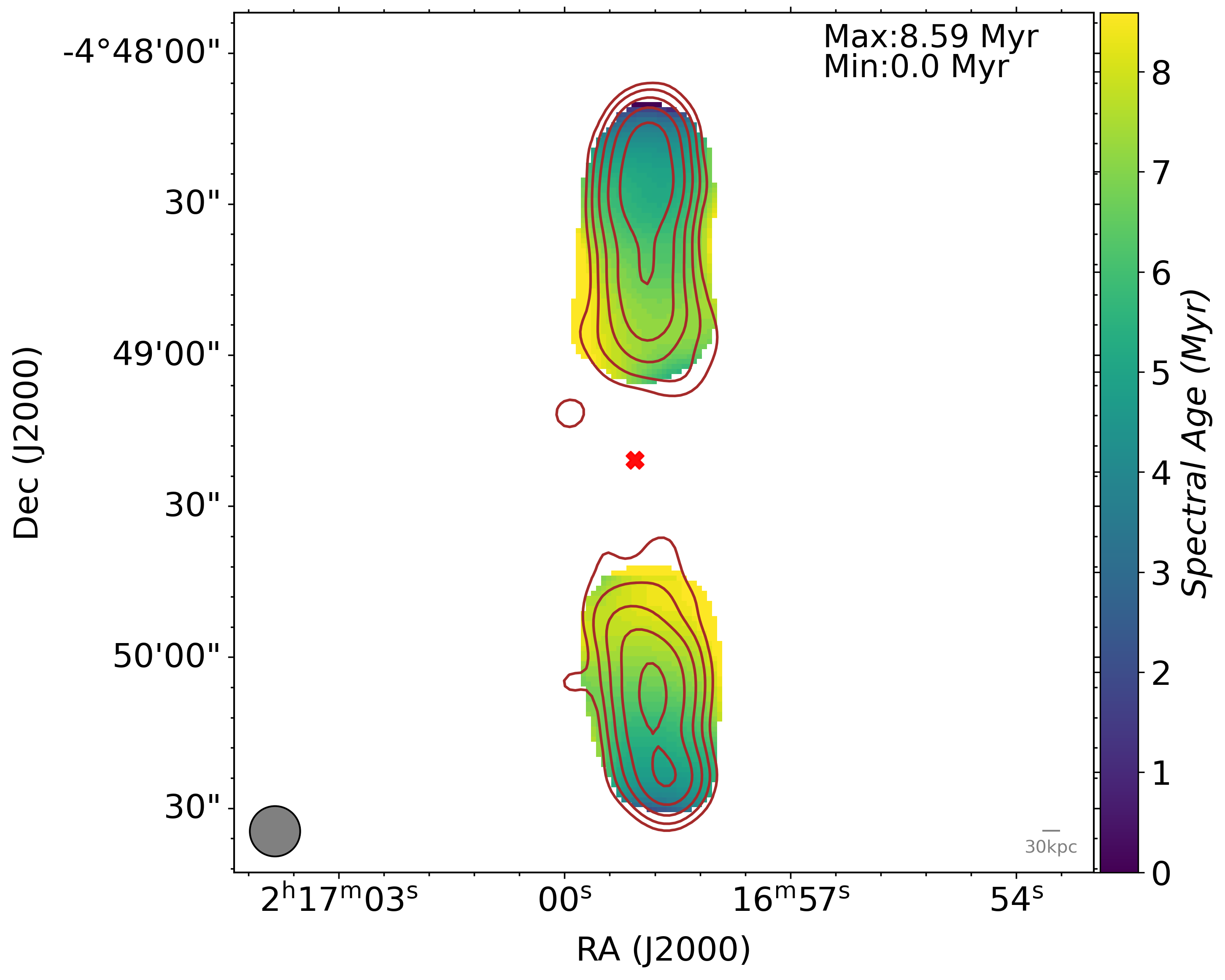}
		\caption{Target 27 (J021658.68-044917.3)}
		\label{t7c}
	\end{subfigure}
	\begin{subfigure}{\columnwidth}
		\includegraphics[width=3.3in]{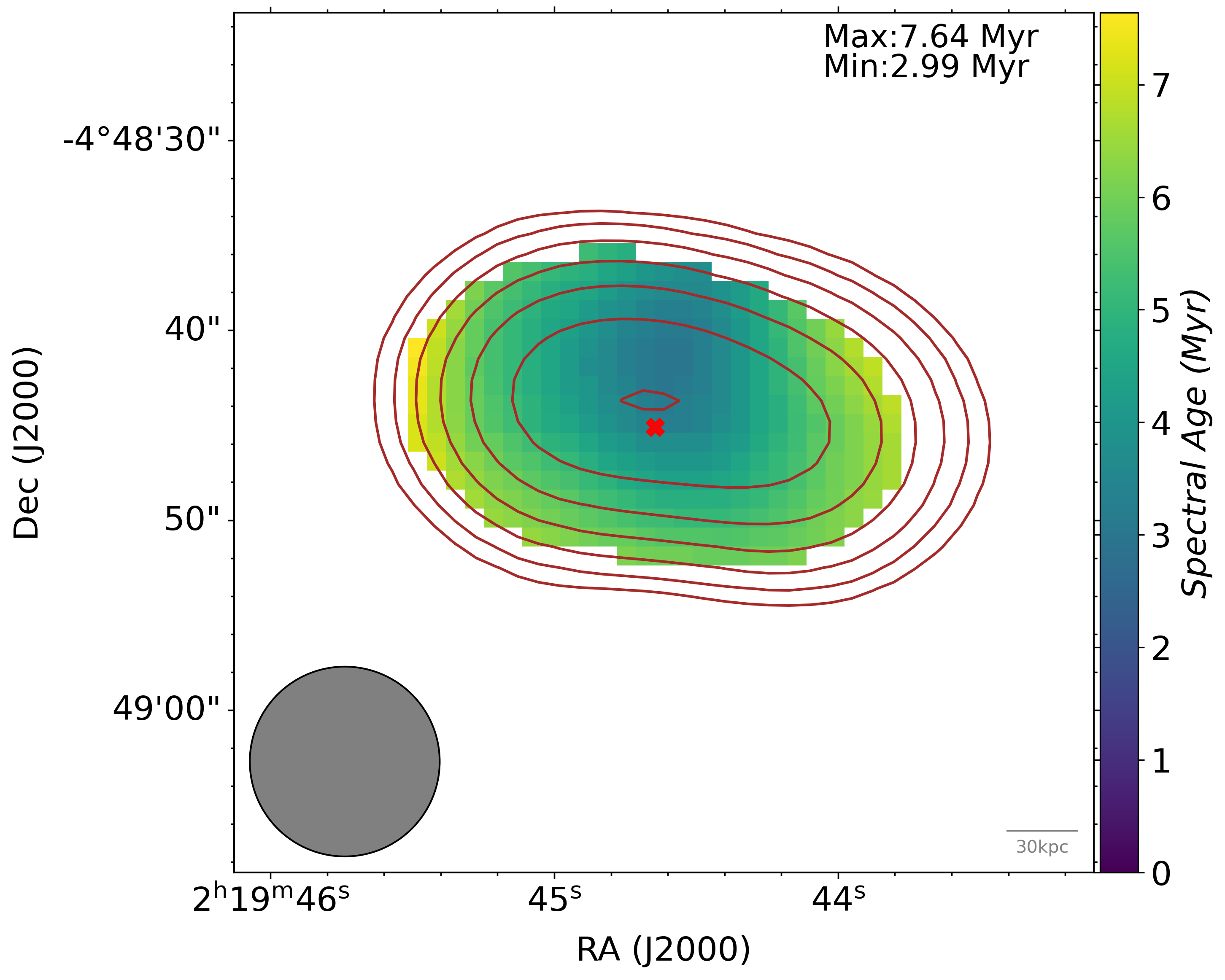}
		\caption{Target 28 (J021944.61-044845.9)}
		\label{t7d}
	\end{subfigure}		
	\caption{Spectral age maps of target 25 to target 28 with contours overlaid from the MeerKAT 1.2 GHz survey at 8.2 arcsec resolution and host galaxy position marked with a cross. The grey solid circle represents circular PSF beam of size 10 arcsec.}
	\label{t7}
\end{figure*}

 
Using the spectral fitting commands in BRATS, we initially set the age range between 0 and 200 Myr, the magnetic field strength (Table \ref{tableinfo}) and the injection index of 0.5 prior to the execution. We investigated the use of the KP and the Tribble models by  fitting both them and the JP model to two large, bright targets (10 and 14) and found (Table \ref{tablemodel}) that the maximum ages returned by the KP and the Tribble model were similar to but slightly larger than JP. Hence, we use the JP model only in what follows. The results of the JP model fitting are summarized in Table \ref{tableresult}. Figure \ref{t1}-\ref{t7} presents the JP spectral age maps for the selected targets, with contours overlaid from the MeerKAT survey (at 8.2 arcsec) and a cross representing the host galaxy position. We choose to use the MeerKAT survey contours as it is the most sensitive for the given data set. From here on we will refer to sources of the sample by their target numbers, given in Table \ref{tableinfo}. Table \ref{tableresult} summarizes the age distribution statistics for the sample with median reduced $\chi^{2}$ values. The table clearly shows that for our sample the average age and the median age are approximately the same for a given source, except for a few targets discussed in detail in Section \ref{AE}. We also observe that the average minimum age is 2.25 Myr, 18 sources show a 0 Myr minimum age. Target 14 (Fig. \ref{t4b}), that is observed to have a higher minimum age, is responsible for weighting the average minimum age towards a higher value, without it the average minimum age value falls to 0.96 Myr. As a check of the model fit we inspected the median reduced $\chi^{2}$ values, as we have independent $\chi^{2}$ values for each pixel for each source. We see that the worst fit is observed for target 22 (Fig. \ref{t6b}). The images used for analysis for this source do not show any artifacts, missing structures in individual maps, or bad data quality. However, we also looked at plots of the flux density as a function of frequency together with the fitted models and we see that the scatter in the data points is substantially larger than what we would expect from the error bars. Other objects that have $\chi^{2}$ value greater than 2 show similar behavior. We cannot accommodate for this by changing any parameters or assumptions, which means that the ages of these sources have systematic uncertainties that we cannot account for. For more typical sources the fits are generally good and any regions of high $\chi^{2}$ are restricted to small regions of the source. The mean average age obtained for the sample is 10.2 Myr with a standard deviation of 17.7 Myr, similarly, the mean maximum age is 23.09 Myr with a standard deviation of 30.35 Myr.\par 
\cite{Tamhaneetal2015} studied morphology, magnetic field strength, and energetics of a giant radio galaxy (target 27 in our sample, Fig. \ref{t7c}) using radio and X-ray data and found that the age of the AGN was 8 Myr for magnetic field strength of 3.3 $\mu$G and injection index of 0.5. These estimates are similar to ours (oldest age of 8.6 Myr with an average age of 6.4 Myr); this consistency between their results and ours further supports the analysis used in this paper.\par
In the next sections we look at the results for our sample in terms of different source aspects and parameters. We also look at the images presented in Fig. \ref{t1}-\ref{t7}, noting any anomalous behavior we observe and their suggested explanation for such abnormalities.

\begin{table*}
	\centering
	\caption{Estimates of the age values obtained after performing the analysis using the JP model for our sample. The table columns consists of values for recorded minimum age (Min age), maximum age (Max age), median age, average age, and median $\chi^{2}$ reduced.} \label{tableresult}
	\begin{tabular}{c c c c c c} 
		\hline
		Target & Min age (Myr) & Max age (Myr) & Median age (Myr) & Average age (Myr) & Median $\chi^{2}_{red}$\\
		\hline\hline
		\\
		\multicolumn{1}{c}{1} & $0.00^{+4.96}_{-0.00}$ & $44.53^{+0.00}_{-5.77}$ & $10.48^{+1.00}_{-0.96}$ & $11.64^{+0.43}_{-0.45}$ & 0.46 \\[1em] 
		\multicolumn{1}{c}{2} & $0.00^{+3.56}_{-0.00}$ & $24.01^{+1.59}_{-1.36}$ & $11.10^{+0.28}_{-0.01}$ & $11.23^{+0.09}_{-0.10}$ & 1.49 \\[1em]
		\multicolumn{1}{c}{3} & $0.00^{+1.80}_{-0.00}$ & $8.41^{+1.75}_{-1.81}$ & $1.19^{+0.14}_{-0.15}$ & $1.65^{+0.07}_{-0.07}$ & 0.89 \\[1em]
		\multicolumn{1}{c}{4} & $0.00^{+26.08}_{-0.00}$ & $115.44^{+7.62}_{-7.39}$ & $83.92^{+0.04}_{-0.01}$ & $82.57^{+0.50}_{-0.53}$ & 0.79 \\[1em]
		\multicolumn{1}{c}{5} & $0.00^{+8.47}_{-0.00}$ & $101.19^{+20.65}_{-12.62}$ & $11.28^{+0.10}_{-1.16}$ & $20.96^{+0.48}_{-0.47}$ & 1.43 \\[1em]
		\multicolumn{1}{c}{6} & $0.00^{+3.47}_{-0.00}$ & $8.71^{+1.12}_{-1.30}$ & $0.59^{+0.45}_{-0.44}$ & $2.16^{+0.09}_{-0.09}$ & 1.21 \\[1em]
		\multicolumn{1}{c}{7} & $0.00^{+8.62}_{-0.00}$ & $46.95^{+10.31}_{-13.02}$ & $0.51^{+0.01}_{-0.01}$ & $4.60^{+0.28}_{-0.26}$ & 0.96 \\[1em]
		\multicolumn{1}{c}{8} & $1.10^{+0.58}_{-1.10}$ & $3.70^{+0.24}_{-0.22}$ & $2.49^{+0.01}_{-0.00}$ & $2.50^{+0.01}_{-0.01}$ & 2.08 \\[1em]
		\multicolumn{1}{c}{9} & $0.00^{+7.04}_{-0.00}$ & $20.49^{+2.17}_{-2.39}$ & $10.98^{+0.03}_{-0.46}$ & $10.59^{+0.17}_{-0.16}$ & 0.9 \\[1em]
		\multicolumn{1}{c}{10} & $0.00^{+3.22}_{-0.00}$ & $24.98^{+0.93}_{-1.28}$ & $13.47^{+0.00}_{-0.45}$ & $13.24^{+0.08}_{-0.08}$ & 1.96 \\[1em]
		\multicolumn{1}{c}{11} & $0.00^{+19.80}_{-0.00}$ & $36.95^{+7.48}_{-9.02}$ & $0.48^{+0.01}_{-0.45}$ & $3.75^{+0.28}_{-0.27}$ & 0.91 \\[1em]
		\multicolumn{1}{c}{12} & $4.25^{+5.81}_{-4.25}$ & $19.26^{+3.37}_{-3.30}$ & $12.01^{+0.23}_{-0.02}$ & $11.87^{+0.14}_{-0.13}$ & 0.26 \\[1em]
		\multicolumn{1}{c}{13} & $4.49^{+1.03}_{-1.27}$ & $8.26^{+1.09}_{-1.07}$ & $5.55^{+0.13}_{-0.01}$ & $5.73^{+0.03}_{-0.03}$ & 0.33  \\[1em]
		\multicolumn{1}{c}{14} & $36.95^{+5.88}_{-6.29}$ & $87.05^{+4.60}_{-3.22}$ & $53.05^{+0.90}_{-0.09}$ & $55.83^{+0.27}_{-0.27}$ & 0.92 \\[1em]
		\multicolumn{1}{c}{15} & $0.00^{+1.25}_{-0.00}$ & $4.09^{+0.71}_{-1.03}$ & $2.79^{+0.01}_{-0.00}$ & $2.43^{+0.05}_{-0.05}$ & 1.27 \\[1em] 
		\multicolumn{1}{c}{16} & $0.00^{+4.22}_{-0.00}$ & $7.36^{+2.24}_{-3.09}$ & $4.65^{+0.14}_{-0.30}$ & $4.16^{+0.13}_{-0.13}$ & 0.51 \\[1em]
		\multicolumn{1}{c}{17} & $3.89^{+0.79}_{-0.90}$ & $5.80^{+0.73}_{-0.77}$ & $4.39^{+0.01}_{-0.08}$ & $4.43^{+0.01}_{-0.01}$ & 0.53 \\[1em]
		\multicolumn{1}{c}{18} & $0.00^{+1.50}_{-0.00}$ & $0.20^{+1.96}_{-0.20}$ & $0.00^{+0.01}_{-0.00}$ & $0.05^{+0.01}_{-0.01}$ & 1.78 \\[1em]
		\multicolumn{1}{c}{19} & $1.80^{+0.26}_{-0.22}$ & $2.80^{+0.20}_{-0.20}$ & $2.19^{+0.01}_{-0.01}$ & $2.17^{+0.01}_{-0.01}$ & 0.44 \\[1em]
		\multicolumn{1}{c}{20} & $0.00^{+0.83}_{-0.00}$ & $3.81^{+0.68}_{-0.86}$ & $0.05^{+0.00}_{-0.04}$ & $0.46^{+0.03}_{-0.03}$ & 1.05 \\[1em]
		\multicolumn{1}{c}{21} & $3.50^{+1.06}_{-1.71}$ & $9.89^{+0.54}_{-0.55}$ & $6.02^{+0.01}_{-0.09}$ & $6.10^{+0.03}_{-0.03}$ & 1.05 \\[1em] 
		\multicolumn{1}{c}{22} & $1.40^{+0.90}_{-1.40}$ & $8.89^{+0.45}_{-0.46}$ & $3.50^{+0.00}_{-0.00}$ & $3.74^{+0.04}_{-0.04}$ & 2.71 \\[1em]
		\multicolumn{1}{c}{23} & $2.70^{+0.76}_{-0.95}$ & $5.80^{+0.35}_{-0.36}$ & $3.49^{+0.01}_{-0.00}$ & $3.61^{+0.02}_{-0.02}$ & 0.55 \\[1em]
		\multicolumn{1}{c}{24} & $0.00^{+2.61}_{-0.00}$ & $8.01^{+1.08}_{-1.38}$ & $3.90^{+0.09}_{-0.01}$ & $3.87^{+0.07}_{-0.07}$ & 0.58 \\[1em]
		\multicolumn{1}{c}{25} & $0.00^{+4.89}_{-0.00}$ & $17.76^{+4.20}_{-3.54}$ & $1.25^{+0.24}_{-0.01}$ & $3.24^{+0.11}_{-0.11}$ & 0.73 \\[1em]
		\multicolumn{1}{c}{26} & $0.00^{+3.31}_{-0.00}$ & $5.99^{+1.55}_{-2.37}$ & $2.09^{+0.14}_{-0.14}$ & $1.90^{+0.05}_{-0.04}$ & 0.50 \\[1em]
		\multicolumn{1}{c}{27} & $0.00^{+1.81}_{-0.00}$ & $8.59^{+0.81}_{-0.91}$ & $6.39^{+0.01}_{-0.17}$ & $6.06^{+0.02}_{-0.02}$ & 0.19 \\[1em]
		\multicolumn{1}{c}{28} & $2.99^{+1.63}_{-2.99}$ & $7.64^{+1.33}_{-1.60}$ & $5.10^{+0.15}_{-0.01}$ & $5.08^{+0.05}_{-0.05}$ & 0.14 \\[1em] 	 
		\hline
	\end{tabular}
\end{table*}

\begin{table*}
	\centering
	\caption{Minimum age, maximum age, average age, median age and median $\chi^{2}_{red}$ values for JP, Tribble, and KP model comparing two sources.} \label{tablemodel}
	\begin{tabular}{ c c c c c c c  } 
		\hline
		Target & Model & Min age (Myr) & Max age (Myr) & Median age (Myr) & Mean age (Myr) & Median $\chi^{2}_{red}$\\
		\hline
		\\
		\multirow{3}{2em}{10} & JP & $0.00^{+3.22}_{-0.00}$ & $24.98^{+0.93}_{-1.28}$ & $13.47^{+0.00}_{-0.45}$ & $13.24^{+0.08}_{-0.08}$ & 1.96\\[0.6em]
		& Tribble & $0.00^{+3.29}_{-0.00}$ & $27.44^{+2.17}_{-2.36}$ & $13.97^{+0.00}_{-0.04}$ & $13.99^{+0.10}_{-0.10}$ & 1.93\\[0.6em]	
		& KP & $0.00^{+2.92}_{-0.00}$ & $31.35^{+16.03}_{-5.03}$ & $12.08^{+0.02}_{-0.01}$ & $12.16^{+0.08}_{-0.08}$ & 1.91\\[0.6em]	
		\hline
		\\
		\multirow{3}{2em}{14} & JP & $36.95^{+5.88}_{-6.29}$ & $87.05^{+4.60}_{-3.22}$ & $53.05^{+0.90}_{-0.09}$ & $55.83^{+0.27}_{-0.27}$ & 0.92\\[0.6em]
		& Tribble & $38.02^{+6.24}_{-6.64}$ & $96.97^{+5.09}_{-4.63}$ & $56.05^{+0.90}_{-0.09}$ & $59.45^{+0.30}_{-0.30}$ & 0.86\\[0.6em]
		& KP & $32.98^{+5.03}_{-6.05}$ & $115.56^{+0.00}_{-19.82}$ & $48.41^{+1.03}_{-0.06}$ & $52.14^{+0.32}_{-0.32}$ & 0.81\\[0.6em]
		\hline
		\\
	\end{tabular}
\end{table*}

\subsection{\normalfont{\textit{Morphology}}}
\label{Morph}
Figure \ref{t1a} shows the spectral age distribution for target 1 where the morphology in the spectral age map shows an increasing flux gradient towards the northern part of the source and indicates that the source morphology is that of a head-tail source, where the core is moving with respect to its environment and causing the ejected plasma to bend towards the movement axis. Hence we can only measure the size of a single lobe. The map shows a hot-spot near the southern region of the source. This region has an age which is consistent with zero which eventually grows to higher values exhibiting an increase in age of the source lobes towards the northern region. Figure \ref{t1b} shows the spectral age distribution for target 2, where we observe hot-spots near the center of each lobe and the age eventually increases radially outwards and towards the core. This type of morphology is indicative of an FR type-II source. We see that the MeerKAT contours converge near the hot-spots with the host galaxy being near the center of the source where we expect the core to be, further confirming the morphology of the source. Similar morphology and features are observed for target 3 (Fig. \ref{t1c}), 4 (Fig. \ref{t1d}), 7 (Fig. \ref{t2c}), 8 (Fig. \ref{t2d}), 9-12 (Fig. \ref{t3a}, Fig. \ref{t3b}, Fig. \ref{t3c}, Fig. \ref{t3d}), 13 (Fig. \ref{t4a}), 15 (Fig. \ref{t4c}), 16 (Fig. \ref{t4d}), 19 (Fig \ref{t5c}), 21-23 (Fig. \ref{t6a}, Fig. \ref{t6b}, Fig \ref{t6c}), 26 (Fig. \ref{t7b}), and 27 (Fig. \ref{t7c}).\par 
Figure \ref{t4b} shows the spectral age distribution for target 14 which has the lowest age values along the ejection axis. The age morphology indicates that the plasma flows out along the ejection axis with no termination point and then eventually spreads outwards perpendicular to the ejection axis, these are characteristics of FR-type I sources. For target 5 (Fig. \ref{t2a}, see Section \ref{AS}), 6 (Fig. \ref{t2b}), 17 (Fig. \ref{t5a}) and 28 (fig. \ref{t7d}) we observe similar morphology and features, hence these sources would also be expected to be FR type I.\par 
\subsection{\normalfont{\textit{Resolution and sensitivity}}}
\label{RS}
For target 3 (in Fig. \ref{t1c}) we observe ages consistent with zero near the edge of the two lobes which indicate the hot-spot regions for the source. Furthermore, regions around the edge of the southern lobe and the top-right part of the northern lobe are missing in the analysis, as our age estimates are limited by the least resolved and lower sensitivity flux density maps. Although the hot-spots do emerge where the contour converges, clearly seen in the northern lobe. Similarly, for target 11 (Fig. \ref{t3c}), due to the varying quality of the data and lower resolution for the maps, we are missing parts of the source near the boundary of the source and observe some age gradient around the centre. Hence, for some sources, we have lost some parts of the structure due to poorly resolved maps and low-sensitivity surveys.\par 

\subsection{\normalfont{\textit{Beam size and source size}}}
\label{BSS}
For sources like target 4 (Fig. \ref{t1d}), 8 (Fig. \ref{t2d}), 12 (Fig. \ref{t3d}), 13 (Fig. \ref{t4a}), 19 (Fig. \ref{t5c}), and 21-23 (Fig. \ref{t6a}, Fig. \ref{t6b}, Fig.\ref{t6c}) we do not observe any zero value region although we do see a lowest age region near the centre of the two lobes, also the region where the contours converge and the age eventually increases radially outwards. They also show much flatter age gradients than any other sources. We do not observe a sharp age gradient for the sources that either have a large beam size or a small source size. In other words, the size of the source is small with respect to the larger beam size, which gives us a lower age gradient, hence the lack of detailed age distribution. We can say the same for the FR type-I, target 17 (Fig. \ref{t5a}) and 28 (Fig. \ref{t7d}). For all of these sources it is plausible that the low resolution of our observations, relative to the source size, means that we cannot detect a consistent zero value region that may really be present. Even where low-age regions are detected, poorly resolved sources with beam size comparable to source size produce a less detailed age distribution map and a flatter age gradient.\par

\subsection{\normalfont{\textit{Source environment}}}
\label{EE}
Target 7 (Fig. \ref{t2c}), at first glance, looks like a head-tail source, but if that were the case we would have observed the host galaxy position at the north-west end where the surface brightness is highest. In fact the most plausible host galaxy is in the middle of the source. Further investigation reveals that there is another galaxy present near the south-east region, where the values are consistent with zero, which may be responsible for underestimating the age values near the south-east edge.\par 
For target 26 (Fig. \ref{t7b}), the two structures seen at the western and the eastern corner of the map can be classified as the lobes of the source with the core sitting in the middle of the two lobes along the ejection axis. For the two lobes we see that the lowest age regions are near the inner edge of the western lobe and on the outer edge of the eastern lobe, from which the plasma ages gradually and moves into the surrounding. The structure of the two lobes does not look similar to those seen in the previous examples, for which one of the reasons could be the external pressure applied by the surrounding ICM and IGM which restricts the uniform expansion of the plasma. We would have to study the host galaxy and its environment in detail to provide more information.\par 

\subsection{\normalfont{\textit{Anomalous sources}}}
\label{AS}
Figure \ref{t2a} shows the spectral age distribution for target 5 from which we obtain age estimates using flux densities from four radio maps. We exclude the radio flux densities from the GMRT survey at 420 MHz, 460 MHz, and 610 MHz as this map has lower signal-to-noise and a lot of artifacts around the source. Excluding the maps for the source allows us to include emission from the entire structure of the lobes and the source. Another noticeable feature of the age map is the giant zero-age region around the center of the lobes. The possible explanation for such a distribution can come from the calibration errors for the maps corresponding to different frequencies or the quality of data obtained during observation. It is difficult for us to point out a specific hot-spot region however, it should lie in the zero-age region. Furthermore, we also notice that the age estimates increase outwards from the zero-age region. The spectral age distribution is consistent with a source with the characteristics of FR-type I. Hence it is safe to say that the source is plausibly an FR type I source. \par
Figure \ref{t5b} shows the spectral age distribution for target 18 where we observe that the entire source is filled with zero age values, corresponding to a flat spectrum everywhere in the source. We looked for emission near the source in the sensitive MeerKAT data but did not find any evidence of contaminating sources that could be responsible for such behavior. As we do not see any clear morphology type, it is difficult to categorise the source. High resolution data would be needed to understand the behavior of this source.\par 
Figure \ref{t5d} shows the spectral age distribution for target 20 which does not show any age distribution with the zero-age region spread across the source. This is not something we expect for any type of morphology and hence points towards problems with the data quality. In order to check the flux maps in our analysis, we excluded maps from the GMRT survey at 610 MHz, 460 MHz, and 420 MHz as the data are of poor quality at these frequencies. We still observe similar behavior in our analysis and thus we can conclude that either the source type is different or the data quality is improper for our analysis.\par
Figure \ref{t6d} shows the spectral age distribution for target 24 which does not show any prominent structure looking at the age distribution. In the MeerKAT contours, we can see the structure of lobes emerging from the core (marked host galaxy position) with visible jet structure. These are characteristics of an FR type-II. From the age distribution map, we do not clearly see the lobe in the southern region and most of the northern lobe is missing. This is most likely due to the data quality from the maps other than MeerKAT. We see that most of the structure seen in the MeerKAT map is missing in other maps i.e. the GMRT survey and the LOFAR survey. For this reason, we only include maps with the best data quality, hence we exclude the data from the GMRT survey with frequencies at 320 MHz, 420 MHz, and 460 MHz. From the age map, we can see a zero-age region in the southern region and the age gradually increases towards the core which further confirms our classification of the source as FR type-II. For this analysis, we include the core as we also want to look at the age distribution around the core which increases radially outwards around the core.\par
Figure \ref{t7a} shows the spectral age distribution for target 25 which shows an increasing age gradient around the core. The morphology is characteristic of a wide-angle tail type source where the age gradient grows away from the core. Another possible reason for such an age distribution could be a short time difference between AGN switch off and switch on where plasma injection was cut off. We also know that data quality plays a vital role so data from 144 MHz (LOFAR) and 610 MHz (GMRT) had to be removed. Data from more representative maps should be able to solve the problem and point out the reason for the anomaly.

\section{Discussion}
\label{D}

\begin{figure*}
\includegraphics[width=3.3in]{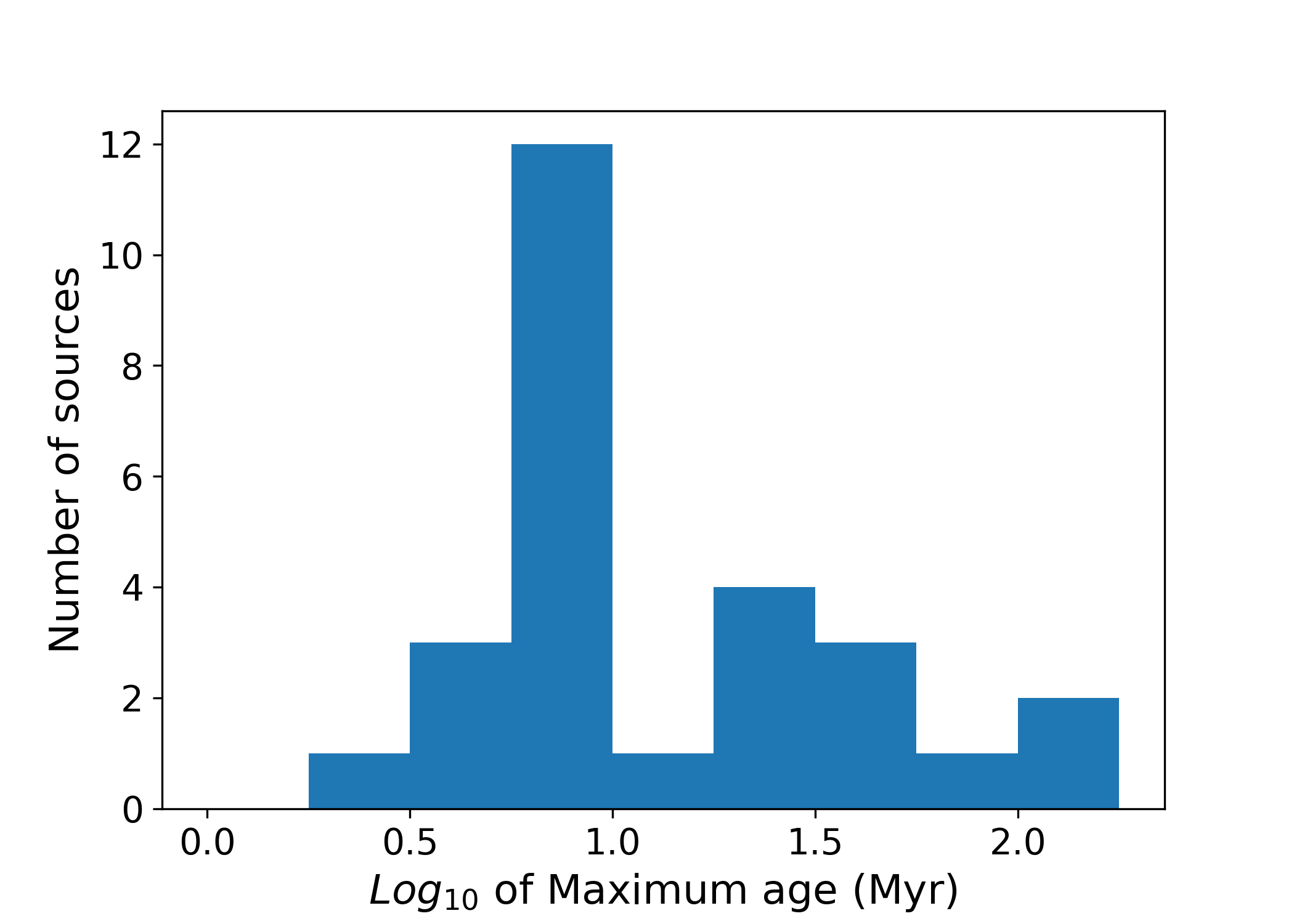}\includegraphics[width=3.5in]{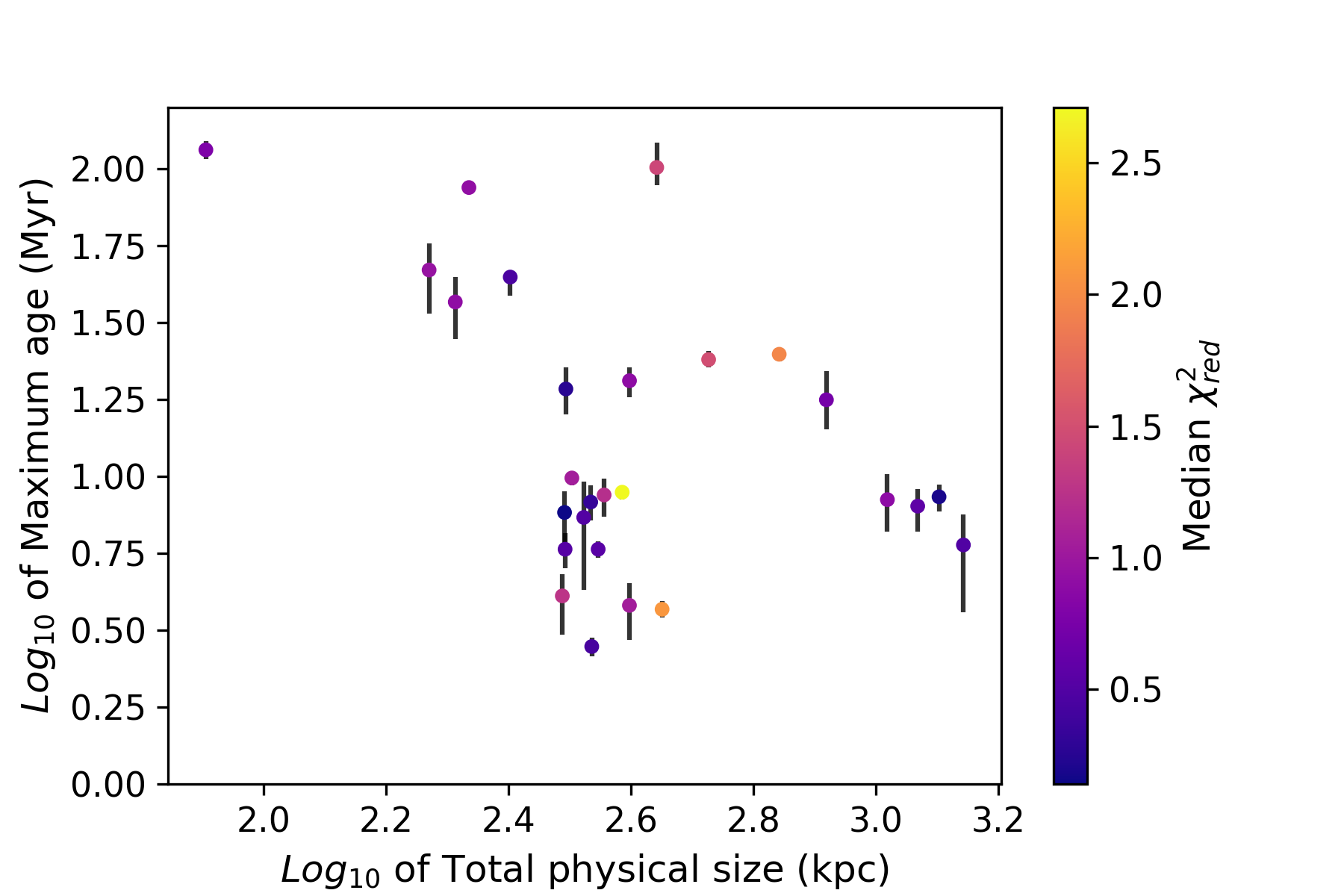}
\caption{Left histogram shows maximum age distribution for the sample with bin size of $10^{0.25}$ Myr, where the x-axis represents logarithmic values of age intervals and y-axis represents the number of sources within the interval. Right plot shows the relationship between the total physical size versus the maximum age of the sources in the sample, color coded by median of $\chi^{2}_{red}$. Target 18 is not included in the plot as the age is not reliable. We note that the errors are uncertainties on the age of the maximum age pixel on the source and as such are just indicative of the uncertainties on the true source maximum age.}
\label{t8}
\end{figure*}

In this study we have analyzed a total of 28 sources in the XMM-LSS field using the MeerKAT survey at 1.28 GHz, the LOFAR survey at 144 MHz, and the GMRT survey at 320 MHz, 370 MHz, 420 MHz, 460 MHz, and 610 MHz. We used the JP model of spectral aging to evaluate the age distribution over other models as this model is computationally least expensive and corresponds well with the physical description of the AGN activity. We assume the injection index for our sample to be 0.5 and obtain the magnetic field values by assuming that they are 0.4 times the equipartition value. In the next few sections we will first discuss the factors affecting our analysis. We explore the different observations made in section \ref{R} and suggest ways in which future observations could be made so as to overcome the limitations of our study. \par 
\subsection{\normalfont{\textit{Resolution limits}}}
\label{RL}
The minimum size of a source we use to select our sample is greater than 10 arcsec and we find that for a beam size of 10 arcsec, the sizes of some sources, such as target 15-17 (Fig. \ref{t4c}, Fig. \ref{t4d}, Fig. \ref{t5a}), 19 (Fig. \ref{t5c}), 21 (Fig. \ref{t6a}), 23 (Fig. \ref{t6c}), and 28 (Fig. \ref{t7d}), are too small to give a detailed age distribution. We recommend that the source size should be around 4 times greater (per lobe) than the beam size to get a detailed age distribution, as we see in our sample for the target 1 (Fig. \ref{t1a}), 3 (Fig. \ref{t1c}), 10 (Fig. \ref{t3b}), 26 (Fig. \ref{t7b}), and 27 (Fig. \ref{t7c}). Also, the lobe size needs to be around 1.5 times greater than the beam size to observe a reasonable age gradient, as we see for target 12 (Fig. \ref{t3d}), 13 (Fig. \ref{t4a}), 15 (Fig. \ref{t4c}), 16 (Fig. \ref{t4d}), and 19 (Fig. \ref{t5c}) (see Table \ref{tableinfo} for lobe sizes). As the resolution is low for these sources, it makes it difficult for us to obtain a detailed age distribution and determine the type of the source, that is, if it is FR type-I or FR type-II. This also means that the resolution of the radio maps plays a significant part in the spectral age analysis where the lowest resolution map becomes the limiting factor. In order to study such small-sized sources in our sample in greater detail, high-resolution maps at a range of different frequencies are required. \par 

\subsection{\normalfont{\textit{Sensitivity and redshift limits}}}
\label{SRL}
We find that for targets 5 (Fig. \ref{t2a}), 6 (Fig. \ref{t2b}), 15 (Fig. \ref{t4c}), 16 (Fig. \ref{t4d}), 20 (Fig. \ref{t5d}), 24 (Fig. \ref{t6d}), 25 (Fig. \ref{t7a}) and 27 (Fig. \ref{t7c}) we have to exclude data from one or more frequencies in order to increase the spatial regions under analysis as the excluded maps lack emission detected clearly at other frequencies. This also means that while selecting sources for our analysis, survey sensitivities and image fidelity impose limitations. High-sensitivity surveys help us detect fainter and more extended emission which allows us to determine the radio source category and provide more spatial regions for analysis. Lack of structure and emission in any one of the maps limits the region under analysis making it difficult for us to obtain a detailed age distribution. Missing redshifts (as discussed in Section \ref{CS}) can also limit our analysis.\par 

\subsection{\normalfont{\textit{Injection index}}}
\label{II}
In our analysis, we have assumed the injection index to be a constant at 0.5. From previous studies, we know that this is not likely to be true for all of the sources in the sample as each source is affected by different ICM, IGM, and pressure conditions which can affect the particle distribution upon injection. For our analysis we did not account for the changing injection index as more low-frequency data would be required to help constrain the injection index value for each source; we used a constant injection index value of 0.5 as this is the lowest possible value we can observe for any source as predicted by shock theory \citep{Bell1978}. Increasing the injection index would mean that we would obtain lower age estimates for a given region which could eventually pull regions to zero age. From Table \ref{tableinjection}, we can see that the age values decrease as we increase the injection index values. Two of the targets (target 2 and 27) are FR-type II and target 14 is FR-type I. The maximum ages of these sources reduce by 6-8\% every time we increase the injection index value by 0.1. Thus a minimum injection index of 0.5 should produce age values that are oldest for the region under analysis which in turn helps us find the oldest possible age for the source.\par 
\begin{table*}
	\centering
	\caption{Minimum age, maximum age, average age, median age and median $\chi^{2}_{red}$ values for increasing injection index of target 2, 14 and 27.} \label{tableinjection}
	\begin{tabular}{ c c c c c c c  } 
		\hline
		Target & Injection index & Min age (Myr) & Max age (Myr) & Mean age (Myr) & Median age (Myr) & Median $\chi^{2}_{red}$\\
		\hline
		\\
		\multirow{4}{2em}{2} & 0.5 & $0.00^{+3.56}_{-0.00}$ & $24.01^{+1.59}_{-1.36}$ & $11.23^{+0.09}_{-0.10}$ & $11.10^{+0.28}_{-0.01}$ & 1.49\\[0.6em]
		& 0.6 & $0.00^{+3.02}_{-0.00}$ & $22.51^{+1.52}_{-1.59}$ & $8.80^{+0.10}_{-0.10}$ & $8.69^{+0.02}_{-0.27}$ & 1.31\\[0.6em]
		& 0.7 & $0.00^{+2.53}_{-0.00}$ & $20.71^{+1.72}_{-1.49}$ & $6.28^{+0.10}_{-0.11}$ & $6.00^{+0.29}_{-0.02}$ & 1.24 \\[0.6em] 
		& 0.8 & $0.00^{+2.14}_{-0.00}$ & $19.19^{+1.54}_{-1.86}$ & $4.02^{+0.11}_{-0.10}$ & $3.28^{+0.03}_{-0.29}$ & 1.37 \\[0.6em]	
		\hline
		\\
		\multirow{4}{2em}{14} & 0.5 & $36.95^{+5.89}_{-6.29}$ & $87.05^{+4.60}_{-3.22}$ & $55.83^{+0.27}_{-0.27}$ & $53.05^{+0.90}_{-0.09}$ & 0.92\\[0.6em]
		& 0.6 & $21.95^{+6.80}_{-8.83}$ & $80.05^{+4.50}_{-4.17}$ & $45.42^{+0.31}_{-0.31}$ & $42.95^{+0.00}_{-0.90}$ & 0.63\\[0.6em]
		& 0.7 & $0.00^{+11.23}_{-0.00}$ & $72.95^{+3.66}_{-4.74}$ & $34.08^{+0.47}_{-0.38}$ & $30.96^{+0.07}_{-0.92}$ & 0.52 \\[0.6em]
		& 0.8 & $0.00^{+10.02}_{-0.00}$ & $65.05^{+4.29}_{-4.81}$ & $21.65^{+0.44}_{-0.42}$ & $17.99^{+0.96}_{-0.04}$ & 0.47 \\[0.6em] 
		\hline
		\\
		\multirow{4}{2em}{27} & 0.5 & $0.00^{+1.81}_{-0.00}$ & $8.59^{+0.81}_{-0.91}$ & $6.06^{+0.02}_{-0.02}$ & $6.39^{+0.01}_{-0.17}$ & 0.19\\[0.6em]
		& 0.6 & $0.00^{+1.32}_{-0.00}$ & $8.01^{+1.03}_{-0.82}$ & $5.48^{+0.02}_{-0.02}$ & $5.79^{+0.00}_{-0.01}$ & 0.17\\[0.6em]
		& 0.7 & $0.00^{+1.20}_{-0.00}$ & $7.61^{+1.06}_{-0.82}$ & $4.89^{+0.03}_{-0.03}$ & $5.19^{+0.00}_{-0.01}$ & 0.15 \\[0.6em] 
		& 0.8 & $0.00^{+0.74}_{-0.00}$ & $7.19^{+0.98}_{-0.87}$ & $4.29^{+0.03}_{-0.03}$ & $4.61^{+0.00}_{-0.02}$ & 0.14 \\[0.6em]  
	\hline
\end{tabular}
\end{table*}

\subsection{\normalfont{\textit{What are the age estimates?}}}
\label{AE}
For our sample, the highest recorded age is 115 Myr while we do observe the lowest age of 0 Myr for a few pixels in 18 out of 28 sources, especially for hot-spots, which is expected as this is where the particles are accelerated and the spectral index is flattest. The average age for the sample is 10.2 Myr and the mean of maximum ages recorded is 23.09 Myr. Using our results from the sample we can construct a histogram of the maximum age from each source and observe the distribution of the oldest age in our sample. The histogram shown in Figure \ref{t8} (left), provides us with the oldest age estimates giving us information about when the AGN was first switched on. We see that the sample peaks at a maximum age range of around 5 Myr to 9 Myr and the number then falls off with increasing age.  \par
From Table \ref{tableresult}, we see that the average age and the median age for a given source is usually the same for our sample with exceptions such as target 5 (Fig. \ref{t2a}), 6 (Fig. \ref{t2b}), 7 (Fig. \ref{t2c}), 11 (Fig. \ref{t3c}), 20 (Fig. \ref{t5d}), and 25 (Fig. \ref{t7a}). We observe a very large zero age region for these sources which is responsible for the difference seen in the estimates of the median age and the average age. We are not sure why we observe these large zero-age regions for these sources. Some of the reasons we point out in terms of data are bad data quality, calibration error, and artifacts which are imposing limitations on our analysis and results. Other effects are physical and include low AGN switch-off switch-on time, or source morphology other than classical FR type-I and FR type-II. This means that the difference in the age estimates of the median age and the average age cannot be narrowed down to one specific reason and neither can it be solved by changing parameter values and assumptions; in general, better data will be required.\par 

\subsection{\normalfont{\textit{What can we say about extended emissions?}}}
We can see in Fig. \ref{t1}-\ref{t7} the age distribution maps and the MeerKAT contours overlaid on the age distribution maps which show radio-emitting regions that have not been accounted for in the spectral age fitting. Although we cannot estimate the ages of material in these regions, since we only have detection at single frequency, they may well be older than the material that we can see. We do not see this excess emission as we are limited by the lowest sensitivity map in our data, which is usually the LOFAR or the GMRT image. This means, in reality, the source can be older than what we obtain through our analysis and we may start seeing a larger number of old sources as we start using high-sensitivity maps, particularly at low frequency, in future work.  \par 
 
\begin{figure*}
\includegraphics{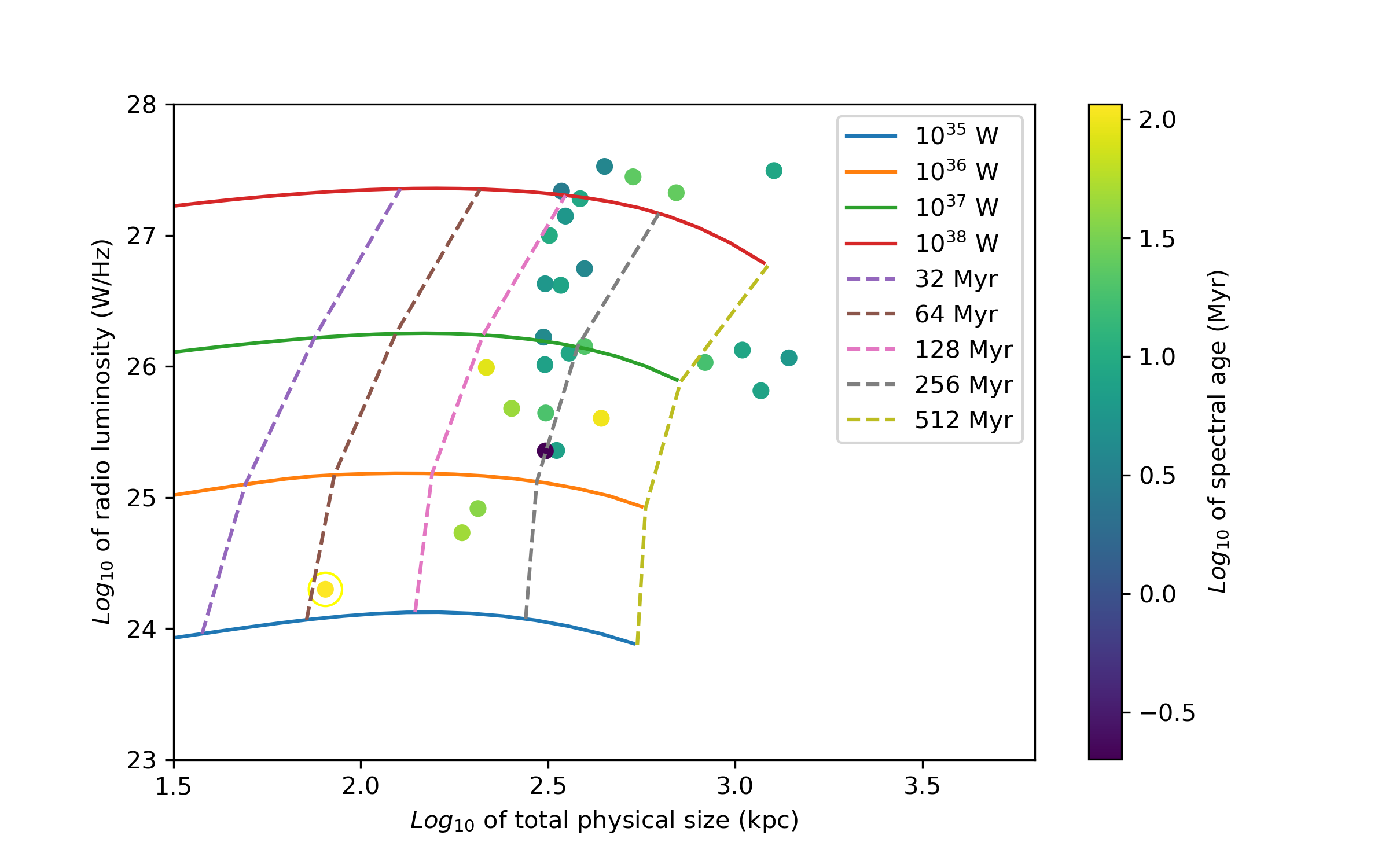}
\caption{Radio luminosity versus total physical size relation color coded by maximum age obtained for our sample where, the ring marks the oldest source. The solid lines are model relations between radio luminosity and physical size for given jet powers between $10^{35}$ and $10^{38}$ W, presented by \citealt{Hardcastle2019}. The dashed lines connect locations of constant dynamical age in these models and so indicate the expected locations of sources with those dynamical ages on the plot. Note that the physical sizes plotted for the sources in our sample are projected sizes since we do not know their orientation with respect to the plane of the sky but this will give rise to only a small offset in their plotted positions.} \label{t9}
\end{figure*} 

\subsection{\normalfont{\textit{What is the total size and age correlation?}}}
The scatter plot given in Figure \ref{t8} (right) shows the relation between the total projected physical size of the sources and their respective evaluated oldest age values. The usual consensus is that the greater the physical size of the source, the higher the age values that should be observed. However, the plot shows if anything, an opposite relation where we see that as the age of the sources increases the total physical size of the source decreases. This could be due to the surface brightness limits for our sample as it can be biased against larger, older sources which are hard to see in all wavelengths. Also due to sensitivity and resolution limits, we cannot determine the age for small sources, less than 10 arcsec for our sample, which creates another bias. This might be the reason why we see empty regions near the top right and bottom left corners of the plot. Again, more sensitive and high-resolution low-frequency data are needed to increase the sample size and overcome these biases.\par 
\subsection{\normalfont{\textit{What can we say about the spectral age model?}}}
\cite{Hardcastle2019} compared the actual size distribution of the sources to the size distribution obtained using dynamical models by assuming a lifetime distribution. In these models there is a direct relationship between the age of the source and its physical size; large RLAGN sources must have older ages. We used their dynamical modeling (which assumes a group environment with $M_{500} = 2.5\times10^{13} M_\odot$, typical for low redshift radio galaxies) to compare with the age of our sources in the sample (Fig. \ref{t9}) and we can see that there is a discrepancy between the spectral age and the ages obtained by dynamical modeling, where the model lines are taken from \cite{Hardcastle2019}. For example, we see that the target 4 (Fig. \ref{t1d}), the oldest source in our sample, in Fig \ref{t9} lies near the 64 Myr curve which is far from the value of 115 Myr that we estimate using spectral analysis. This kind of discrepancy is also observed for other sources in our sample. In general, we see that the spectral ages that we measure are significantly lower than would be expected from their position on the plot of Fig. \ref{t9}. This disagrees with the inferences drawn by \cite{Hardcastle2019} and indicates that there might be processes that either spectral or dynamical or both types of modeling do not account for, as also concluded by previous studies such as \cite{Harwoodetal2016, Turneretal2018, Mahatmaetal2019}, although in our work we have not been able to account for the different environments of our sources, since in general we have no information on the host galaxy environment. This discrepancy potentially has important implications for determining the jet powers for powerful RLAGN which will affect the modeling of feedback. \par 

\subsection{\normalfont{\textit{What is the relationship between the age distribution maps and the source morphology?}}} 
From the age distribution map, we see that most of the maximum ages evaluated are near the outskirts or near the boundary of the lobes where one can suspect that the maximum age might be overestimated or highly uncertain due to missing data from low resolution and less sensitive maps or bad data quality. On the other hand, almost all of the sources show regions that are only detected in the sensitive MeerKAT data which means that there could be older material that is not included in our analysis. Furthermore, the age distribution map of the sources can also be used to infer the Fanaroff-Riley classification of the source as shown in Appendix \ref{A1}.\par

\section{Summary and Conclusions}
We have used the data from the early science release MIGHTEE survey, the GMRT survey, and the LOFAR survey to evaluate the age of the sources for our sample, where we used the Jaffe-Perola (JP) model \citep{JaffeandPerola1973} to perform spectral age analysis, incorporating data from frequencies of 144 MHz, 320 MHz, 370 MHz, 420 MHz, 460 MHz, 610 MHz, and 1284 MHz. For the first time, we have been able to create and evaluate the age of a relatively large size of 28 sources in a single survey field, due to MeerKAT's high resolution and high sensitivity observations which were used as a reference to find extended sources and perform the spectral age study. We also determined the age distribution of the sources and the distribution of the maximum age for our sample. \par 
Our sample's age distribution maps show sources of various sizes and structures, some of which exhibit peculiar characteristics. Some of this can be a result of the quality of the data. For two sources parts of the structure were missing from the analysis, suggesting limitations arising due to lower resolution and less sensitive flux density maps. Around 10 sources show no zero age regions in the age distribution maps along with some showing a flatter age gradient; these sources are all poorly resolved and so we conclude that the low resolution of our study is preventing us from isolating the regions of current particle acceleration. We had to exclude information from some frequencies as they were of poor quality and restricted us from analyzing emissions from the entire structure of the source. We have observed anomalies for 5 different sources in our sample, most of which correspond to the use of bad quality data or poor detection in one or more images. Hence, we note that our sample is limited by the least resolved maps, the availability of redshifts, the survey sensitivities, and the size of the sources.\par 
We summarize the key findings related to the questions posed in Section \ref{OQA}, in the order they appear \textbf{in} that section:
\begin{enumerate}[i]
	\item we see that the oldest source in our sample is observed to be 115 Myr old and the youngest source has an age of 2.8 Myr, the mean of the sample is 23.09 Myr and the median is 8.71 Myr. We observe the maximum age distribution to peak at values between 5 Myr to 9 Myr which correlates well with the median age values for our sample.
	\item Most sources (17 sources) in our sample can be classified as FR-type II but there are 5 sources that can be classified as FR-type I and one source which is classified as a Head-tail source, using the spectral index plots and drawing inferences from the age distribution plots.
	\item As we overlaid the MeerKAT contours over the age distribution maps, we observe excess emission that has not been accounted for in the analysis, which means there is most likely older material beyond our region of analysis and our estimates of the maximum age are very probably giving us lower limits of age for the sources. We suggest use of high sensitivity low-frequency maps in future analysis which can help estimate much older age values for a given source. We also observe that from a detailed age distribution observation, we can clearly classify the morphology of a source (see Appendix \ref{A1}). 
	\item We can conclude from this study that the beam size and the source size plays a vital role in giving us a detailed age distribution map along with well constrained age estimates. The size of the source should be at least 4 times bigger than the beam size to observe detailed age distribution or the beam size should be such that we can accommodate a minimum of 4 but preferably more than 5 beams over the entire source.
	\item For our sample, we observe no clear relation between the total size of the sources and their age, contrary to what would be expected in dynamical models. We point out the use of small sample, resolution, sensitivity, and surface brightness limits as the factors that may give rise to this observation. Furthermore, we have observed a discrepancy between spectral age and dynamical age analysis where we see that spectral ages are significantly lower than what we would expect for dynamical ages. This can be an indication that there might be processes that either or both the models do not account for but further investigation is necessary as these discrepancies in ages could invalidate analysis used to infer the magnitude of jet power and therefore of AGN feedback.
\end{enumerate}  

Overall, the model and the analysis also require us to make assumptions about the parameters of the spectral ageing model (such as the injection index and the magnetic field strength) that depend on the data availability and the data quality. We have pointed out different limiting factors that are affecting the analysis and discussed the possible ways to overcome the limitations. The present study tries to draw its conclusion from a sample of 28 sources, which is larger than the samples used before in any other study and has highlighted some important aspects to consider when attempting such analysis on large samples of radio galaxies in the future. The superMIGHTEE survey in the full XMM-LSS area with the uGMRT (Lal, Taylor, et al., submitted) will provide multi-frequency radio data for a much larger sample, while in the long term the Square Kilometer Array (SKA) is expected to generate very large quantities of radio images that can be used in this way.

\section*{Acknowledgments}
The research has used the high performance computing facility (\url{http://uhhpc.herts.ac.uk/}) located at the University of Hertfordshire. We wish to thank National Center for Radio Astrophysics for their help and guidance along with the GMRT collaboration. This research has made use of the NASA/IPAC Extragalactic Database (NED), which is operated by the Jet Propulsion Laboratory, California Institute of Technology, under contract with the National Aeronautics and Space Administration. IHW acknowledge generous support from the Hintze Family Charitable Foundation through the Oxford Hintze Centre for Astrophysical Surveys. The MeerKAT telescope is operated by the South African Radio Astronomy Observatory, which is a facility of the National Research Foundation, an agency of the Department of Science and Innovation. We acknowledge use of the Inter-University Institute for Data Intensive Astronomy (IDIA) data intensive research cloud for data processing. IDIA is a South African university partnership involving the University of Cape Town, the University of Pretoria and the University of the Western Cape. The authors acknowledge the Centre for High Performance Computing (CHPC), South Africa, for providing computational resources to this research project. This paper makes use of \textsc{TOPCAT} (\citealt{topcat}) and \textsc{Astropy} \citep{astropy} for the analysis.

\section*{Data Availability}
This work is primarily based on publicly available MIGHTEE data \citep{Heywood2022}. Other data used in this study can be obtained on reasonable request to the lead author. 


\bibliographystyle{mnras}
\bibliography{reference}



\appendix
\section{Spectral age distribution versus distance}
\label{A1}
\begin{figure*}
	\includegraphics[width=1.4in,height=1.3in]{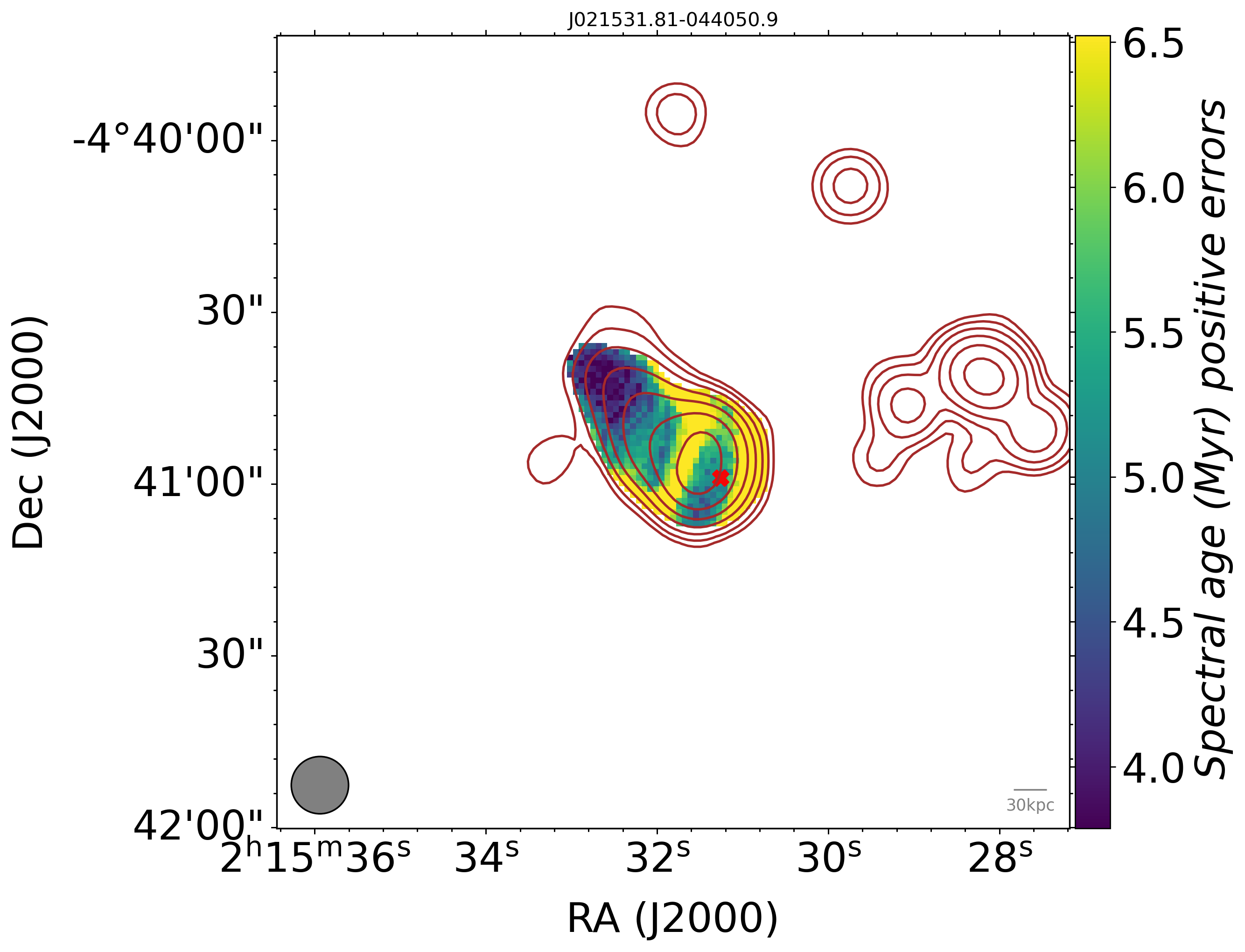}\includegraphics[width=1.4in,height=1.3in]{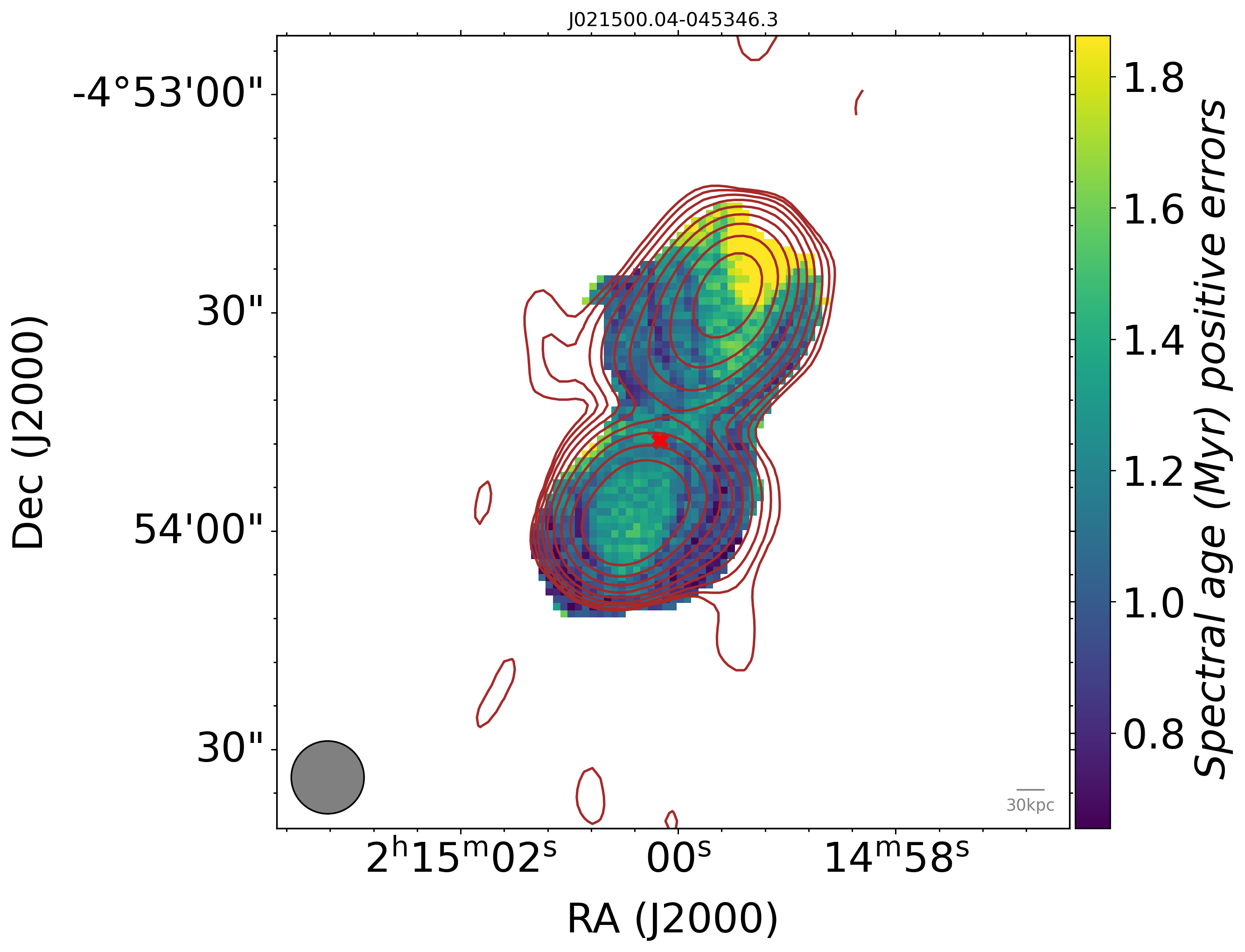}\includegraphics[width=1.4in,height=1.3in]{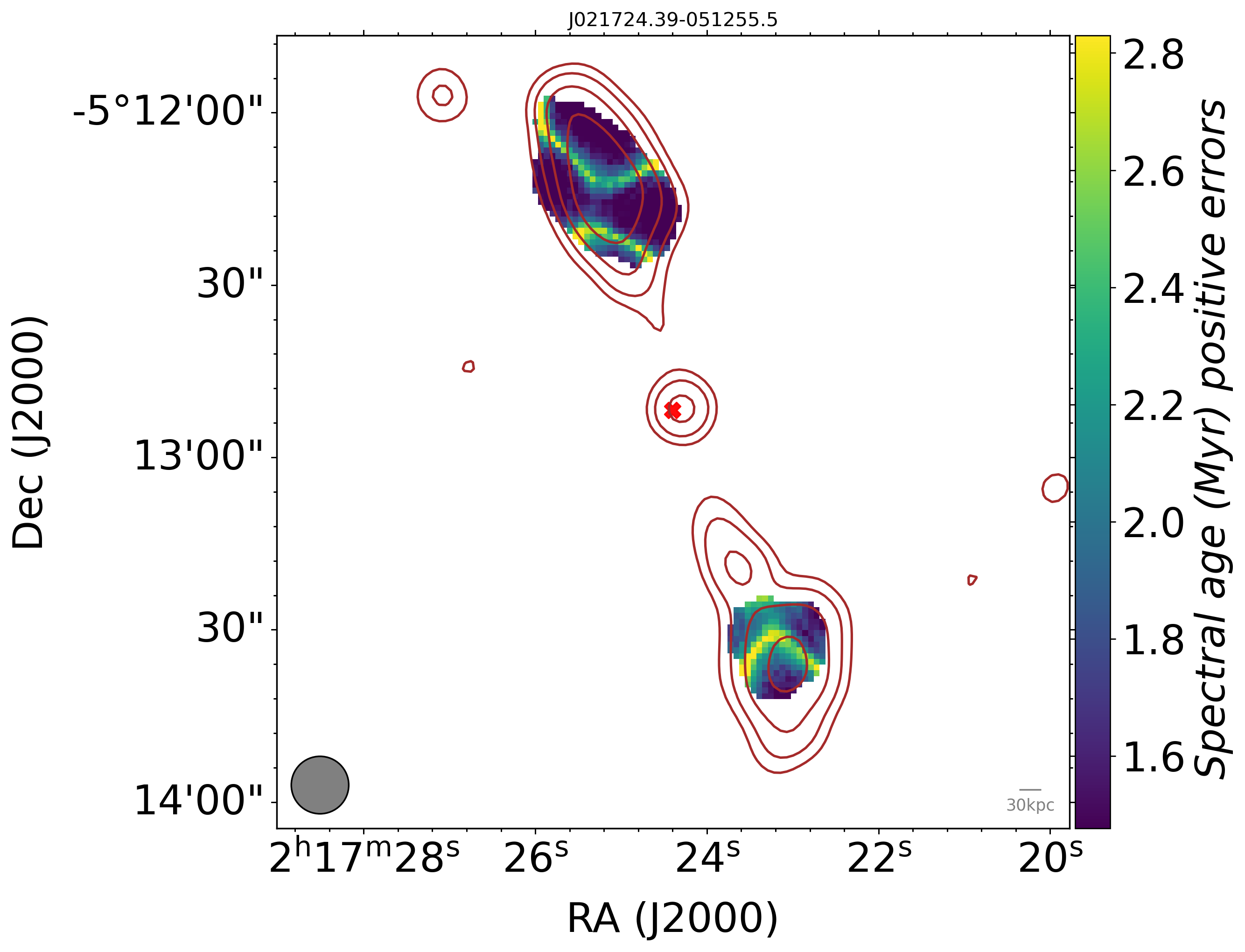}\includegraphics[width=1.4in,height=1.3in]{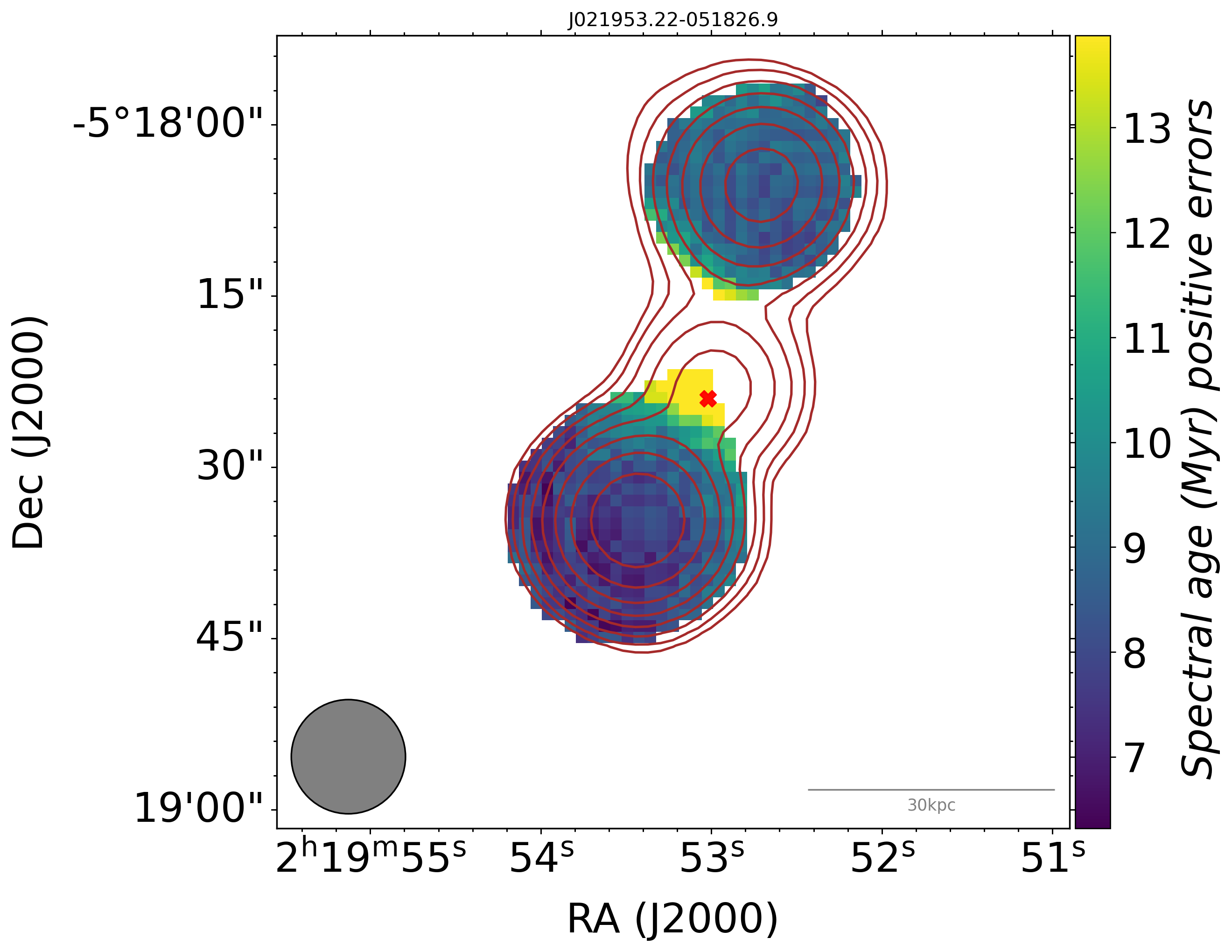}
	\includegraphics[width=1.4in,height=1.3in]{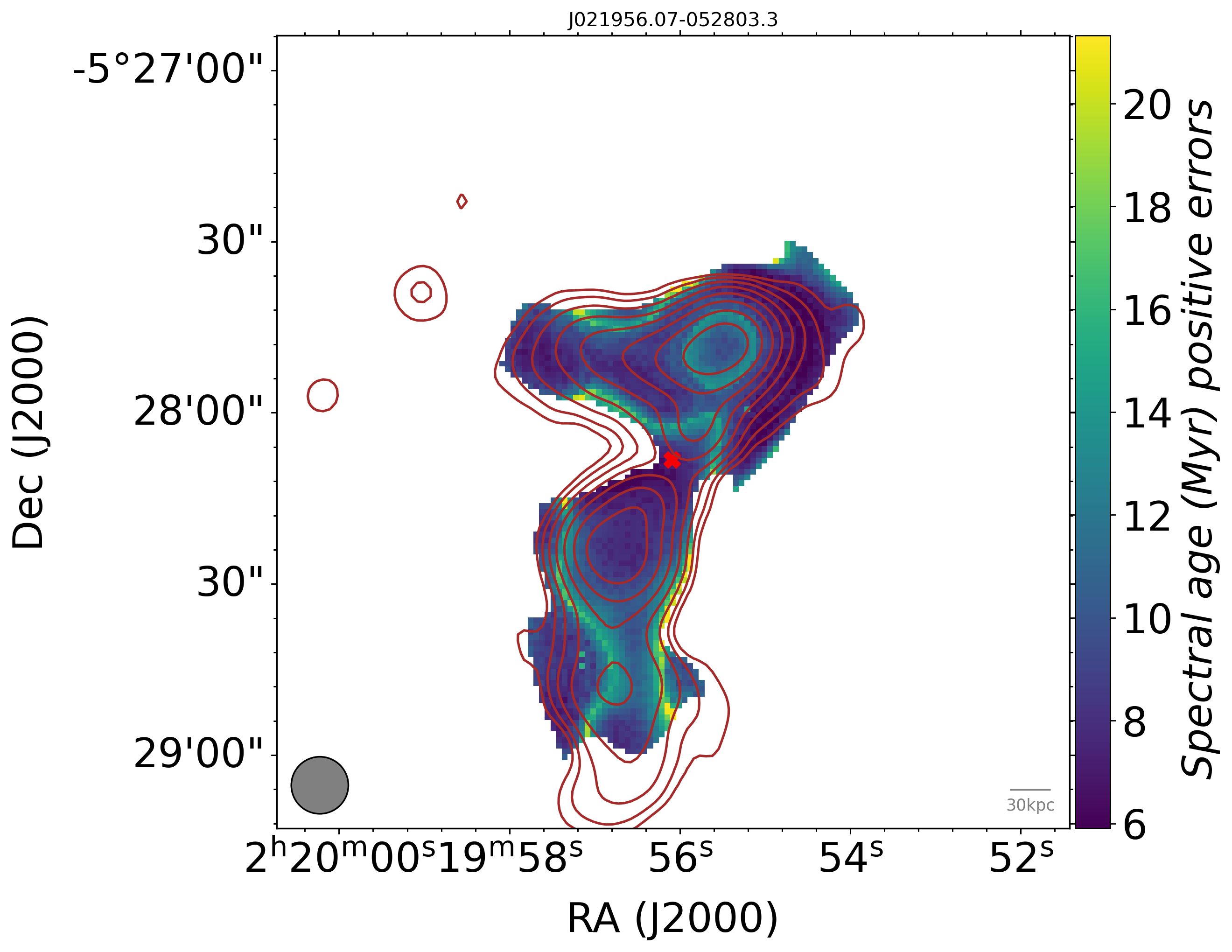}\includegraphics[width=1.4in,height=1.3in]{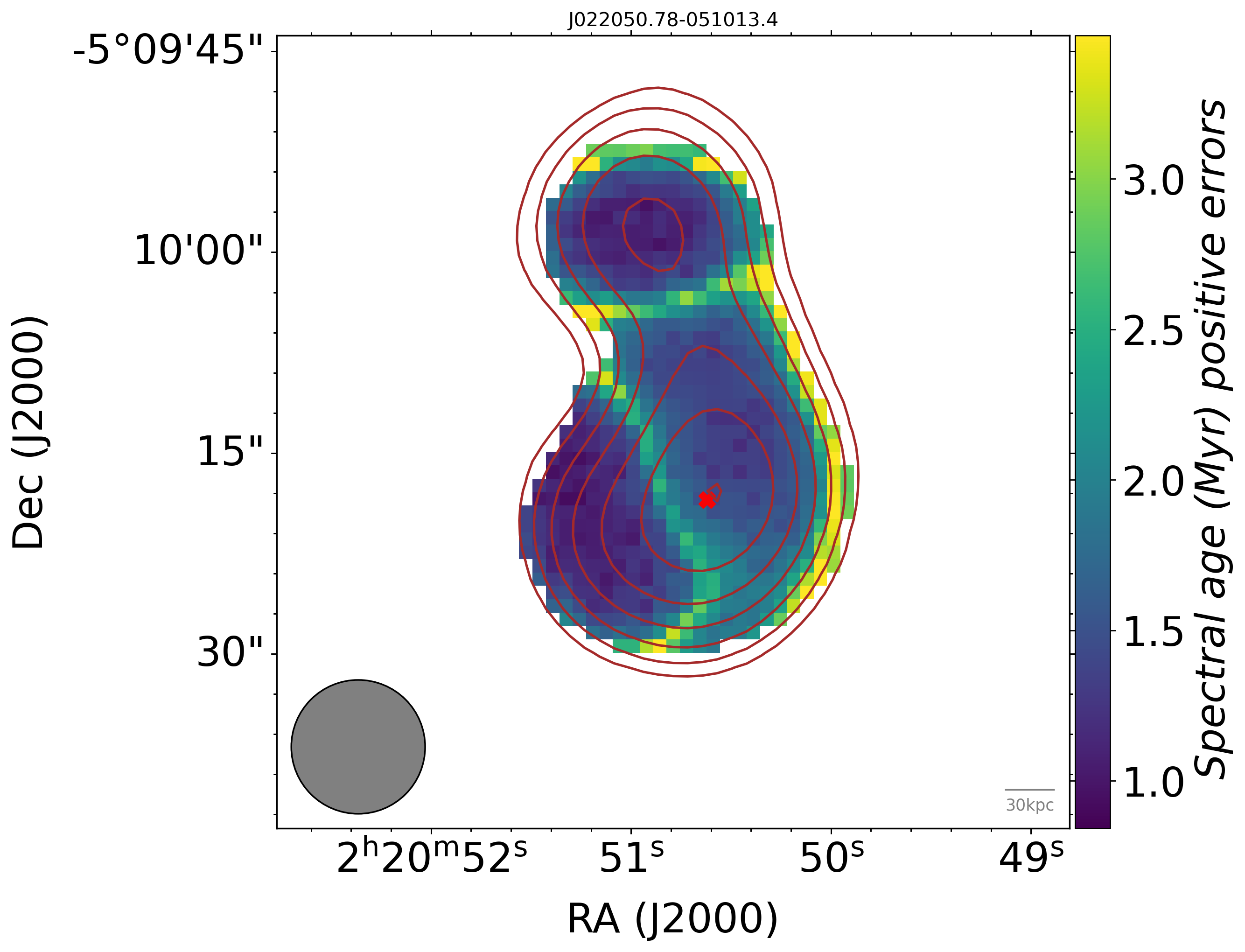}\includegraphics[width=1.4in,height=1.3in]{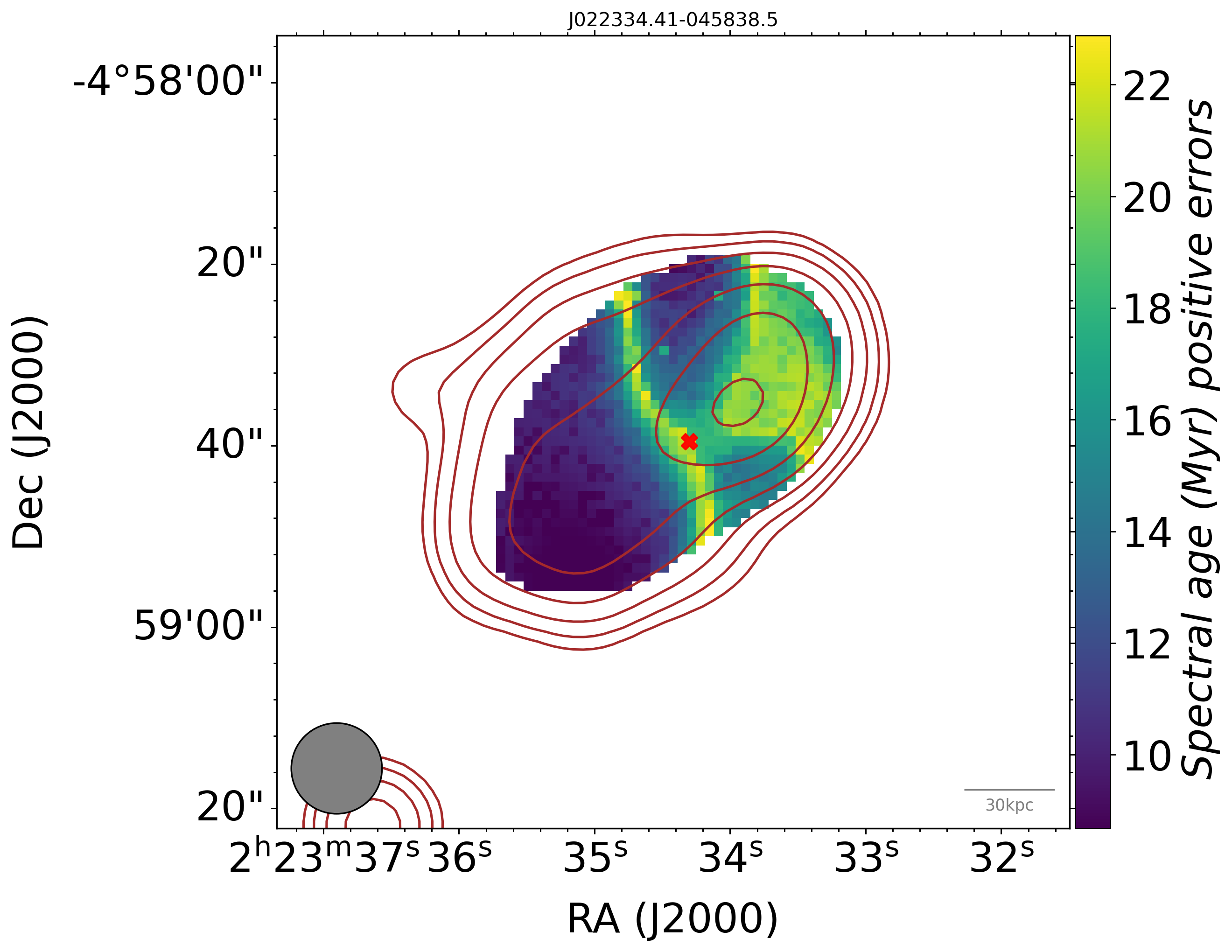}\includegraphics[width=1.4in,height=1.3in]{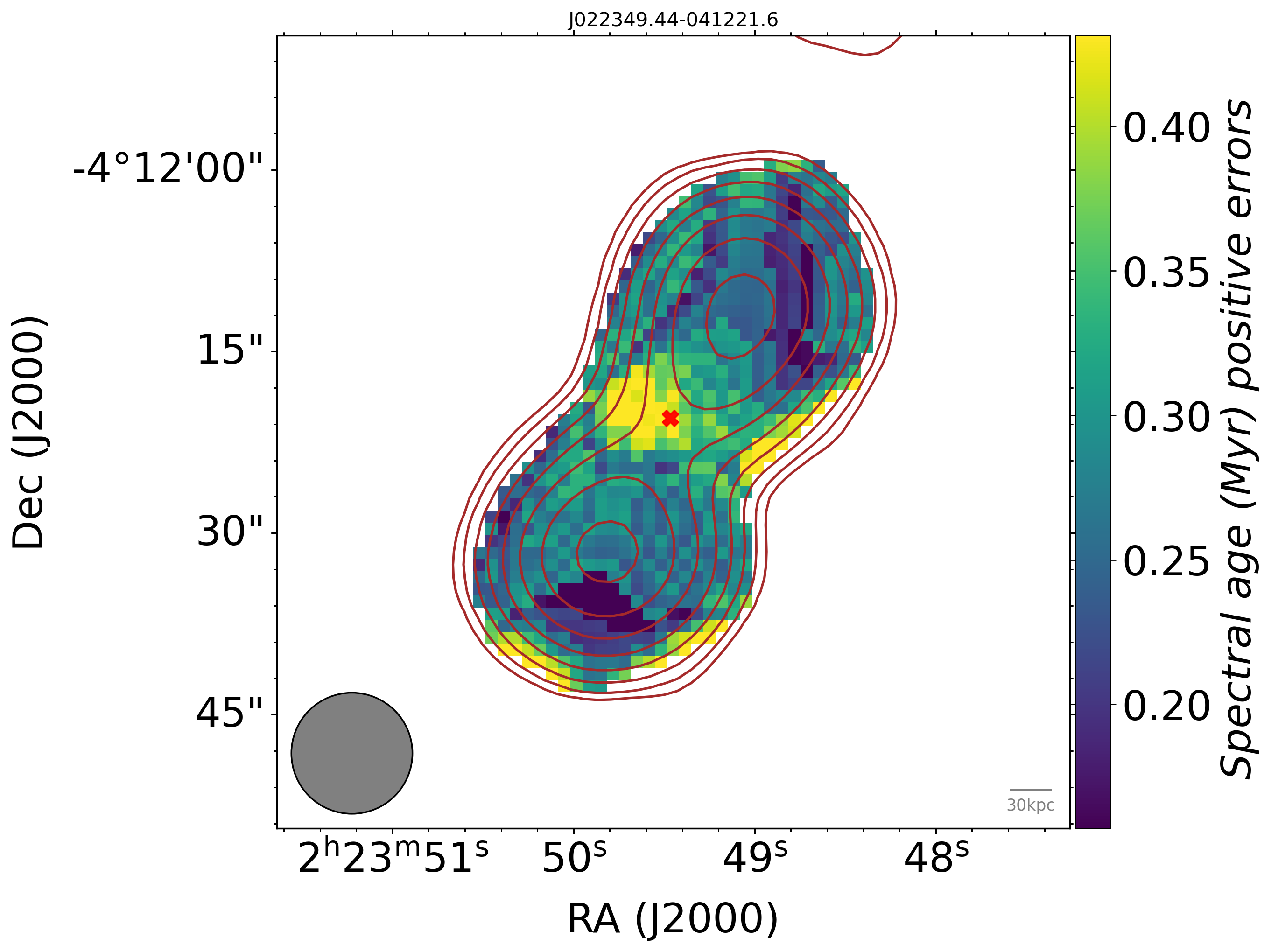}
	\includegraphics[width=1.4in,height=1.3in]{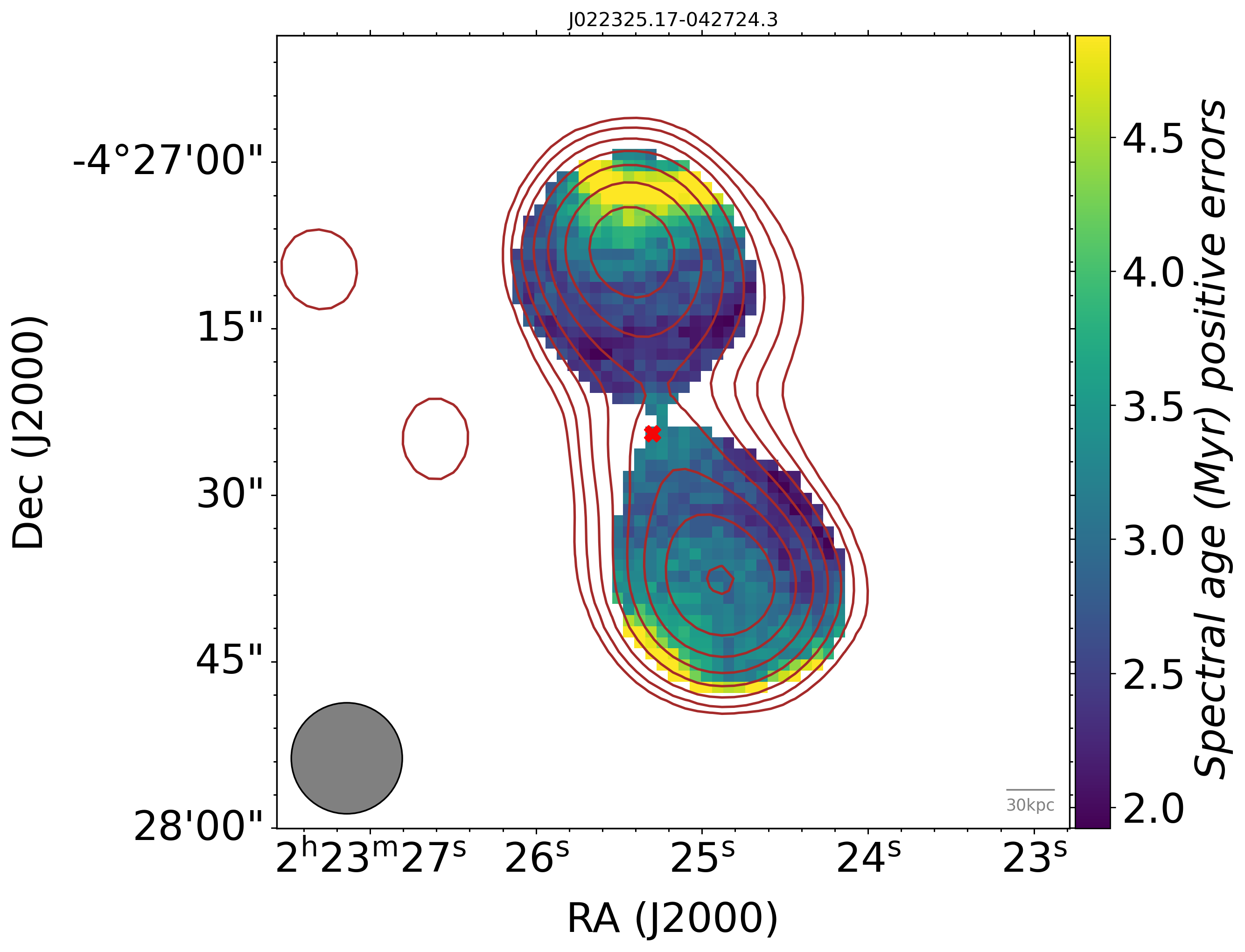}\includegraphics[width=1.4in,height=1.3in]{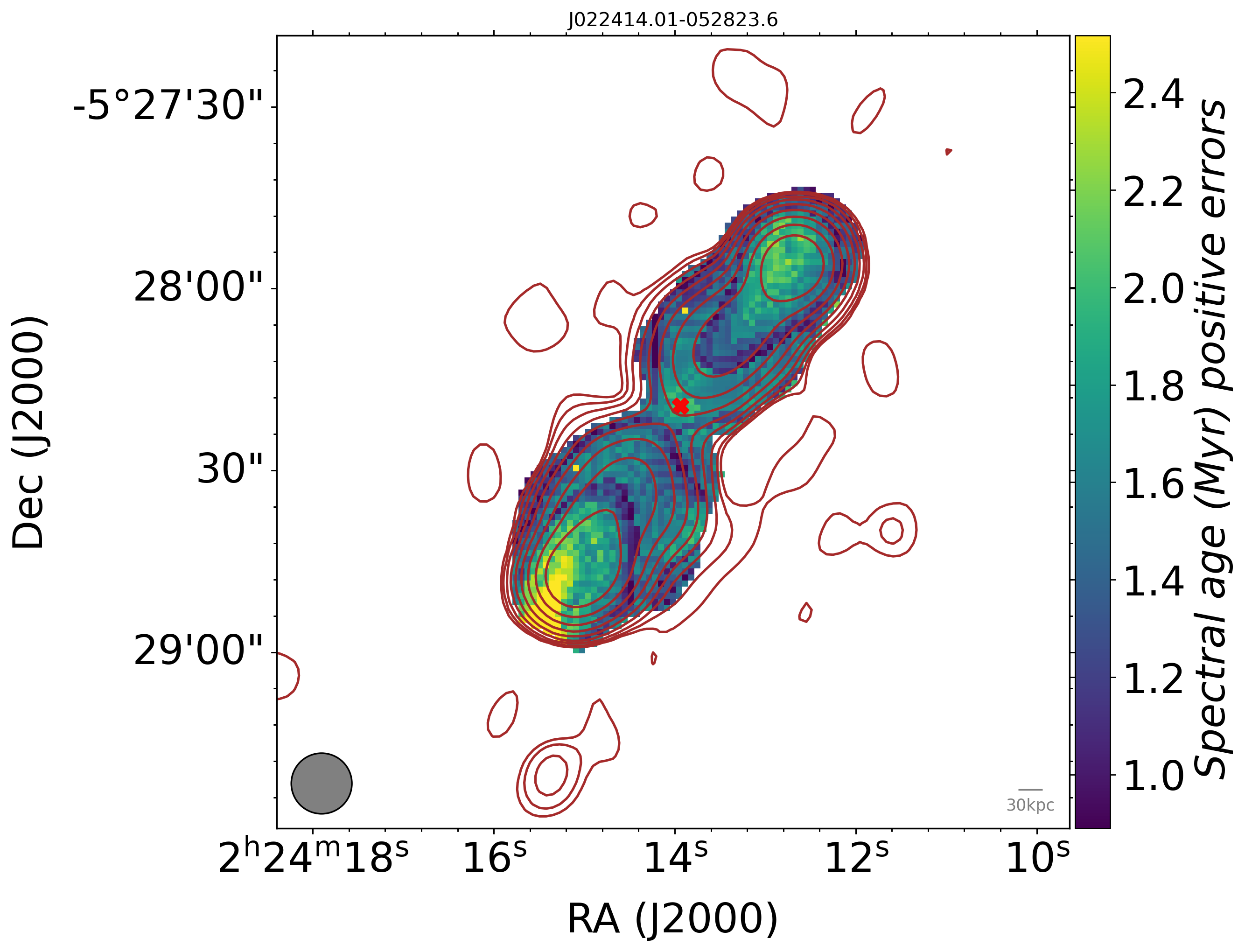}\includegraphics[width=1.4in,height=1.3in]{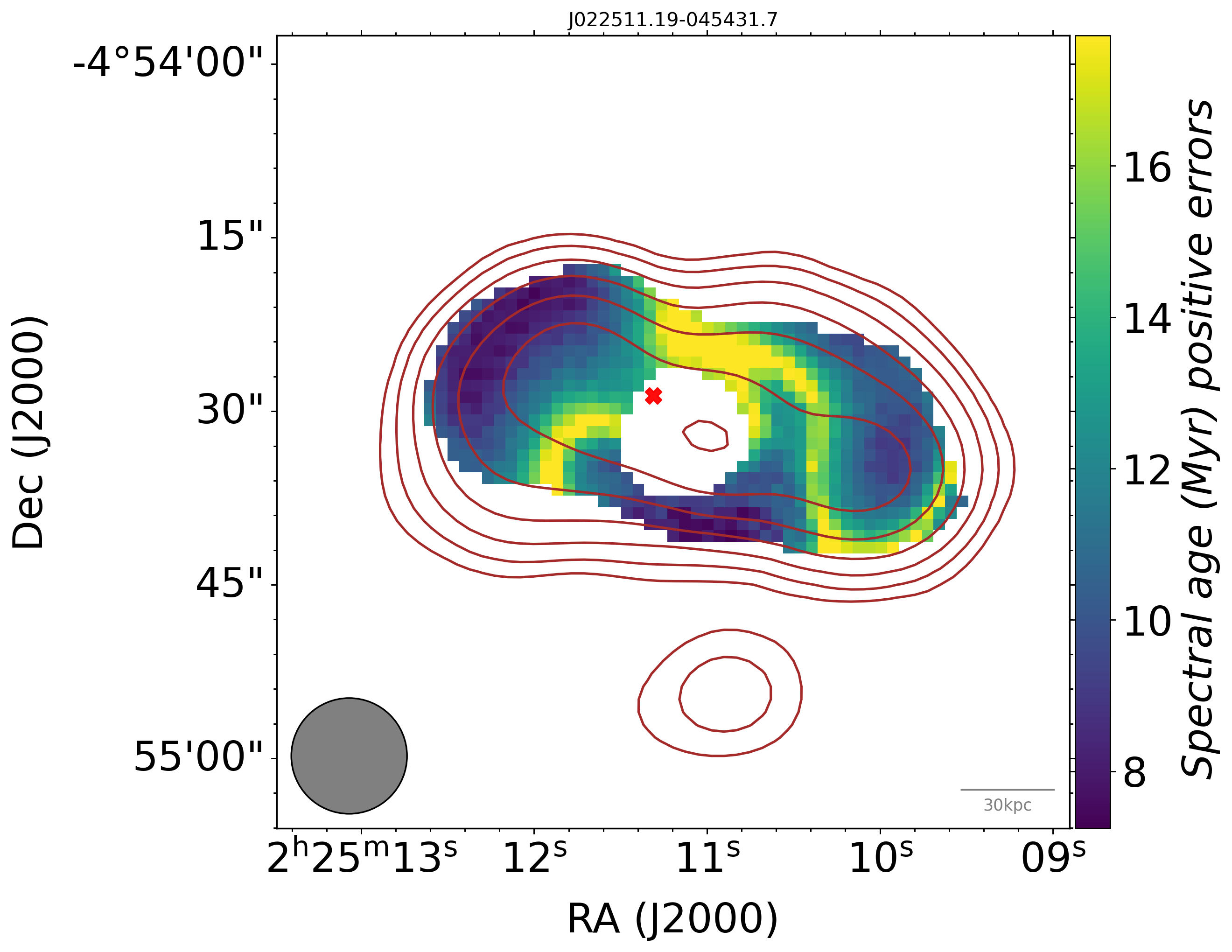}\includegraphics[width=1.4in,height=1.3in]{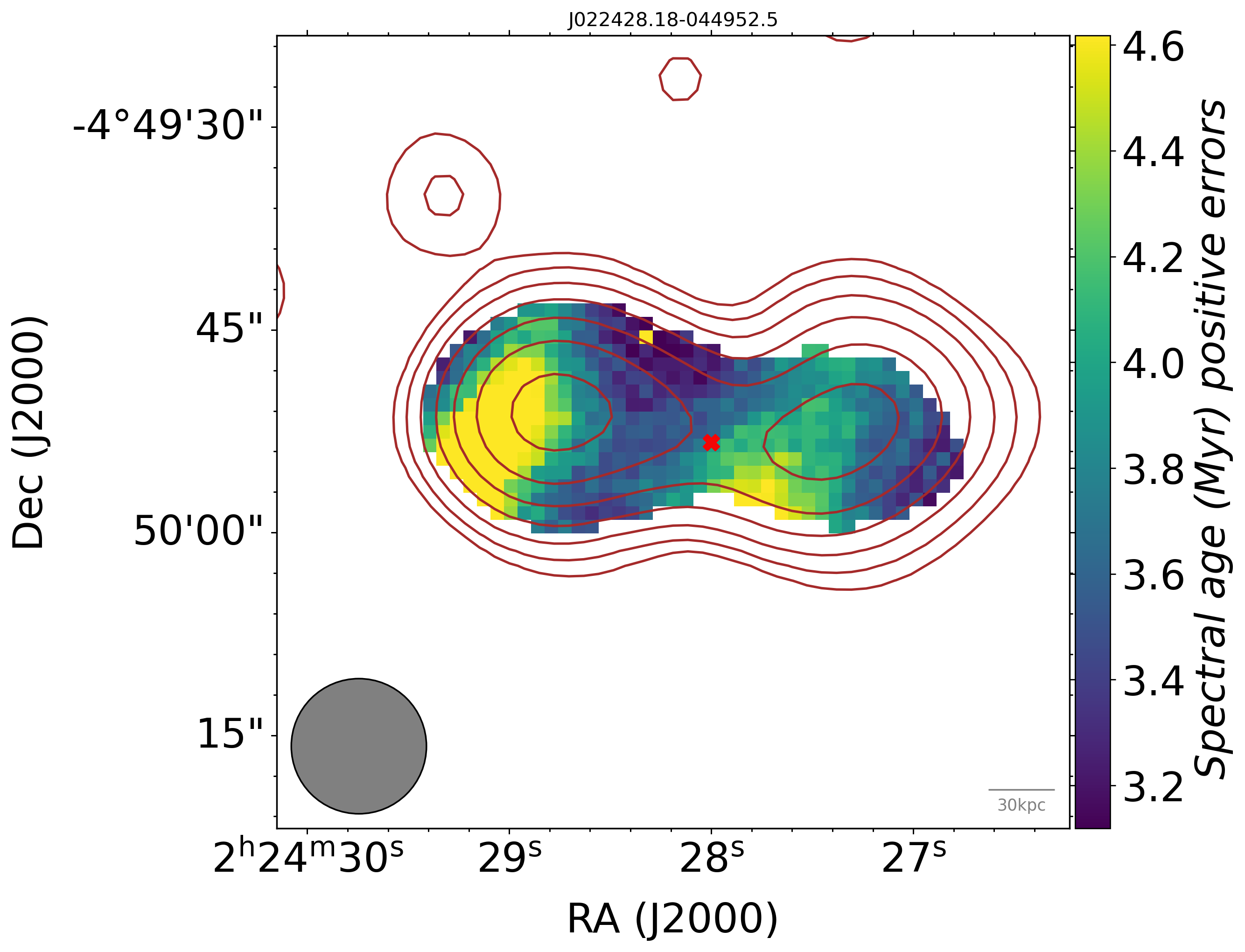}
	\includegraphics[width=1.4in,height=1.3in]{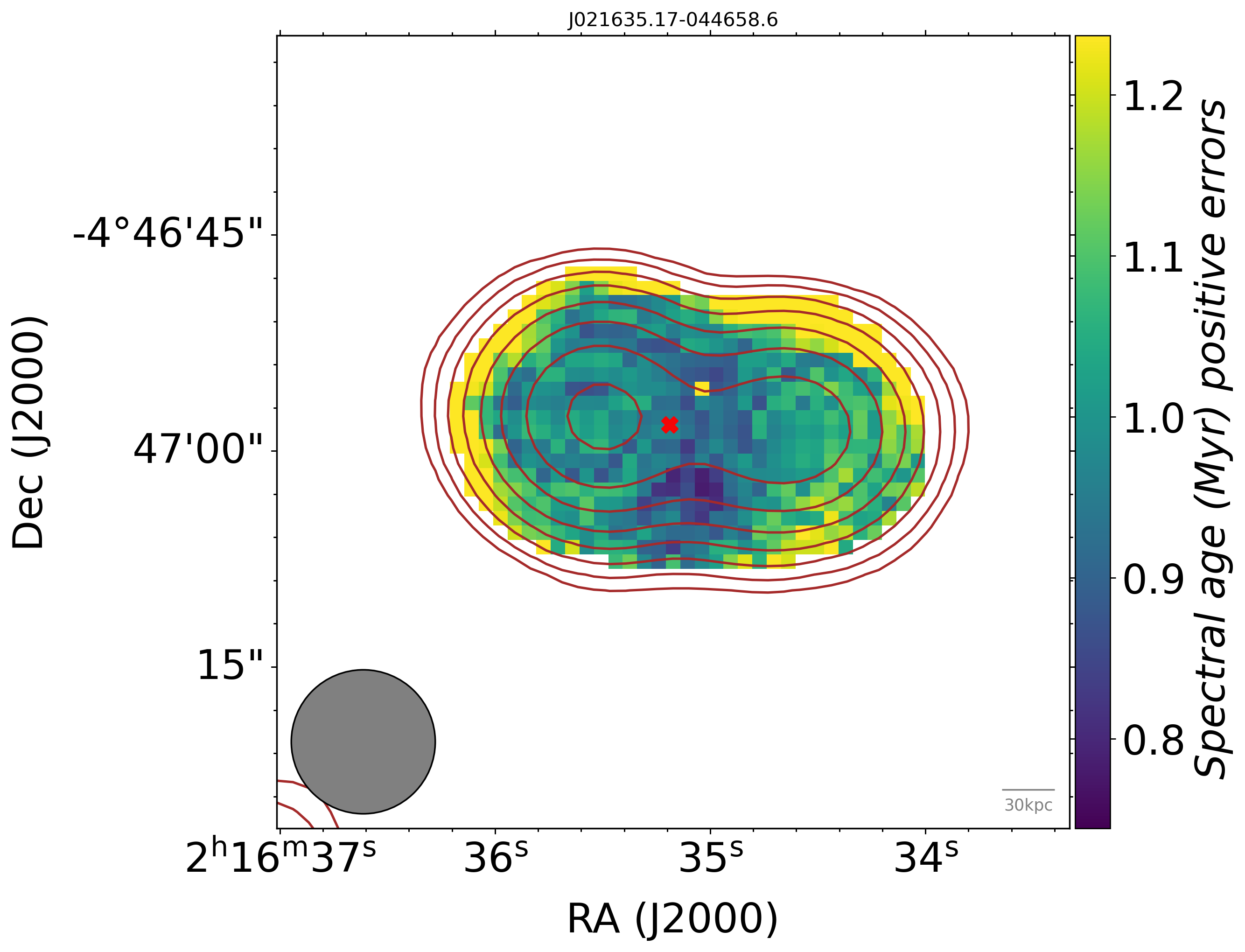}\includegraphics[width=1.4in,height=1.3in]{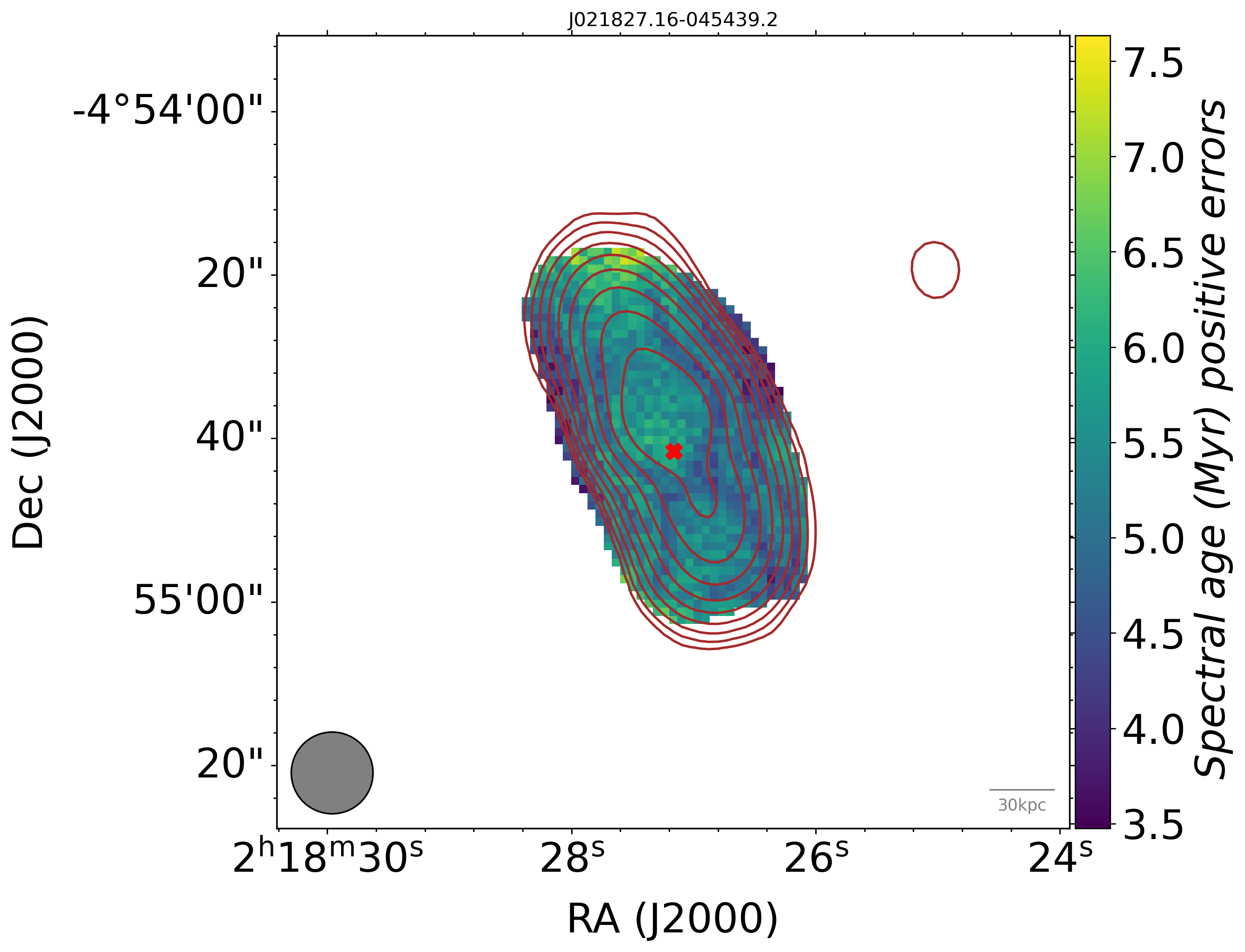}\includegraphics[width=1.4in,height=1.3in]{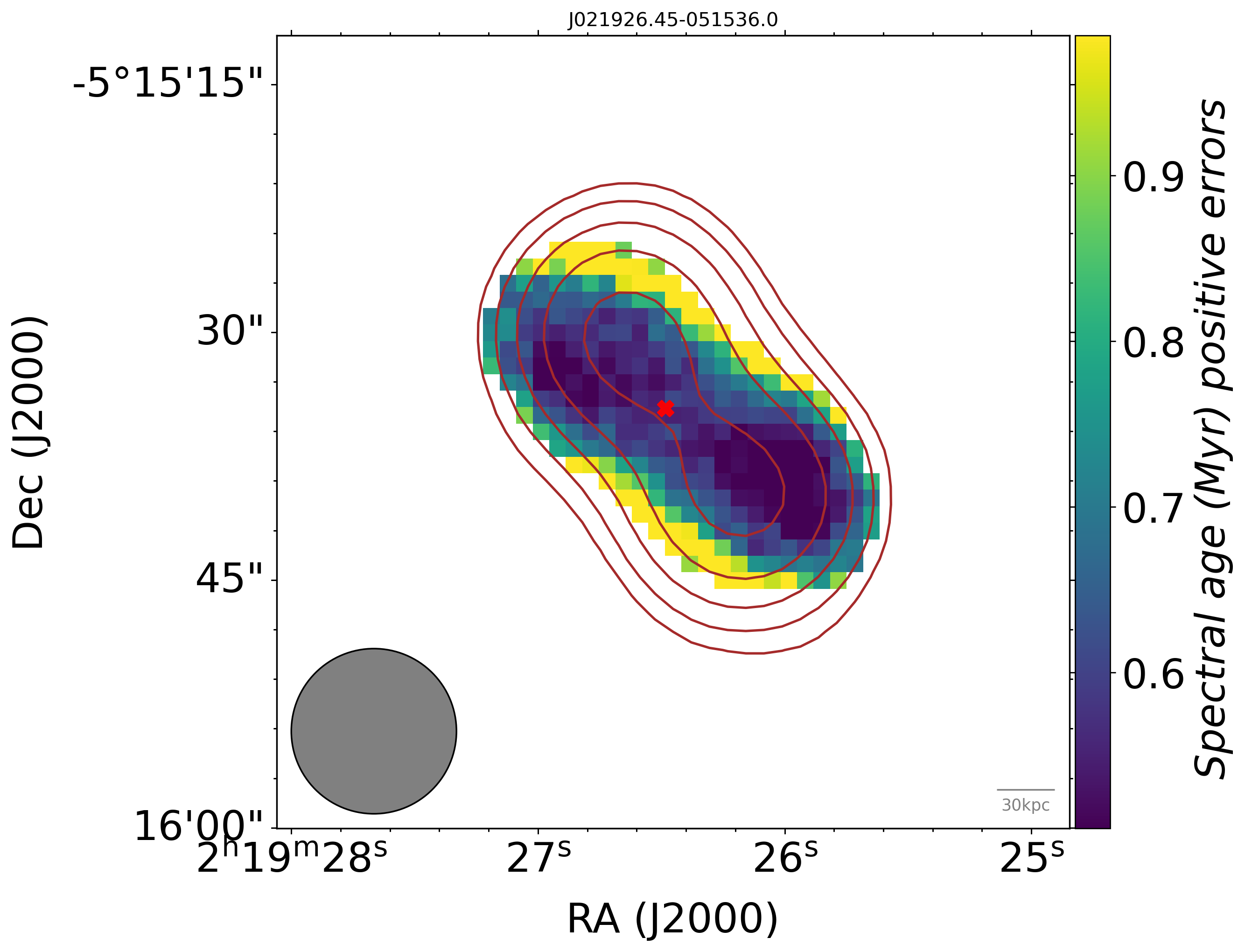}\includegraphics[width=1.4in,height=1.3in]{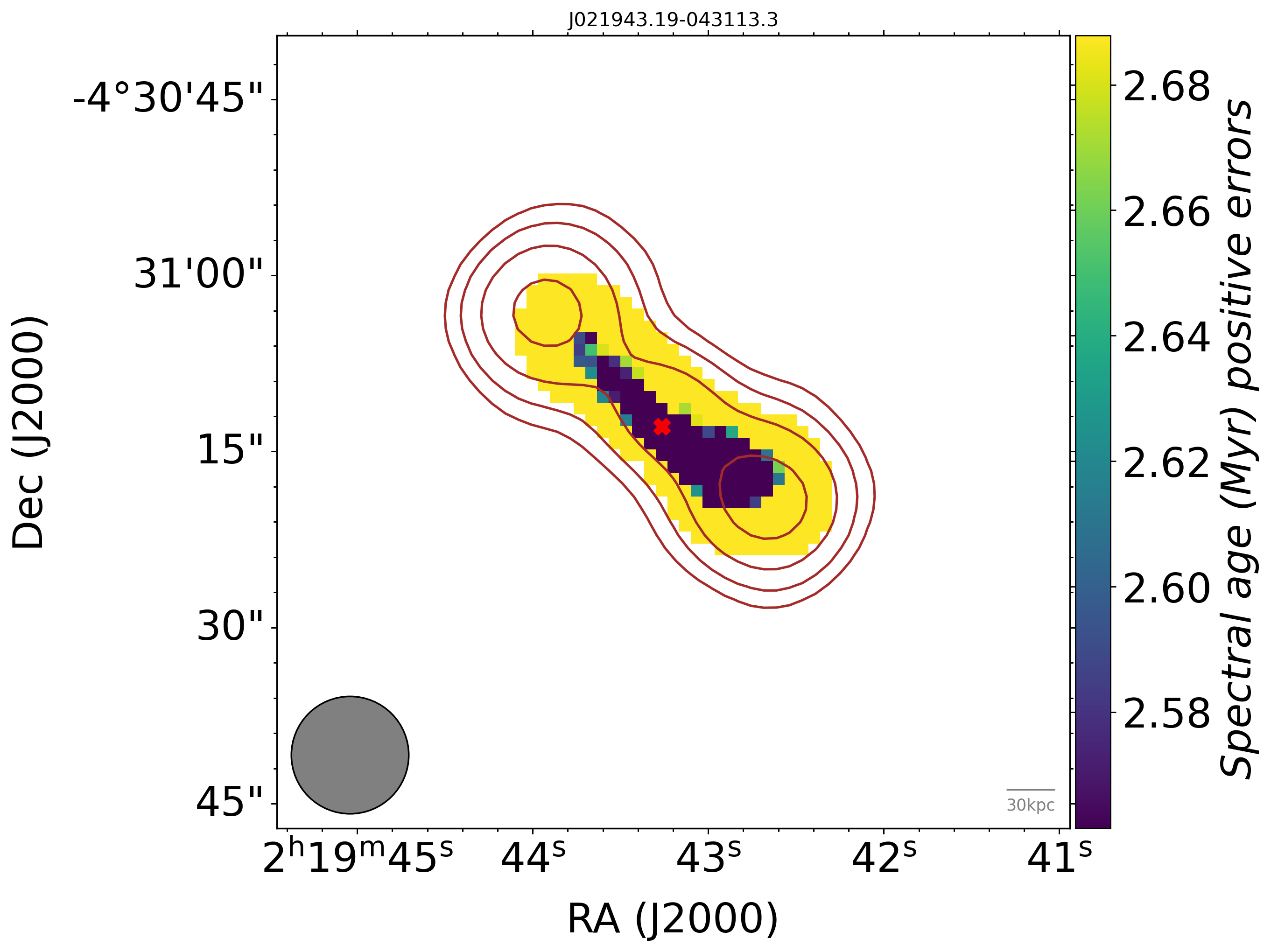}
	\includegraphics[width=1.4in,height=1.3in]{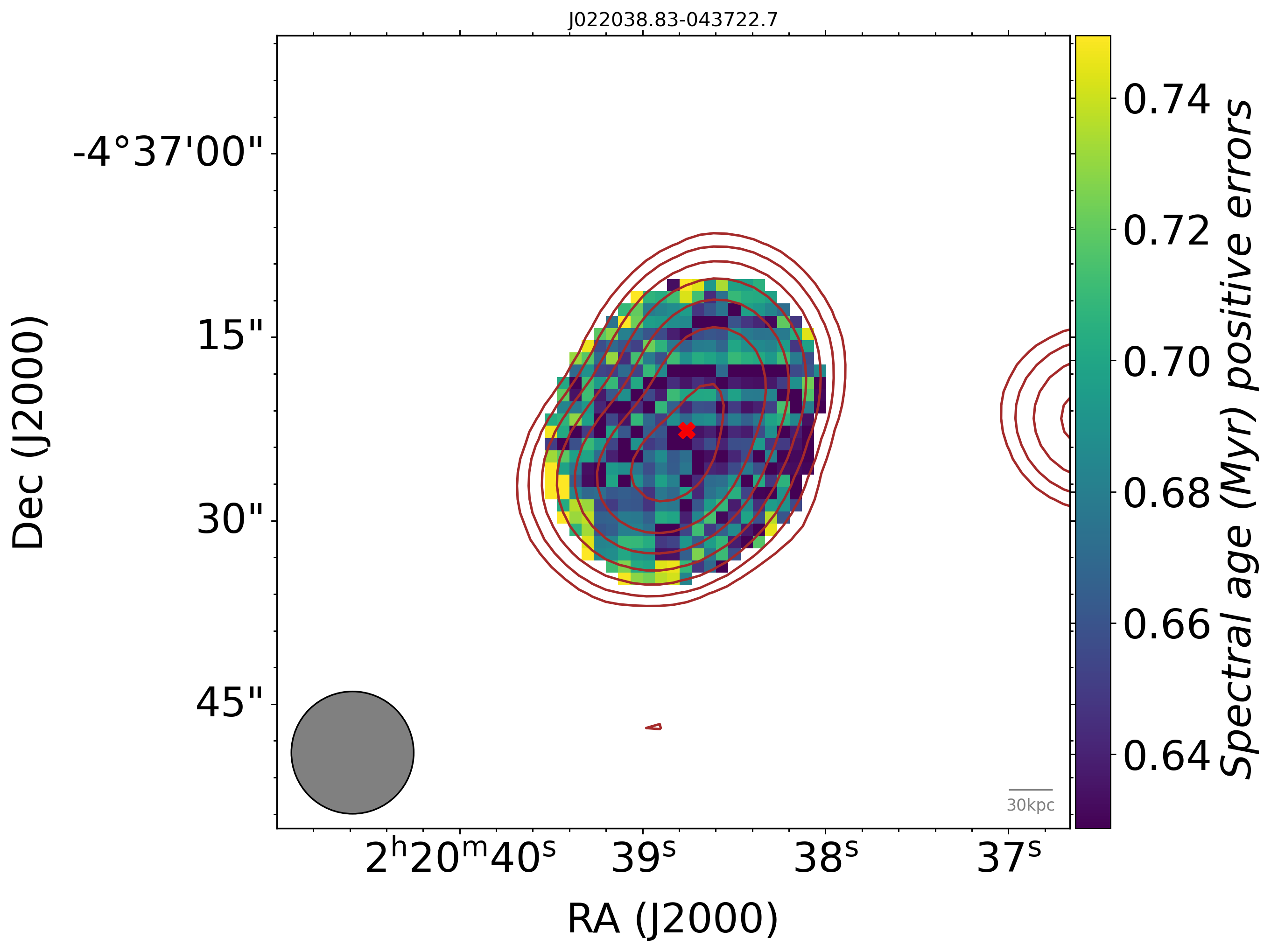}\includegraphics[width=1.4in,height=1.3in]{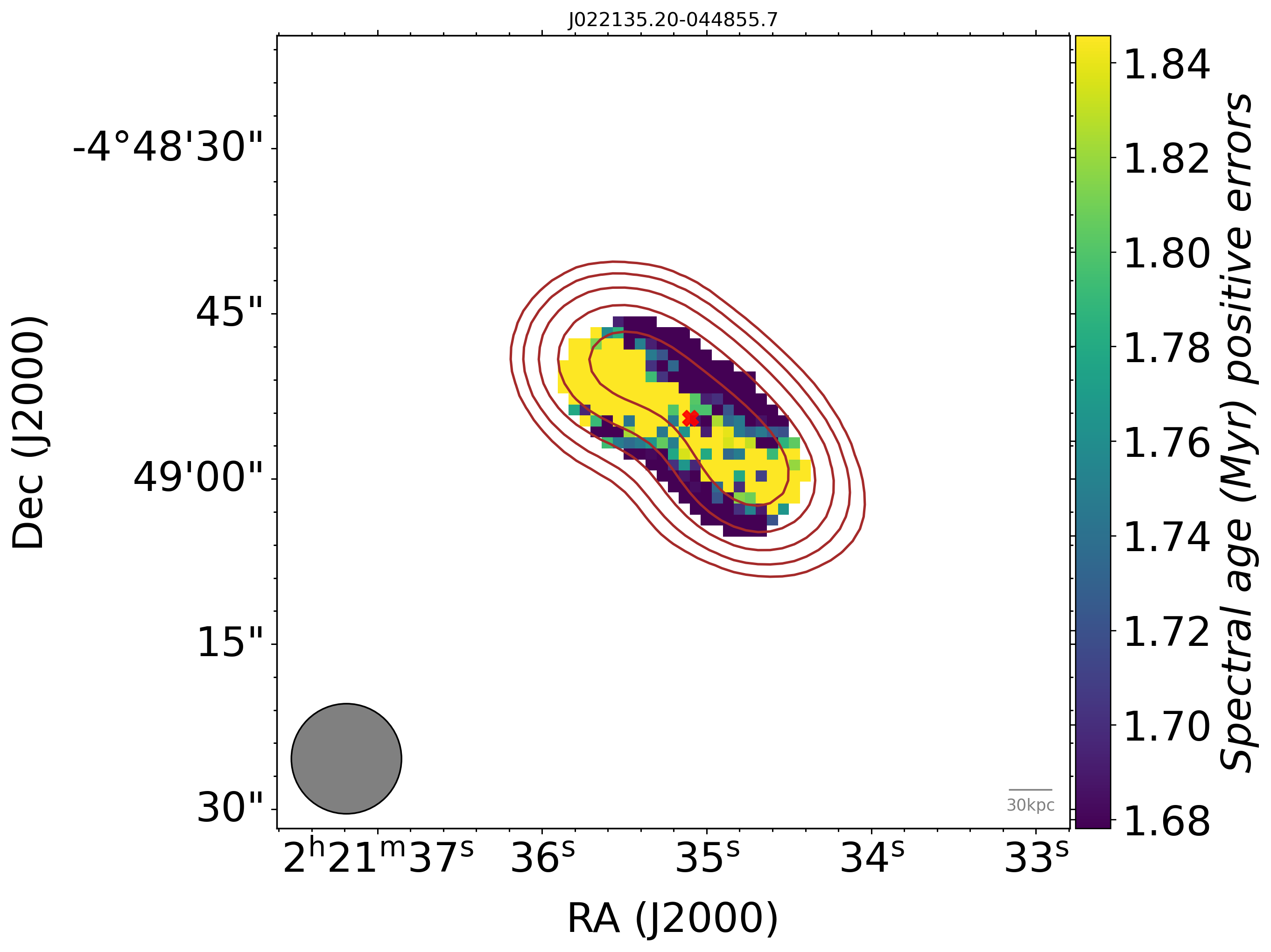}\includegraphics[width=1.4in,height=1.3in]{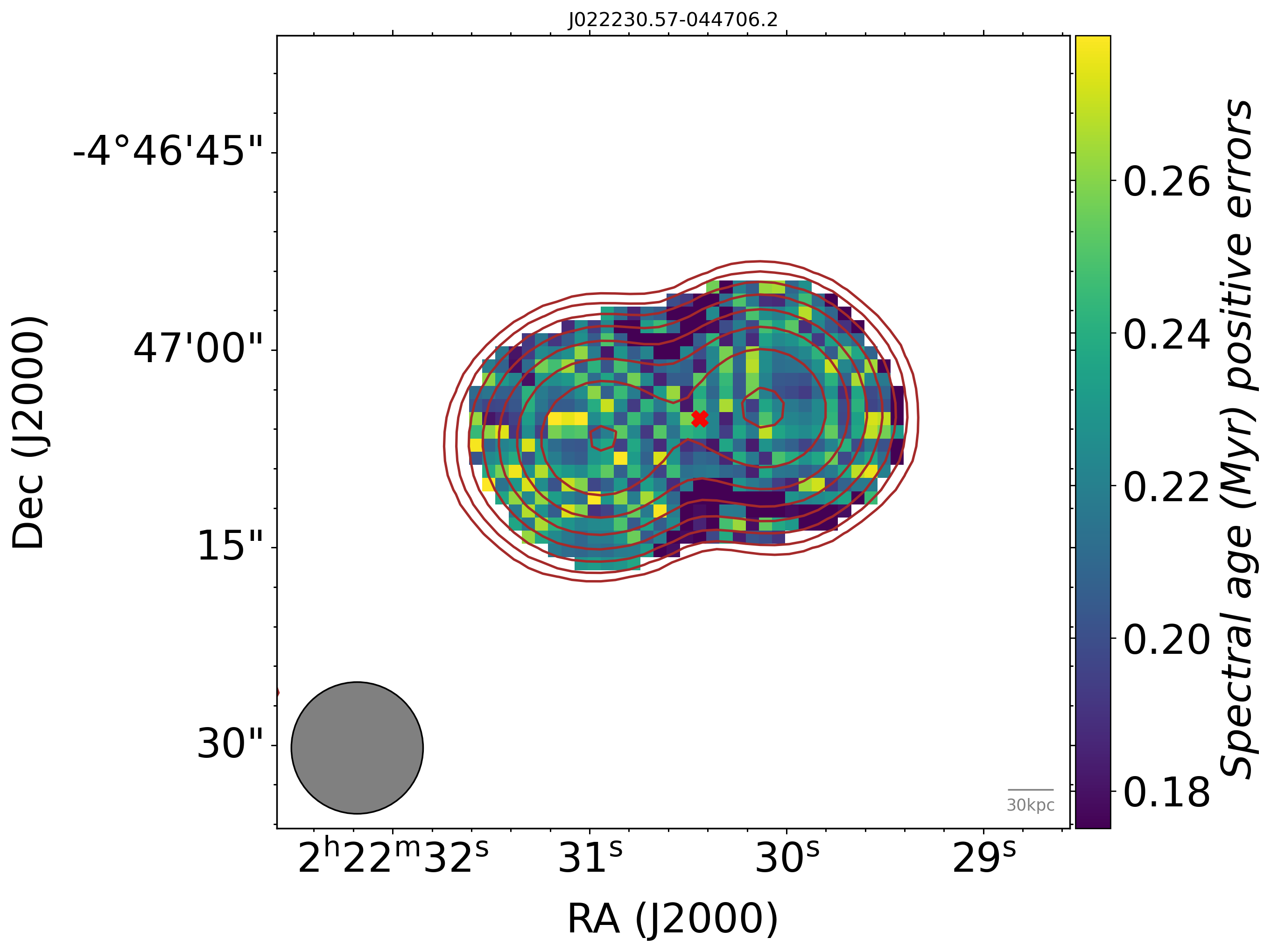}\includegraphics[width=1.4in,height=1.3in]{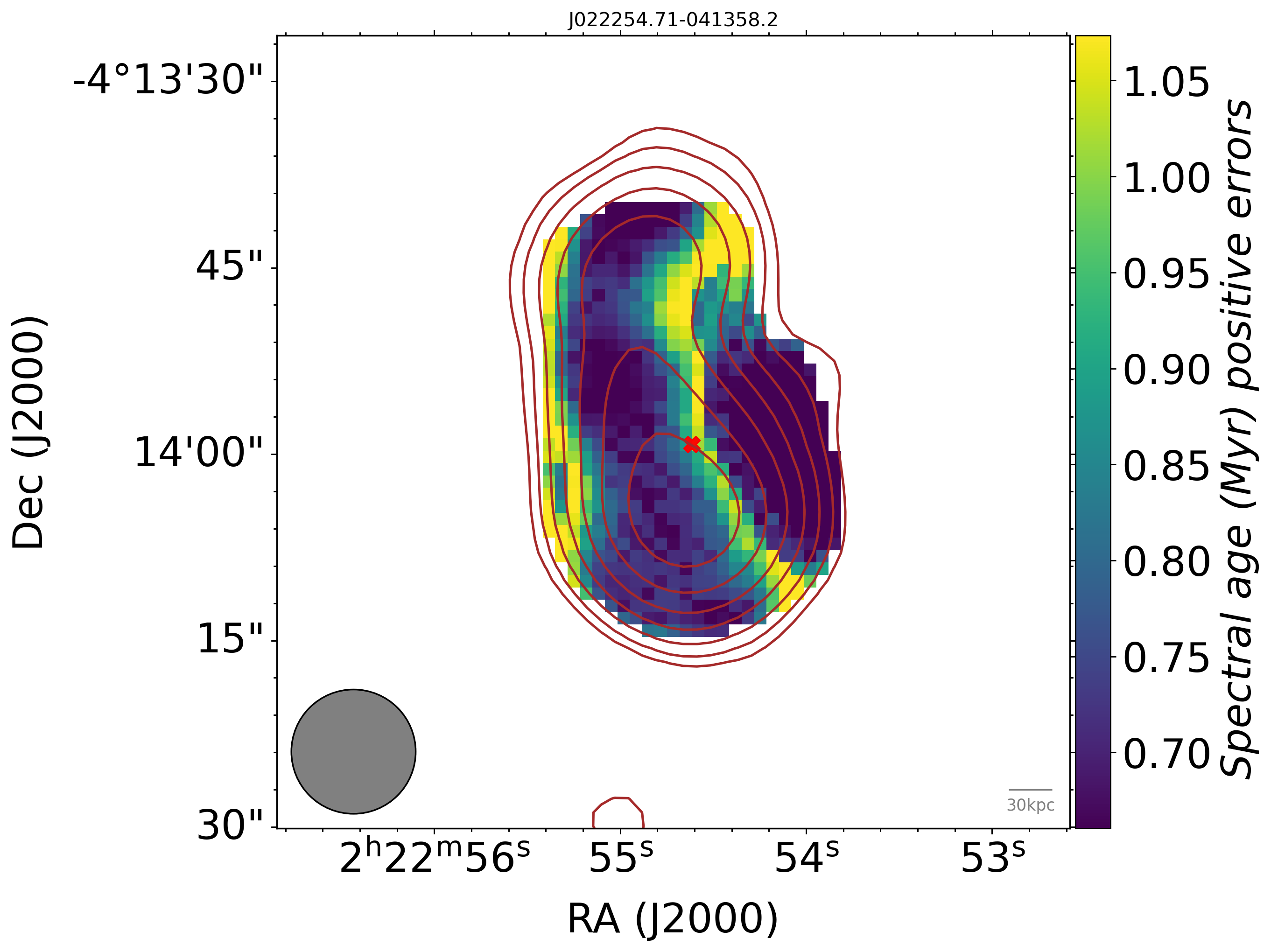}
	\includegraphics[width=1.4in,height=1.3in]{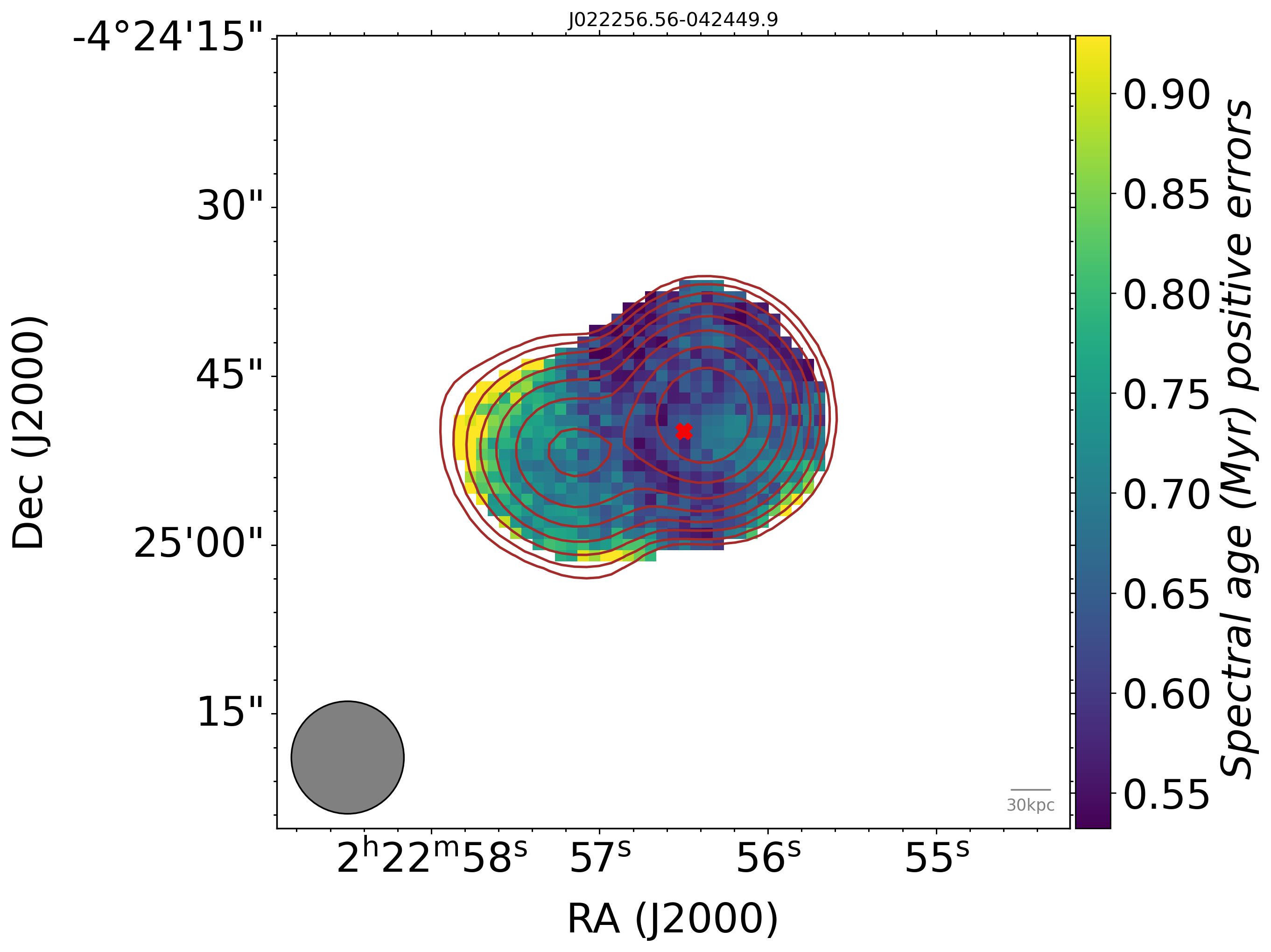}\includegraphics[width=1.4in,height=1.3in]{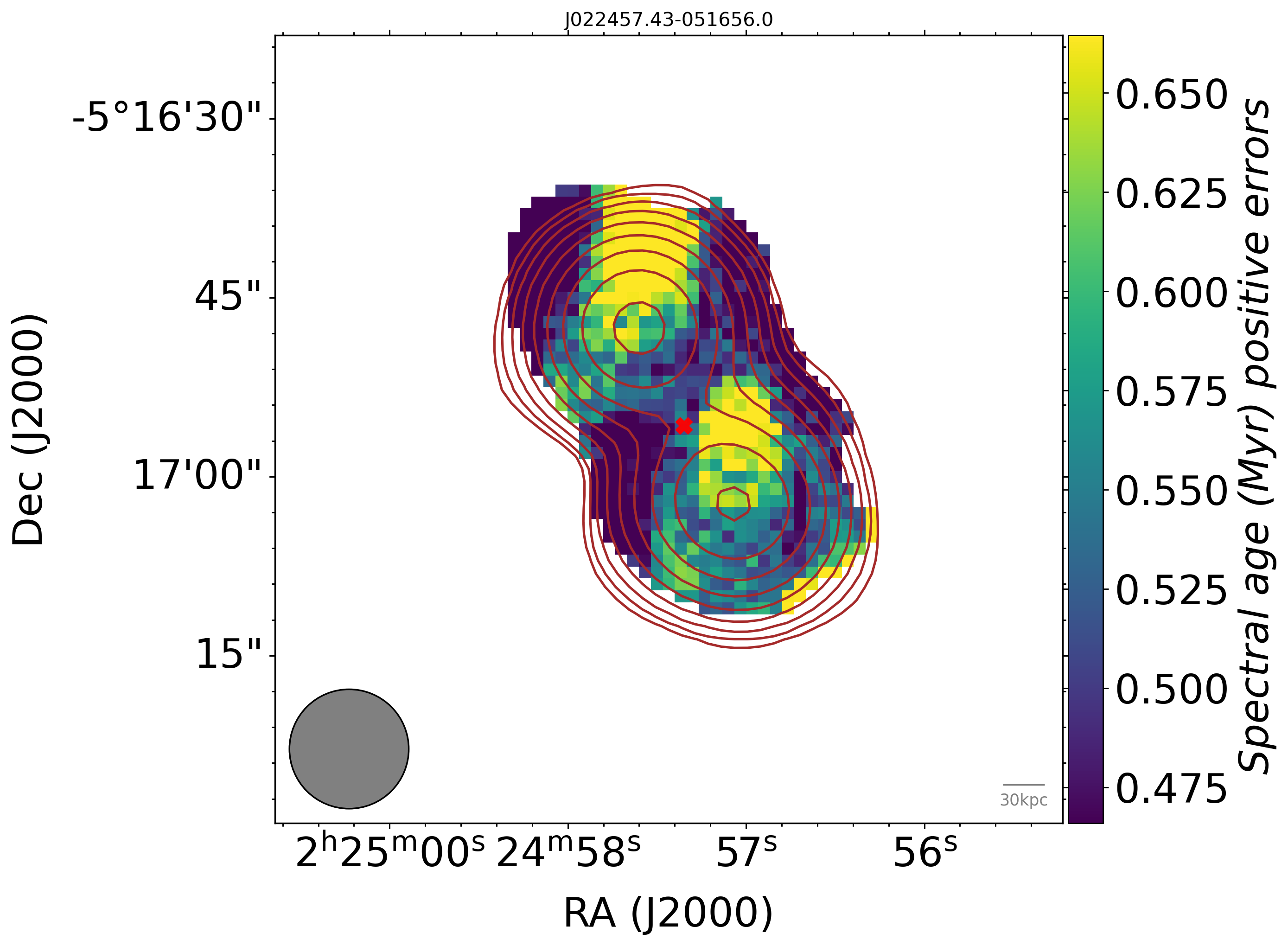}\includegraphics[width=1.4in,height=1.3in]{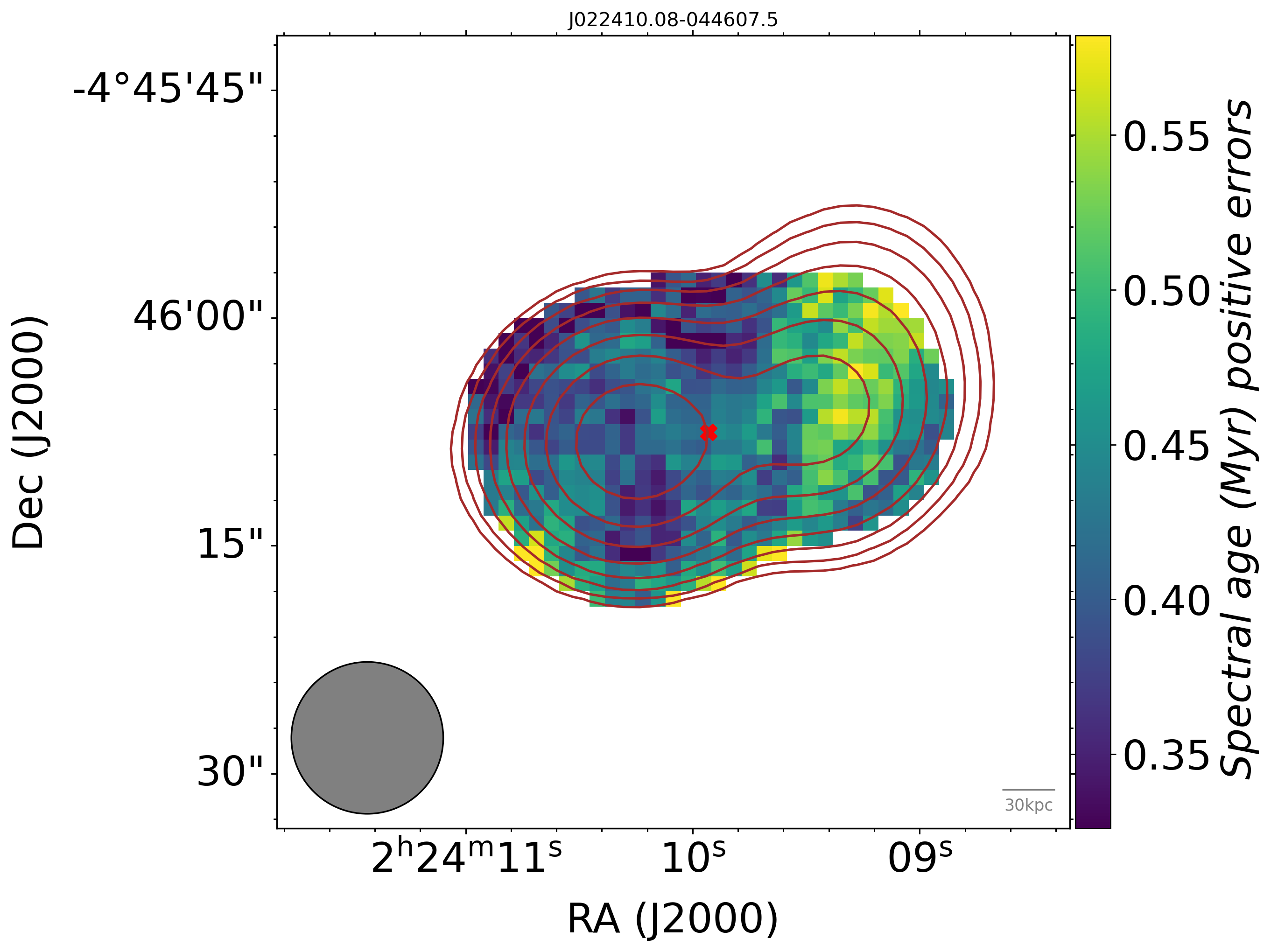}\includegraphics[width=1.4in,height=1.3in]{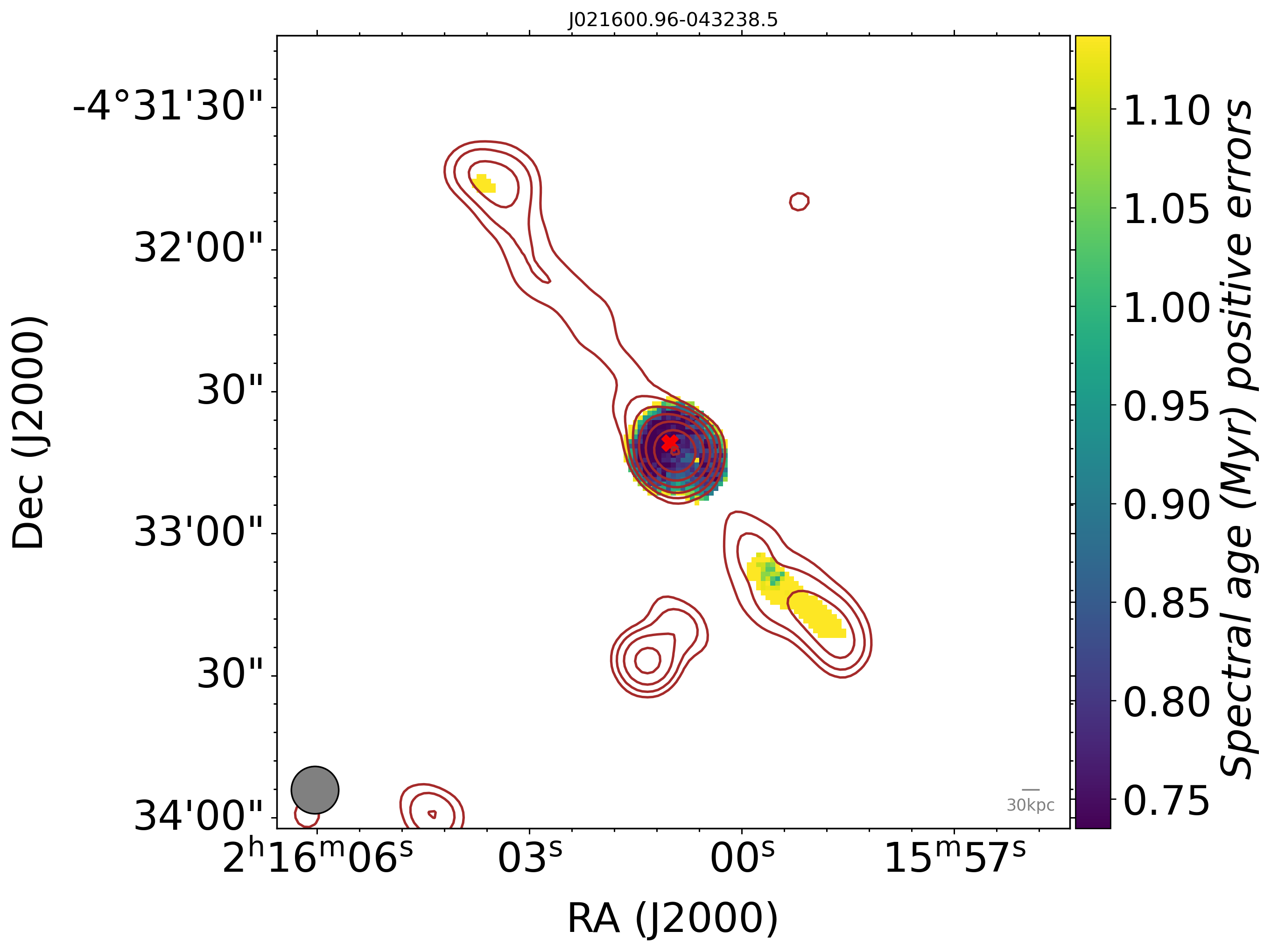}
	\includegraphics[width=1.4in,height=1.3in]{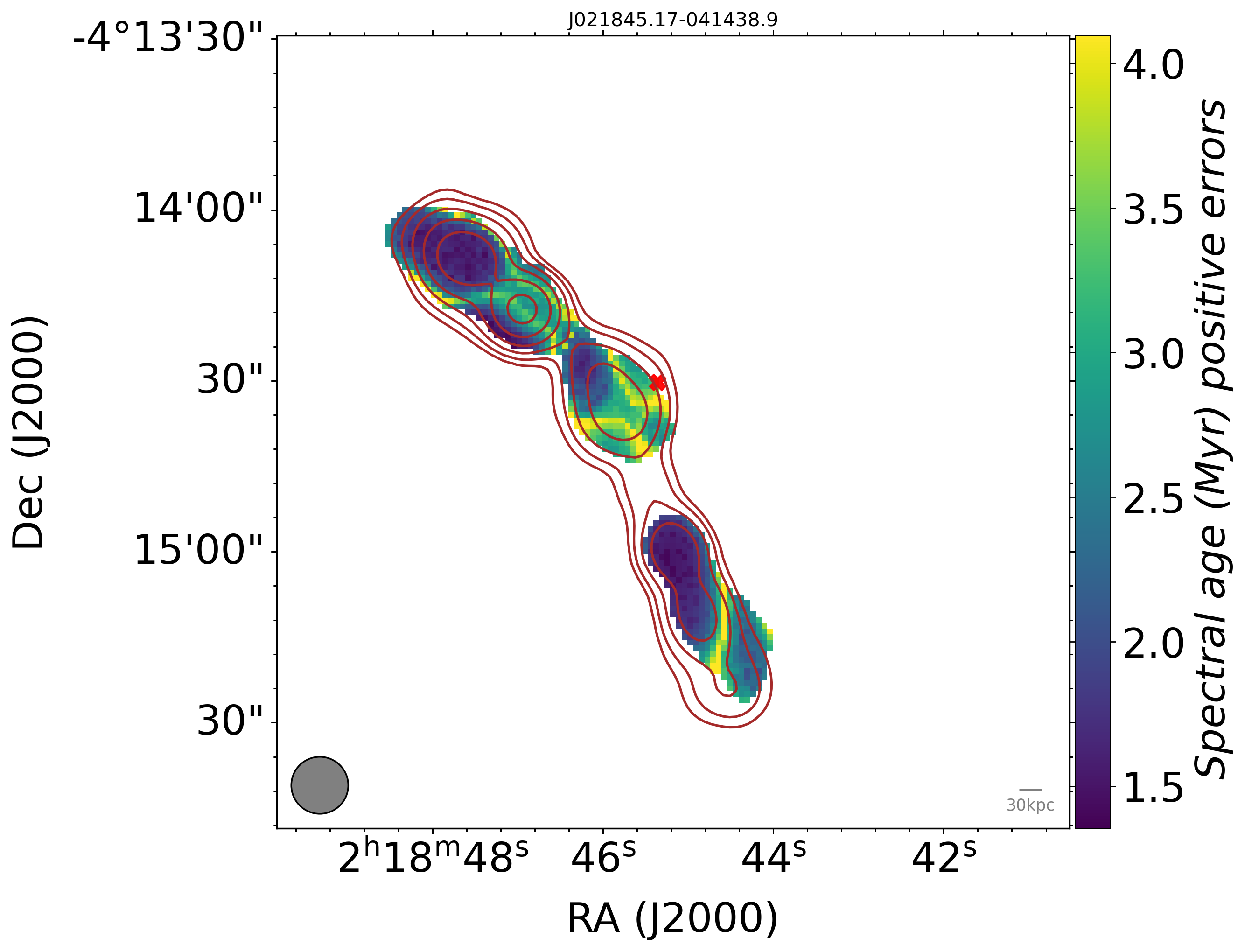}\includegraphics[width=1.4in,height=1.3in]{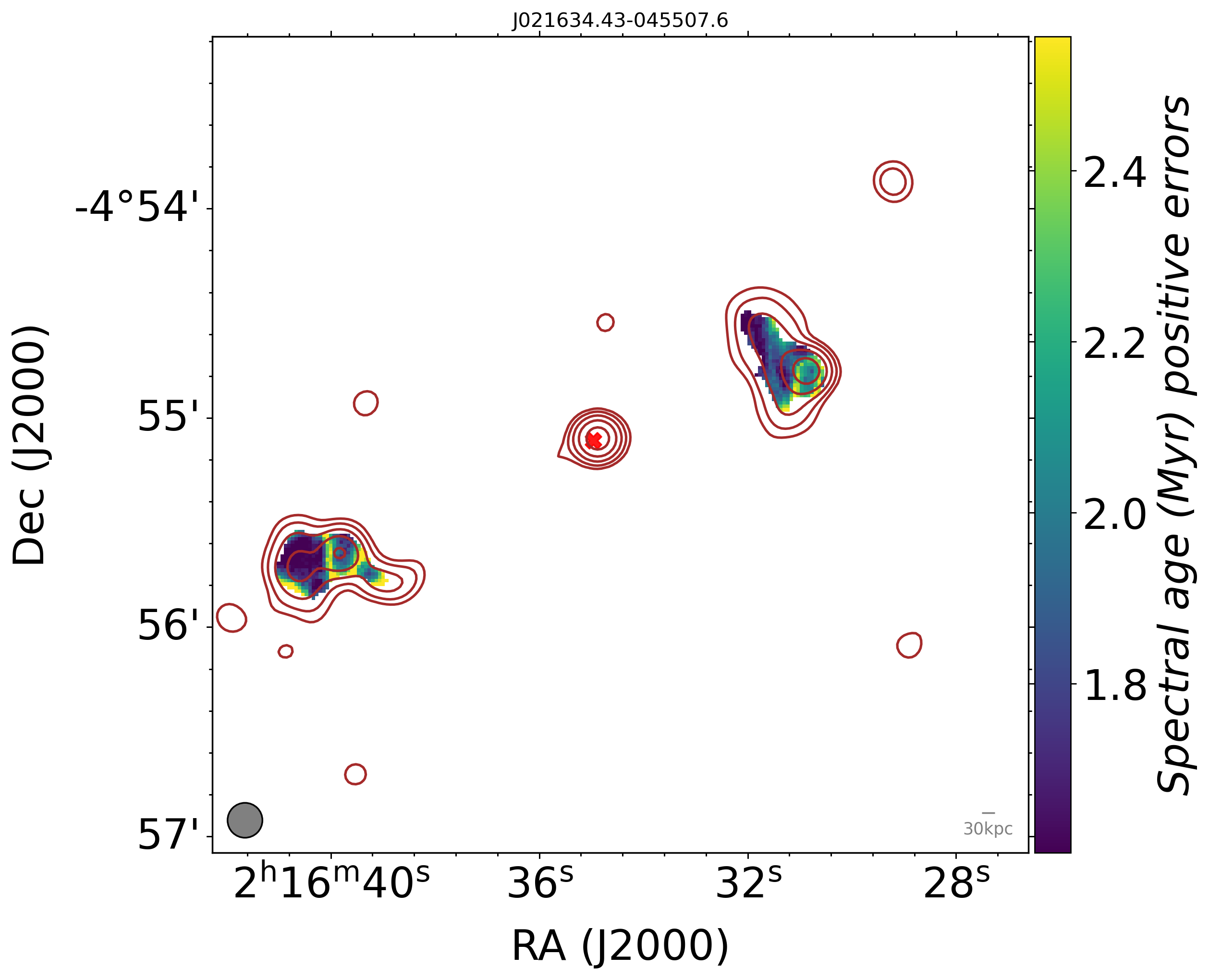}\includegraphics[width=1.4in,height=1.3in]{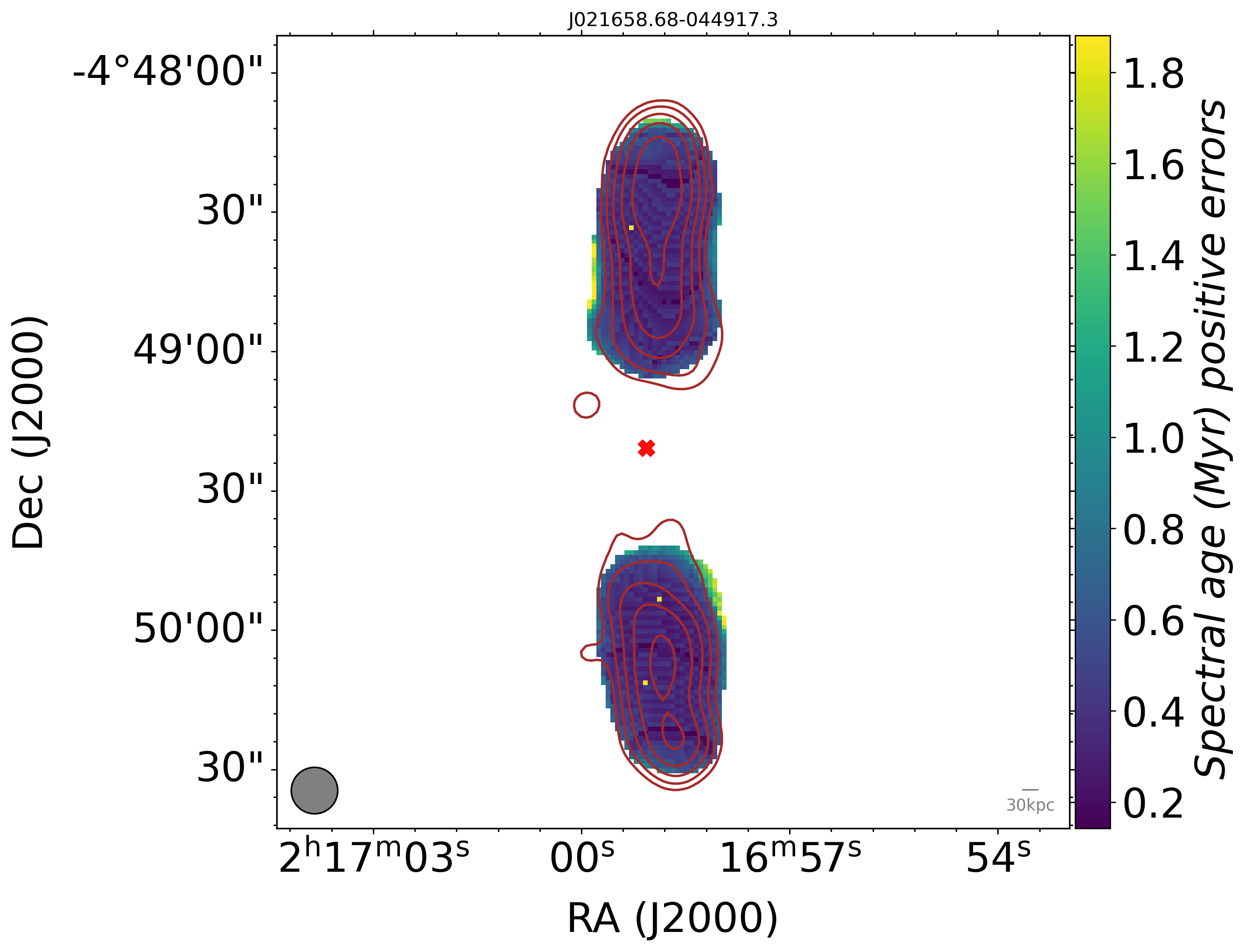}\includegraphics[width=1.4in,height=1.3in]{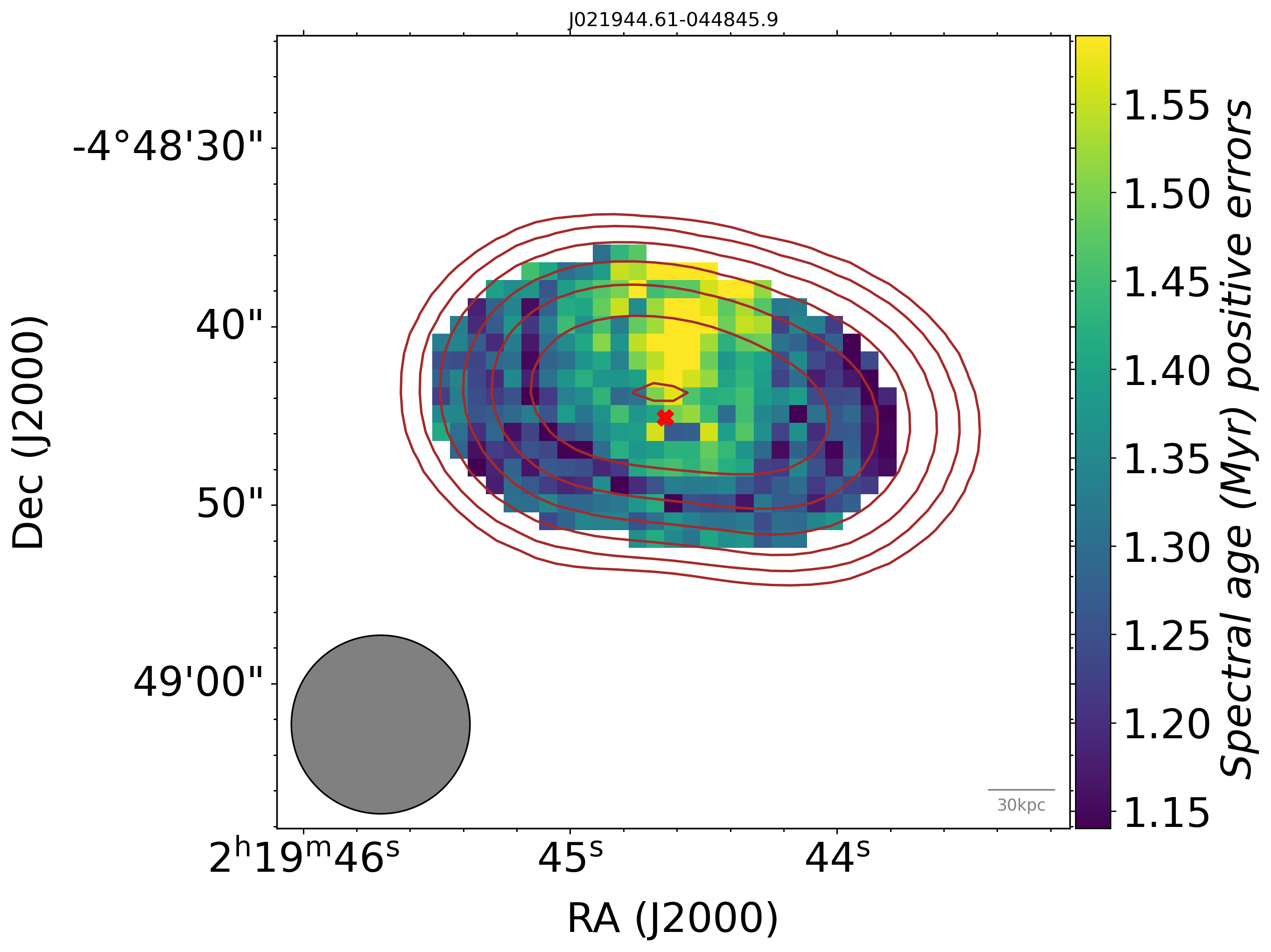}
	\caption{Positive error distribution of age for each source.}
	\label{ap}
\end{figure*}

\begin{figure*}
\includegraphics[width=1.3in,height=1.3in]{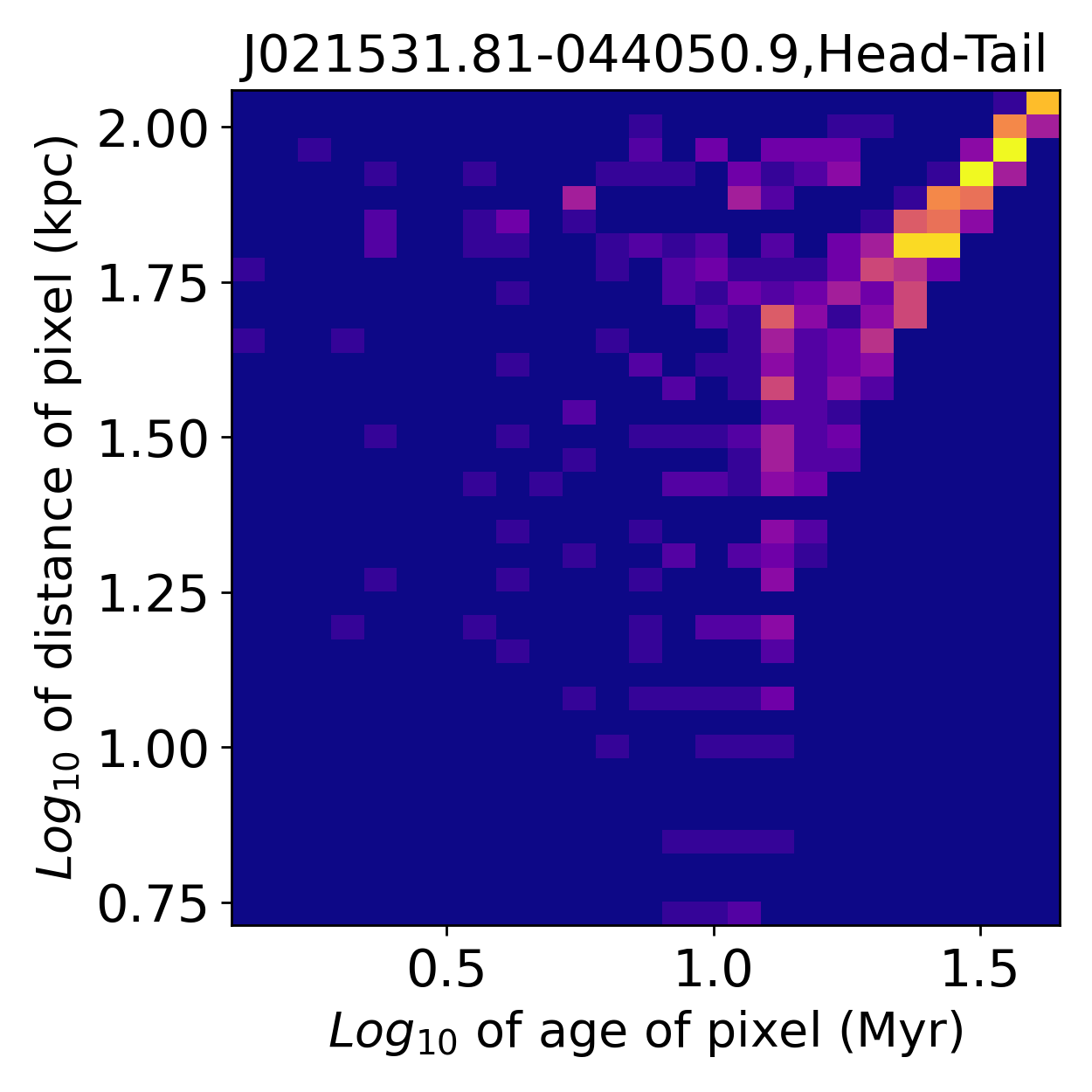}\includegraphics[width=1.3in,height=1.3in]{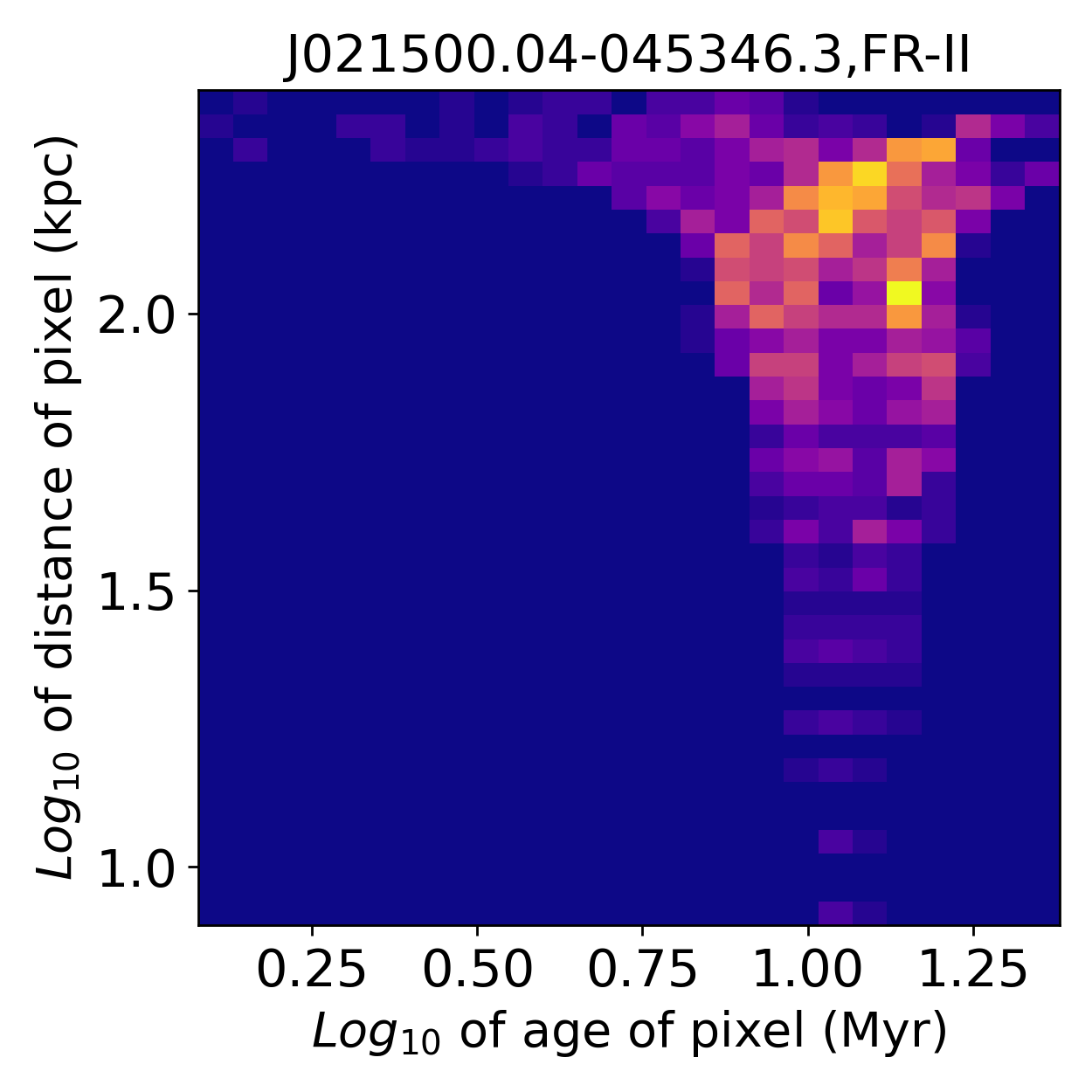}\includegraphics[width=1.3in,height=1.3in]{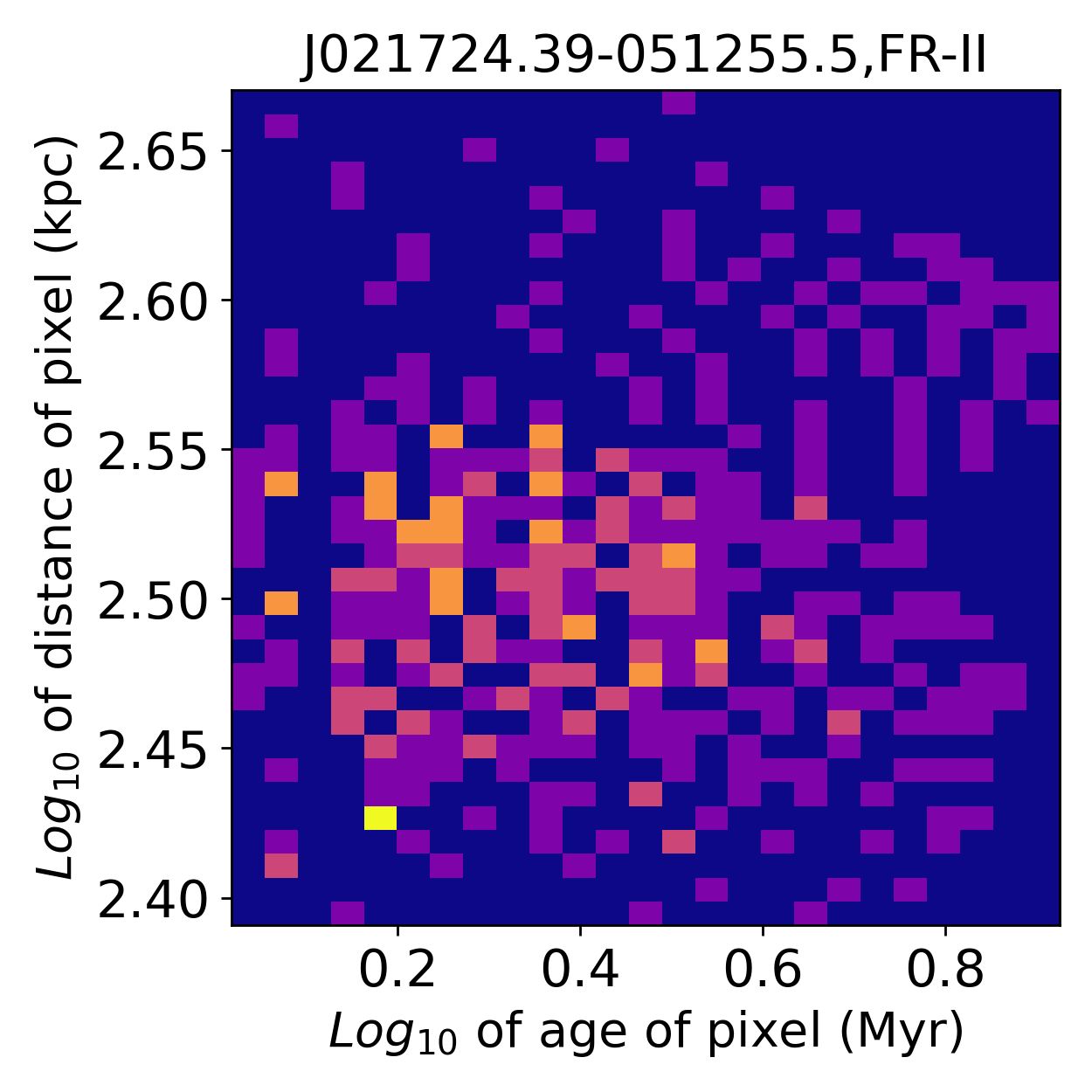}\includegraphics[width=1.3in,height=1.3in]{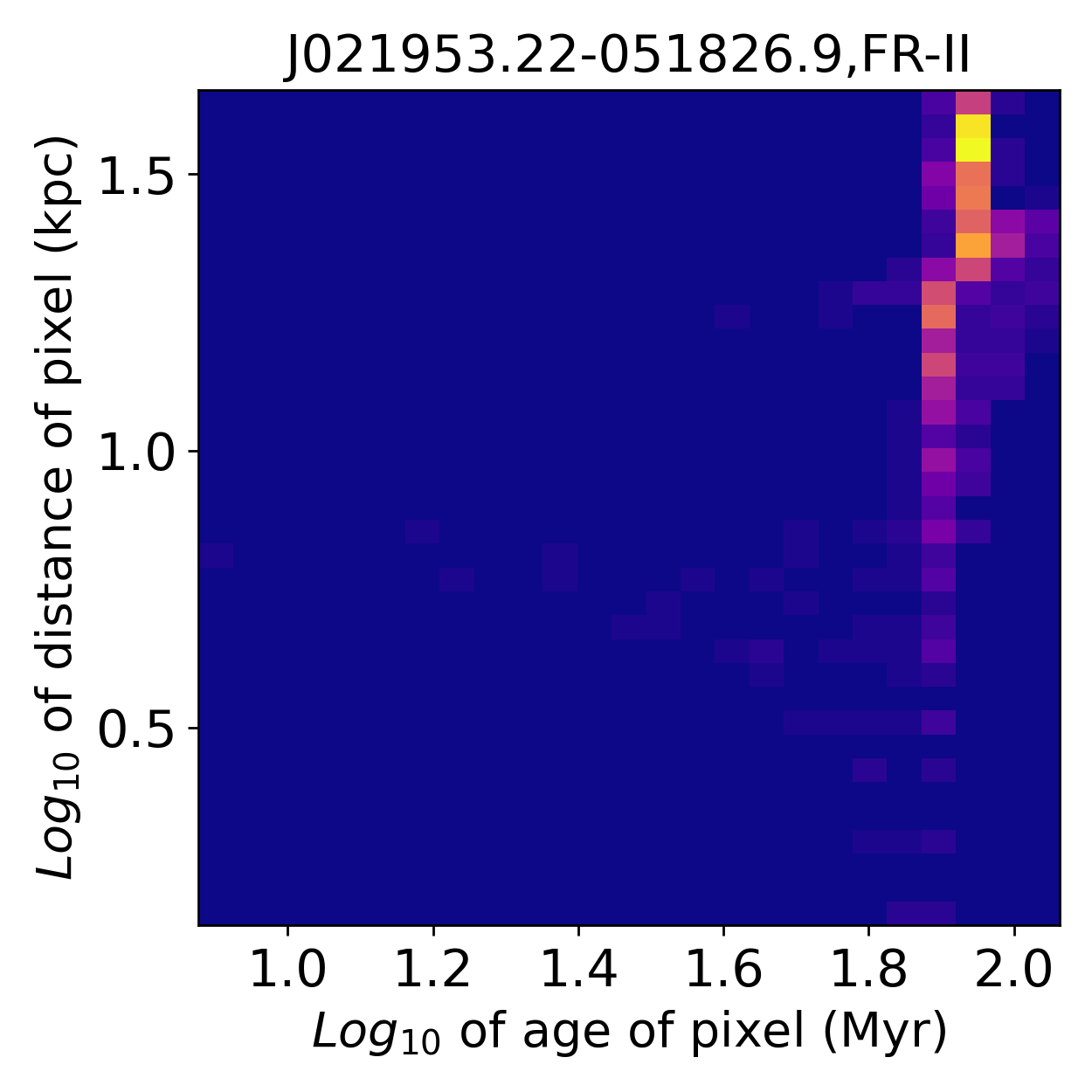}
\includegraphics[width=1.3in,height=1.3in]{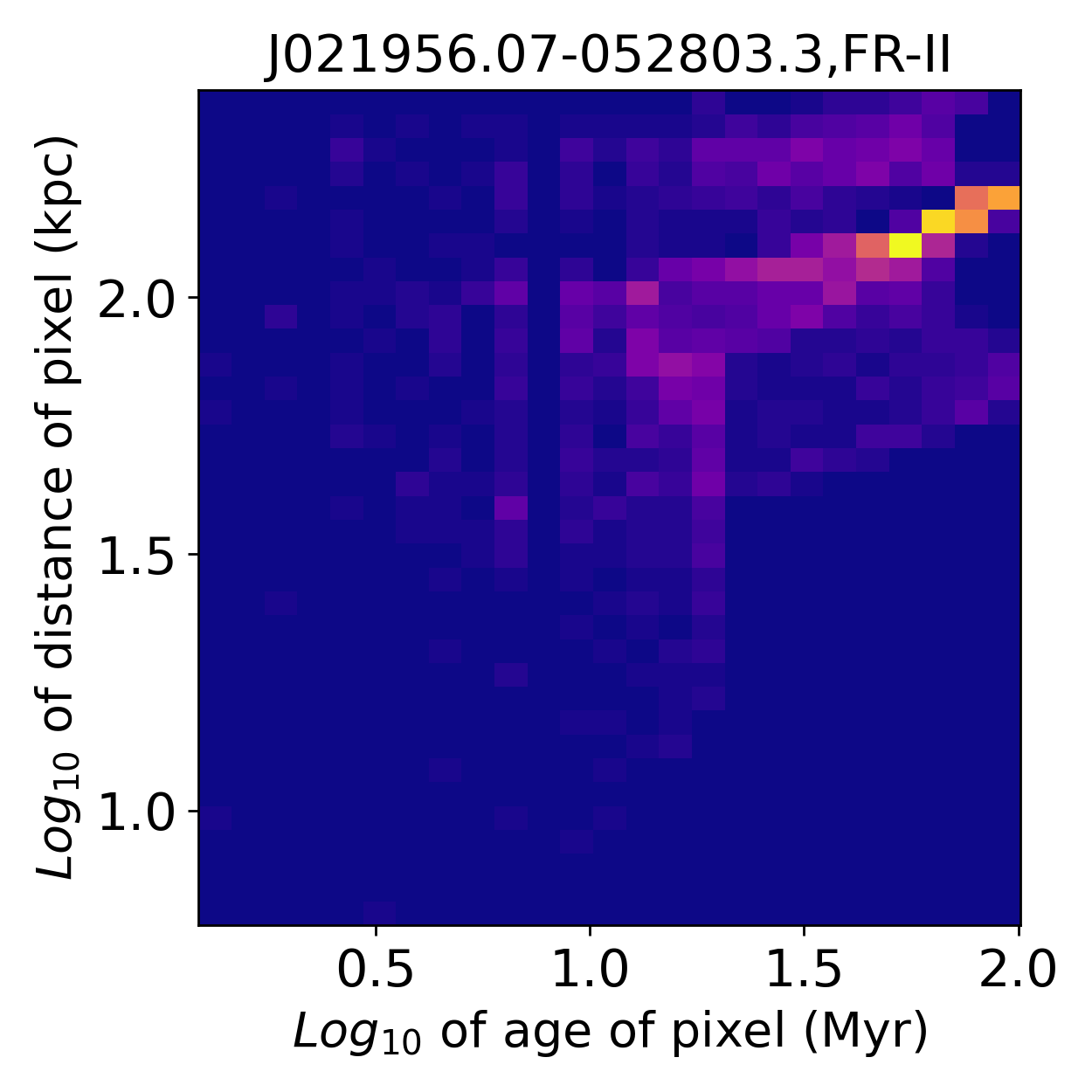}\includegraphics[width=1.3in,height=1.3in]{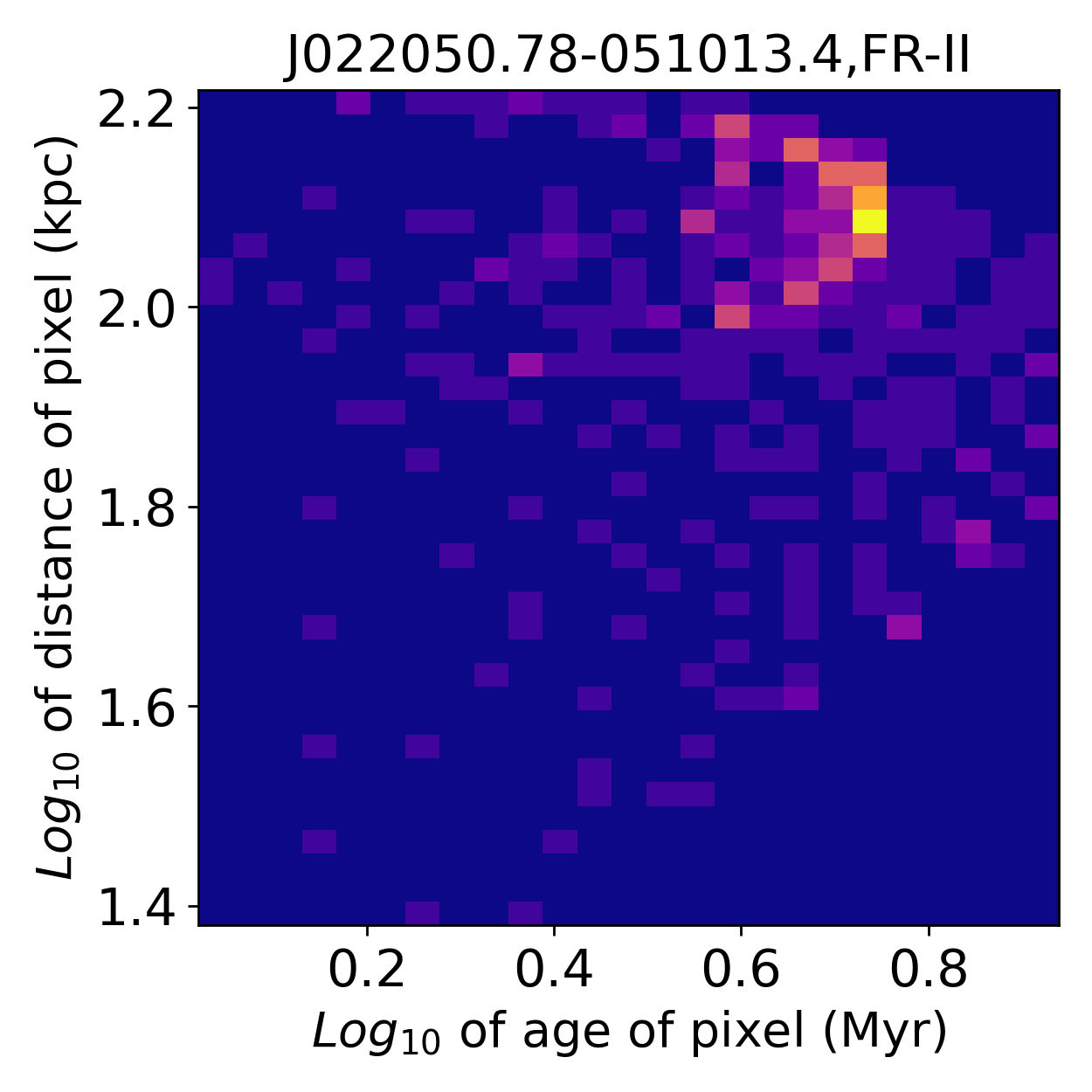}\includegraphics[width=1.3in,height=1.3in]{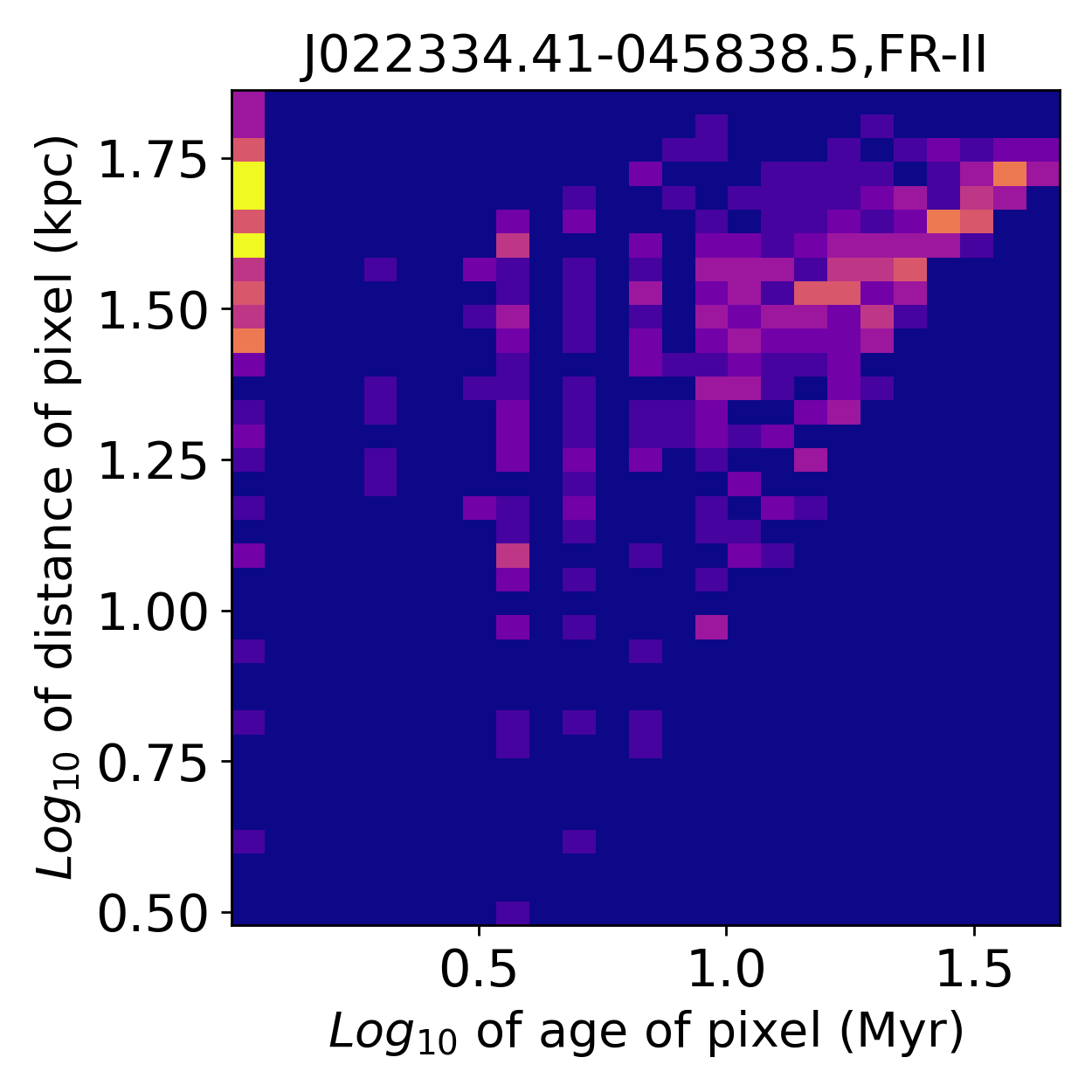}\includegraphics[width=1.3in,height=1.3in]{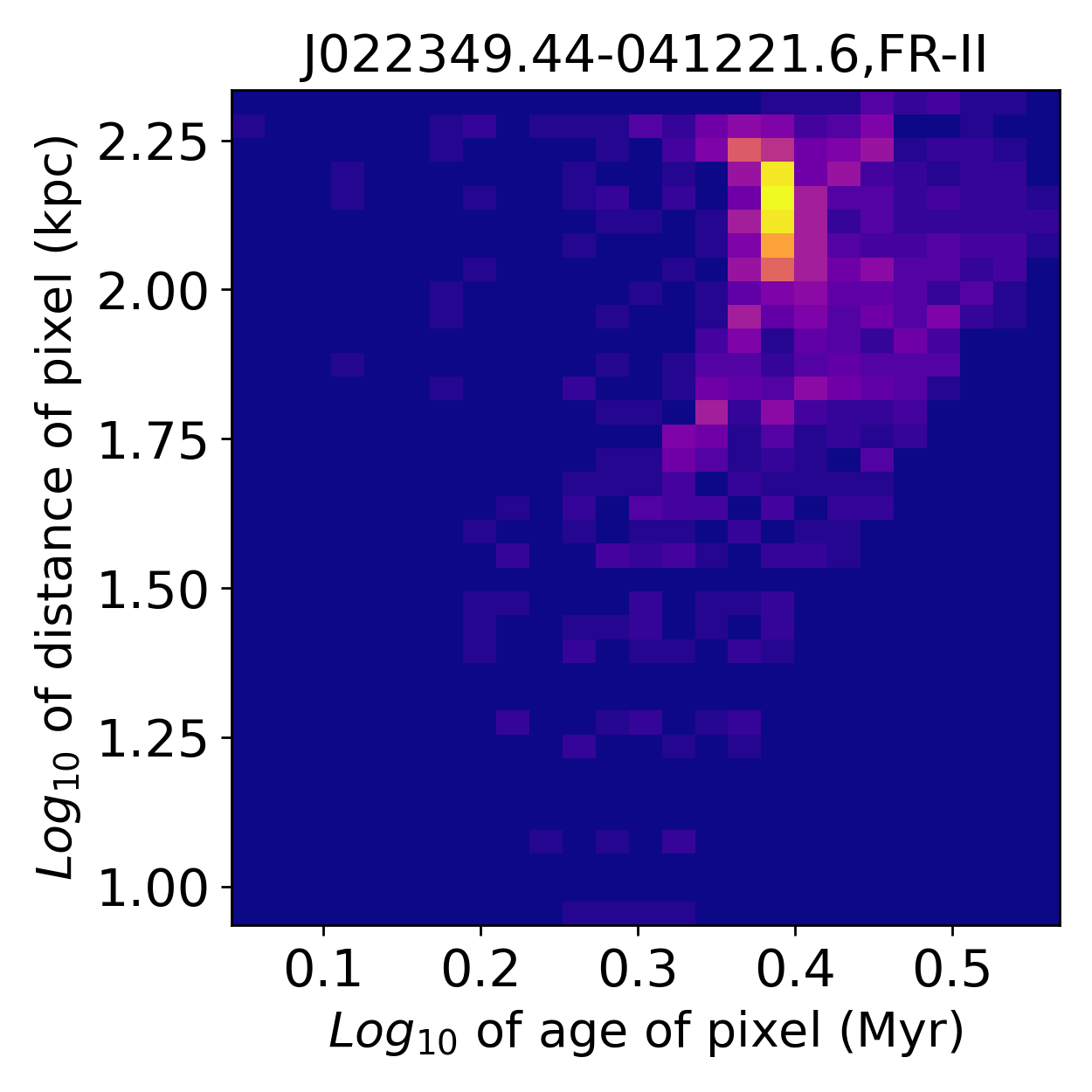}
\includegraphics[width=1.3in,height=1.3in]{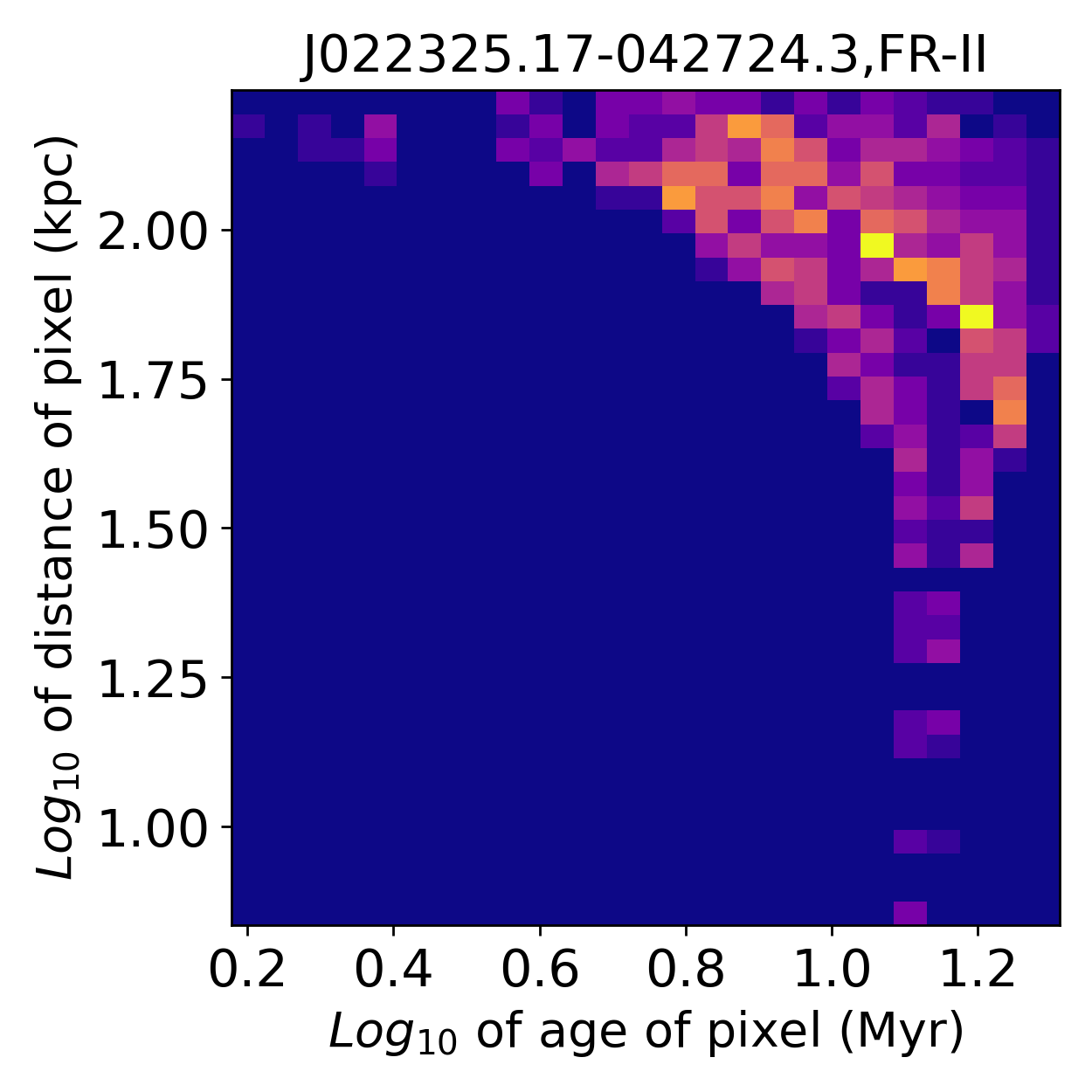}\includegraphics[width=1.3in,height=1.3in]{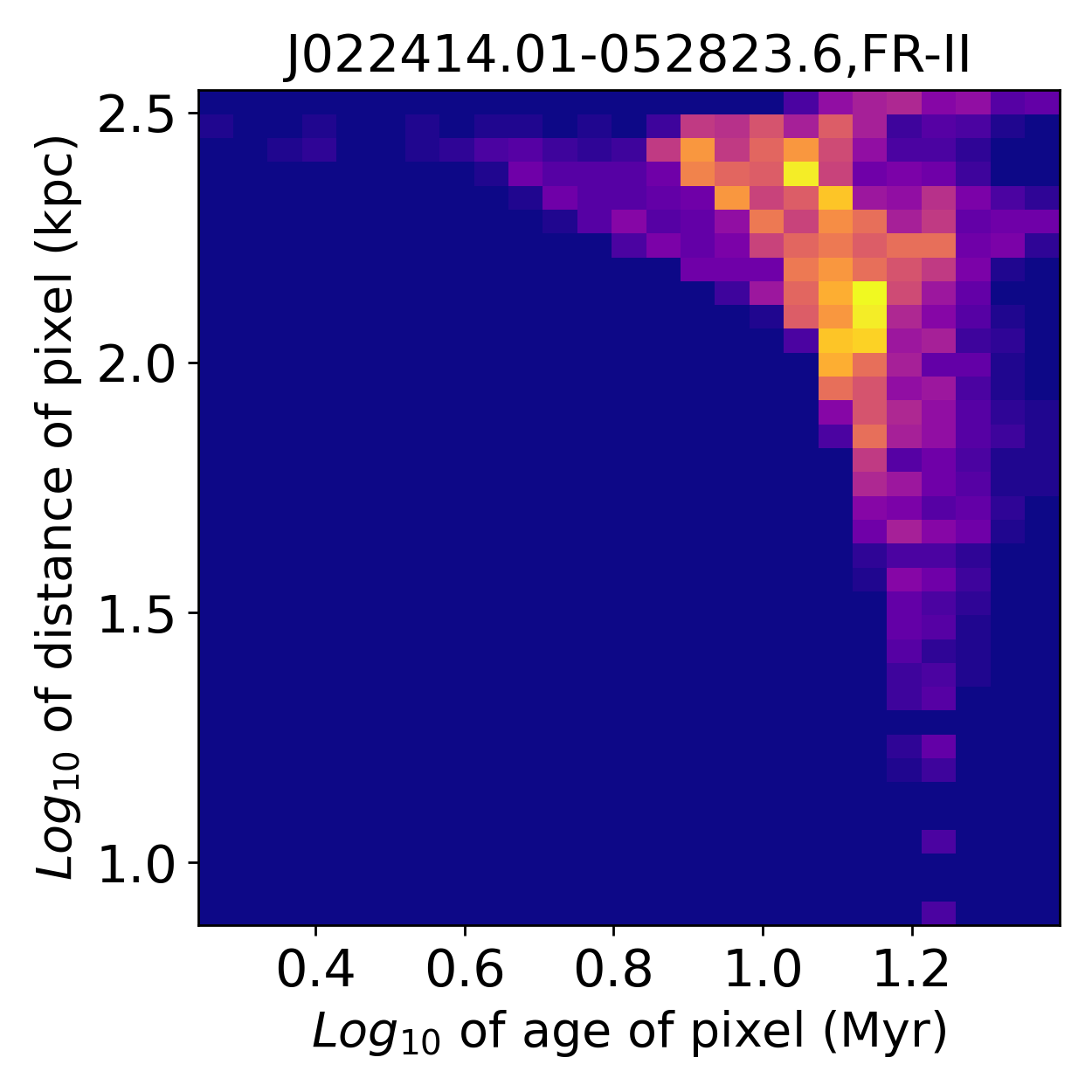}\includegraphics[width=1.3in,height=1.3in]{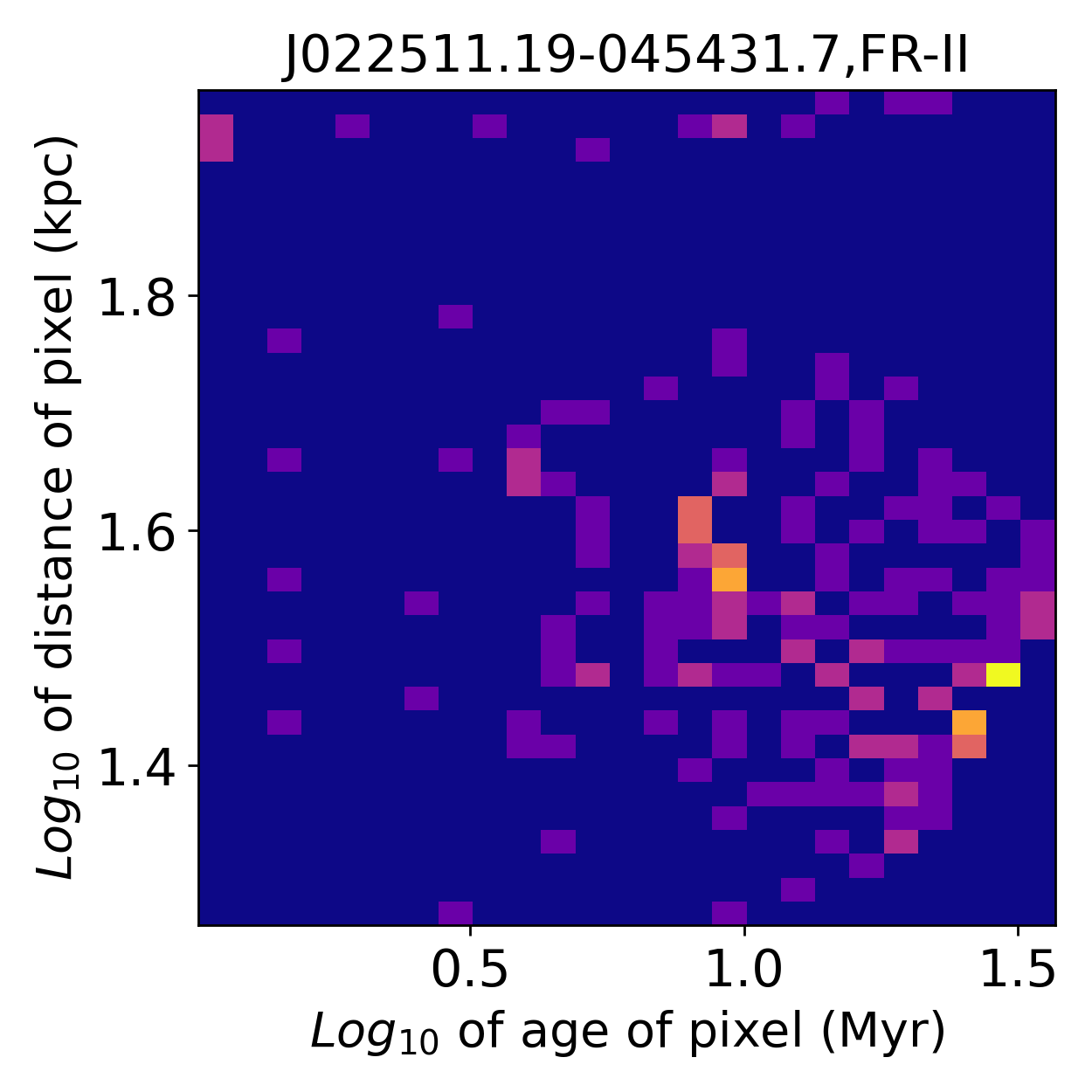}\includegraphics[width=1.3in,height=1.3in]{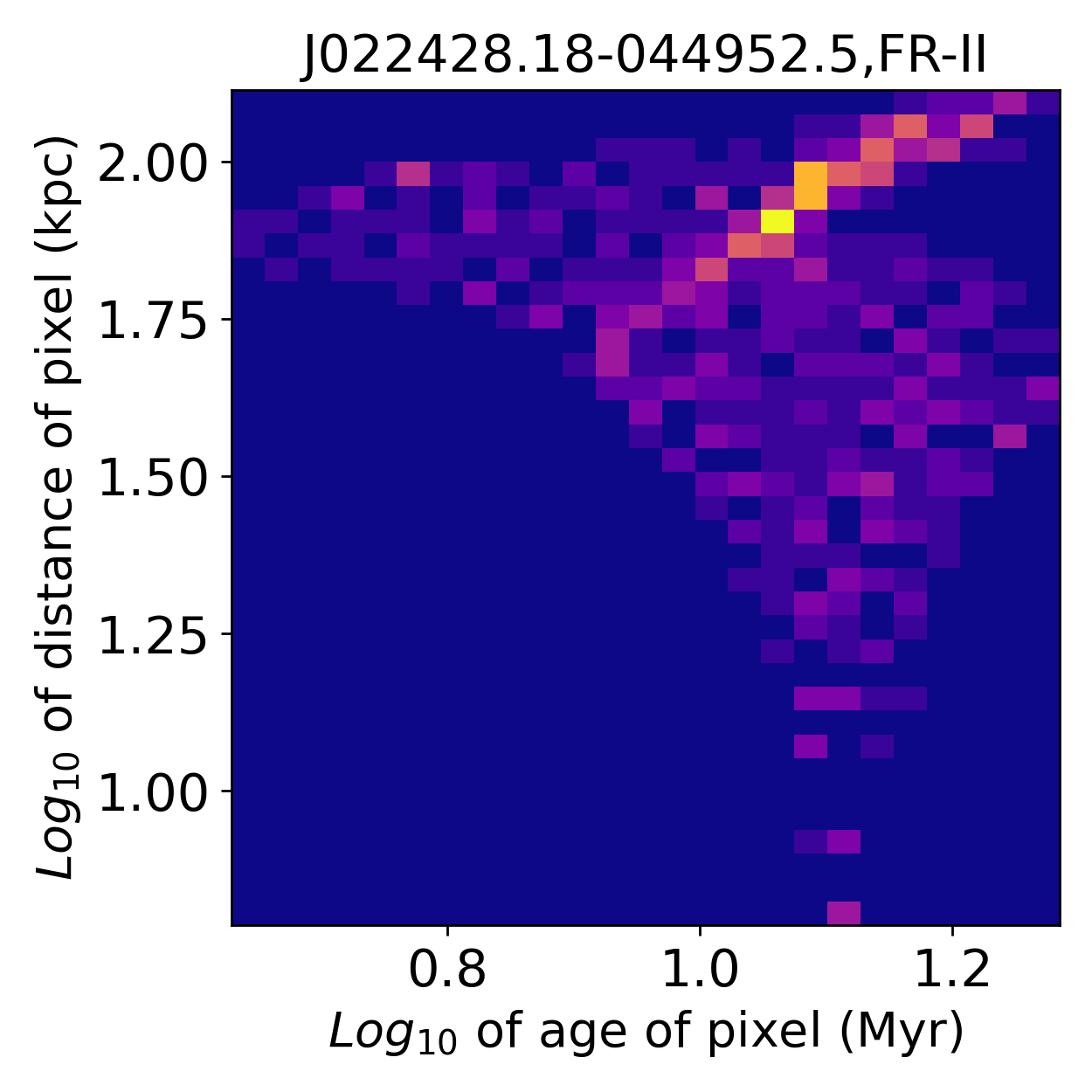}
\includegraphics[width=1.3in,height=1.3in]{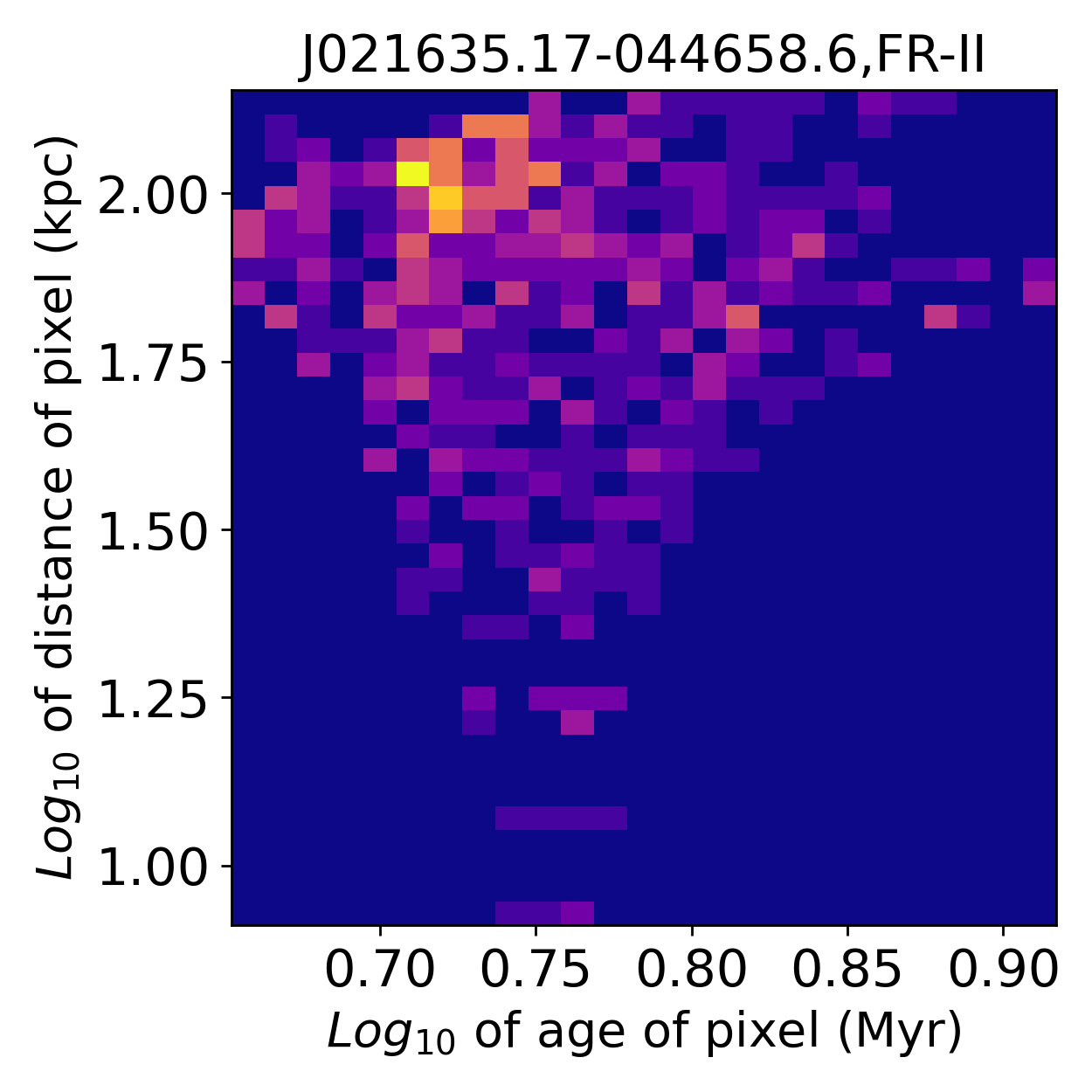}\includegraphics[width=1.3in,height=1.3in]{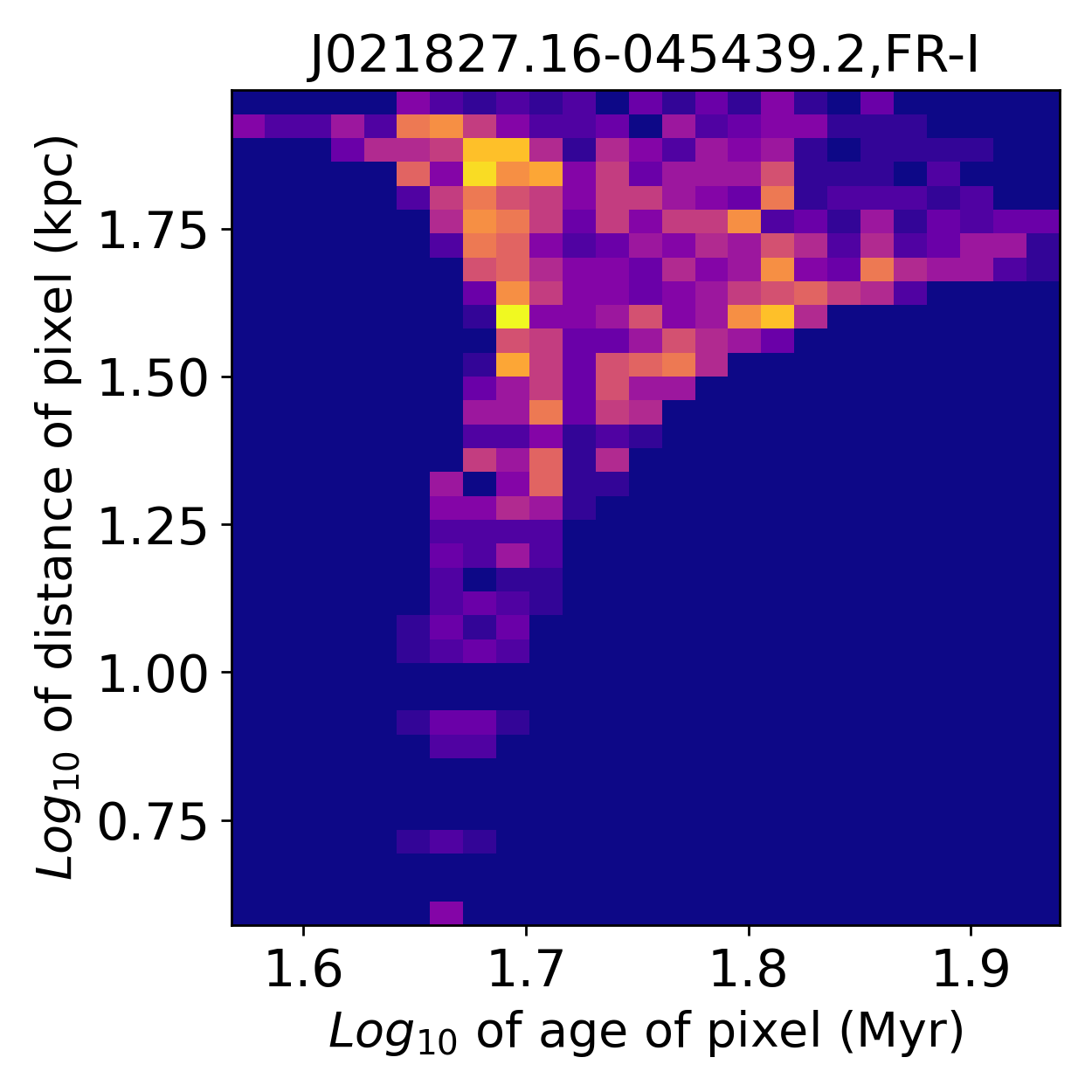}\includegraphics[width=1.3in,height=1.3in]{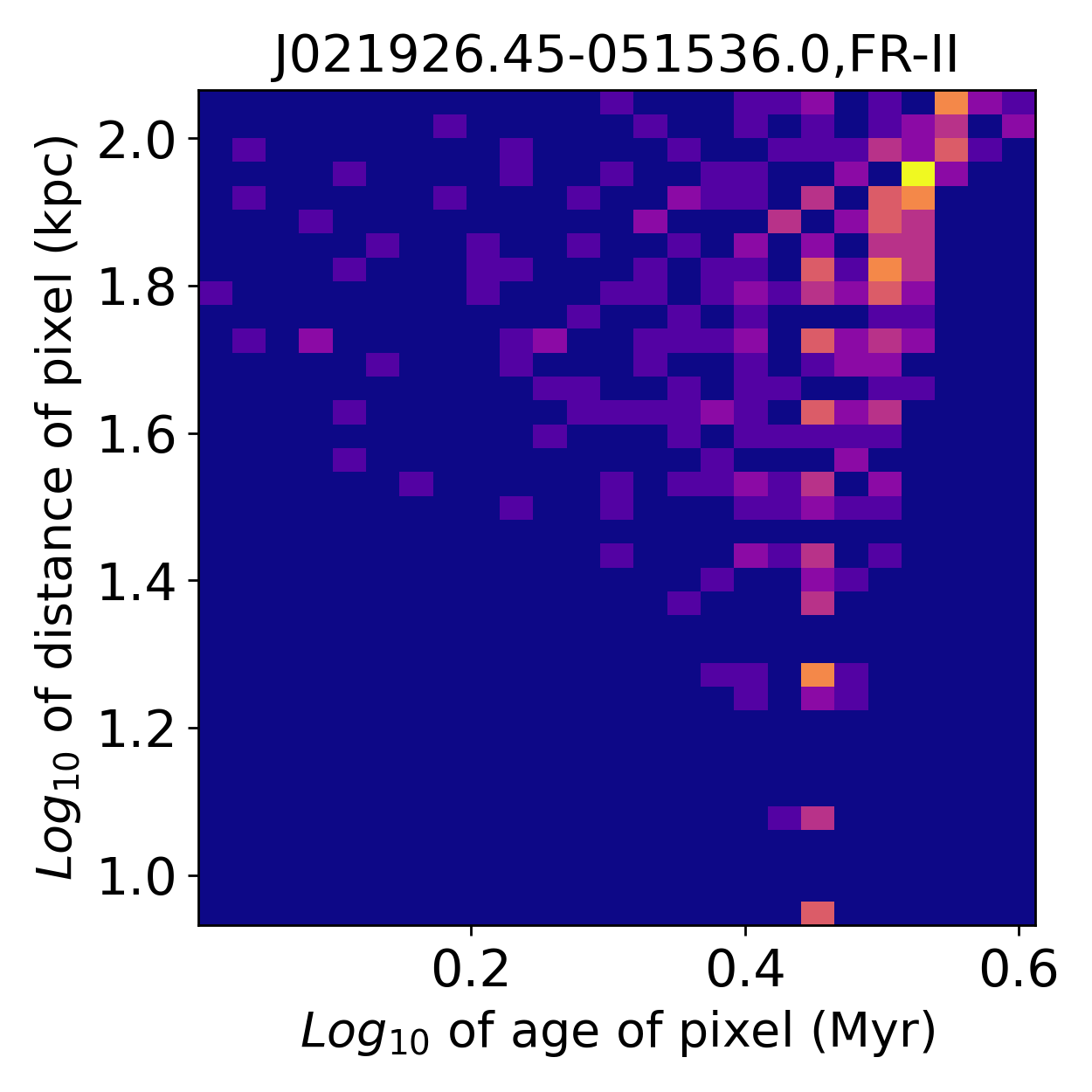}\includegraphics[width=1.3in,height=1.3in]{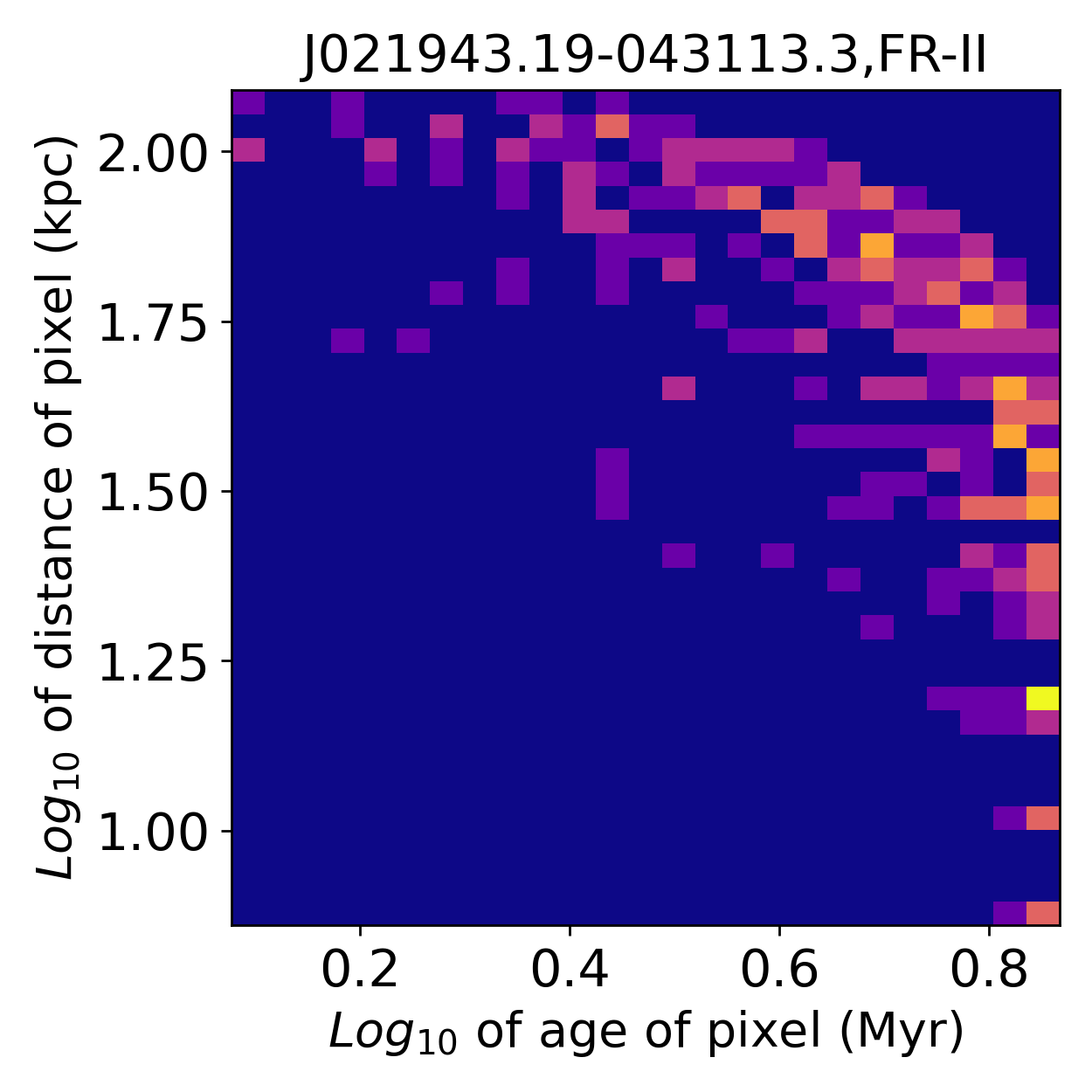}
\includegraphics[width=1.3in,height=1.3in]{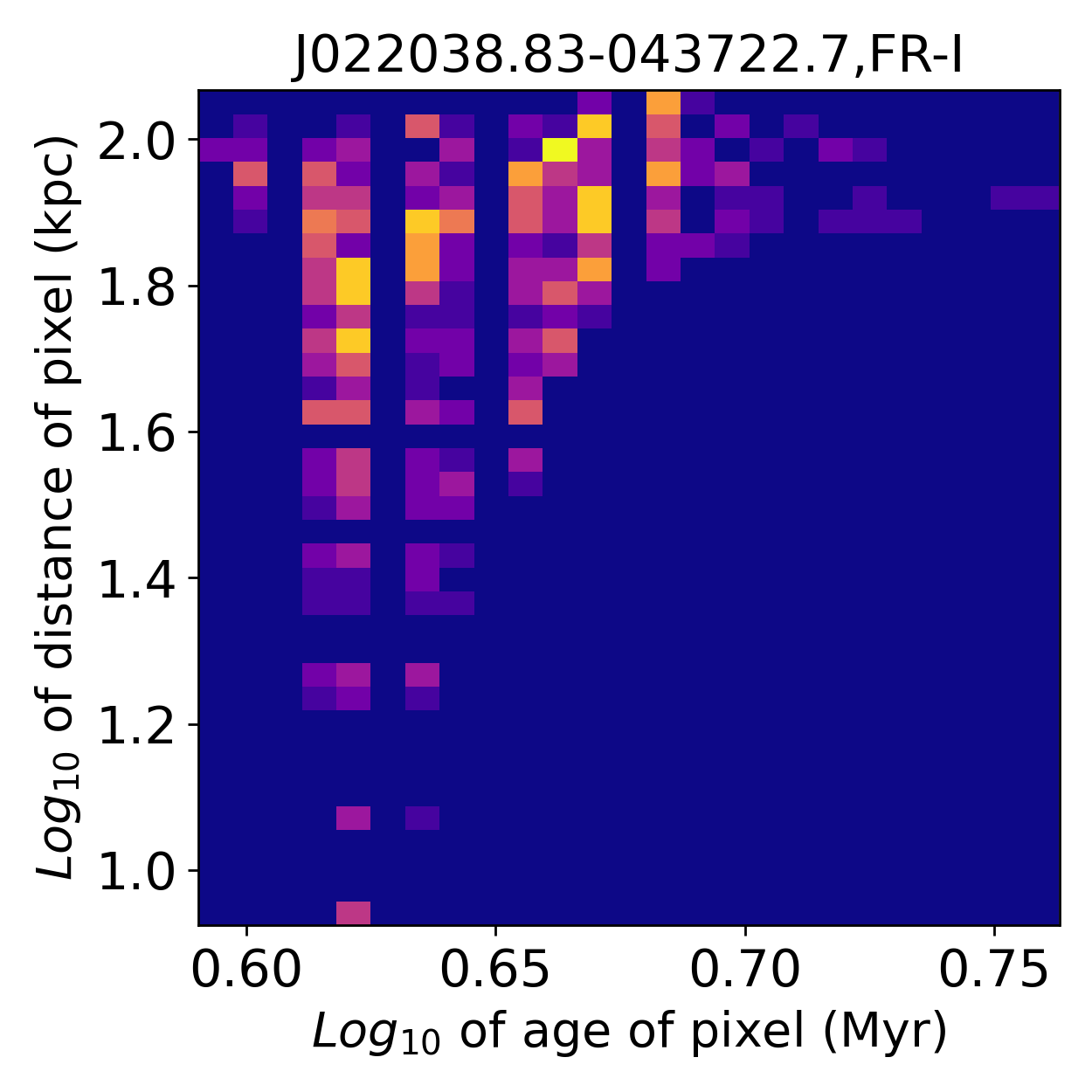}\includegraphics[width=1.3in,height=1.3in]{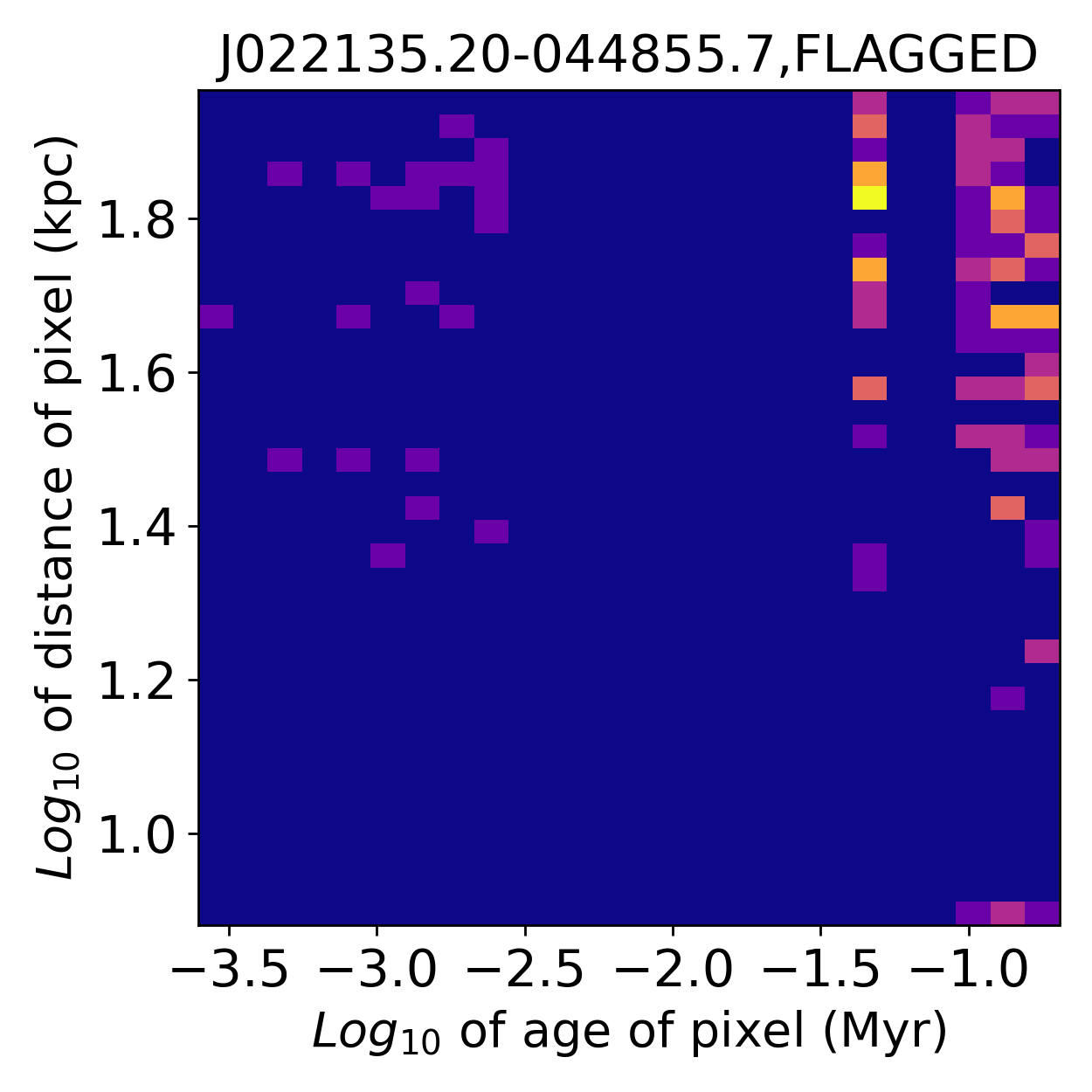}\includegraphics[width=1.3in,height=1.3in]{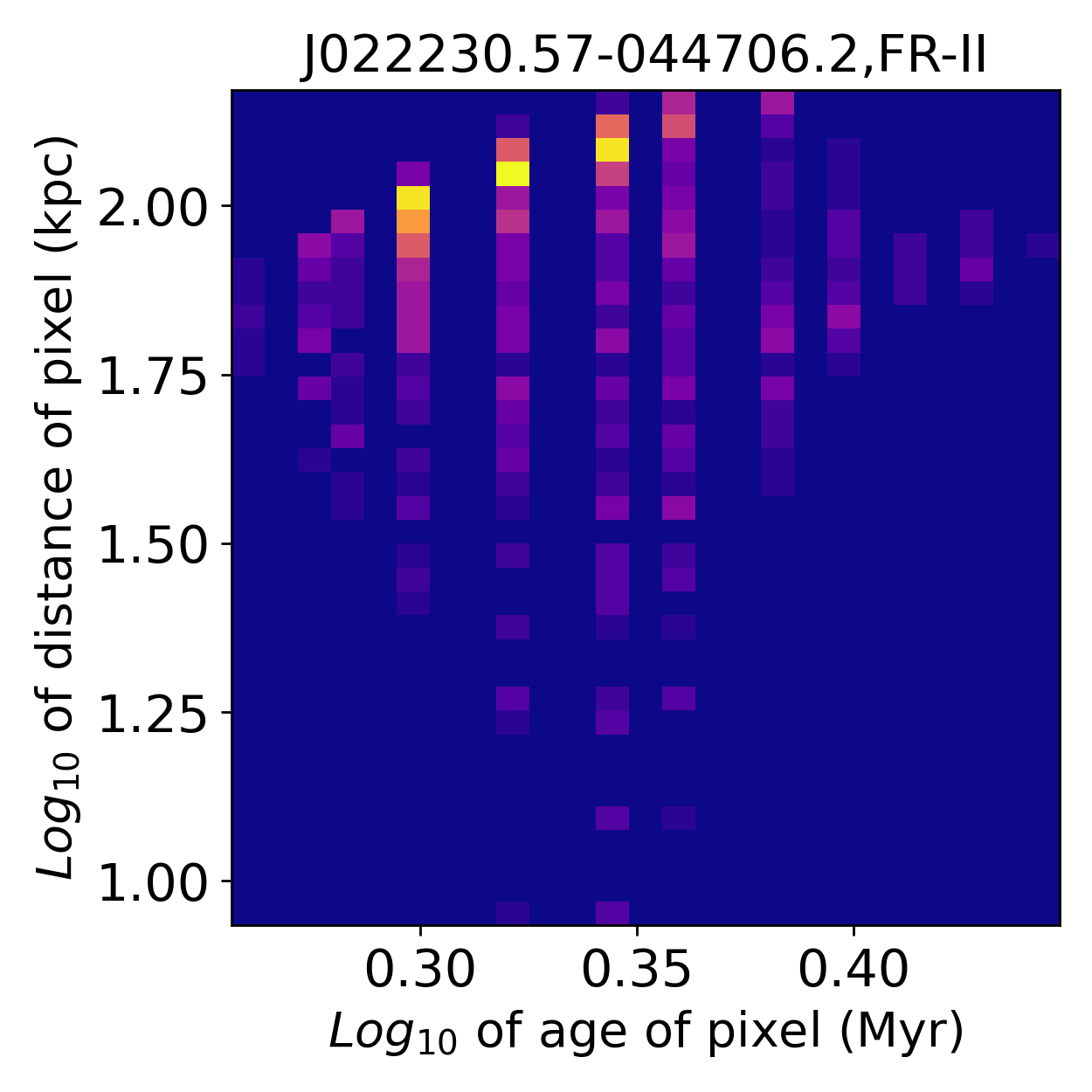}\includegraphics[width=1.3in,height=1.3in]{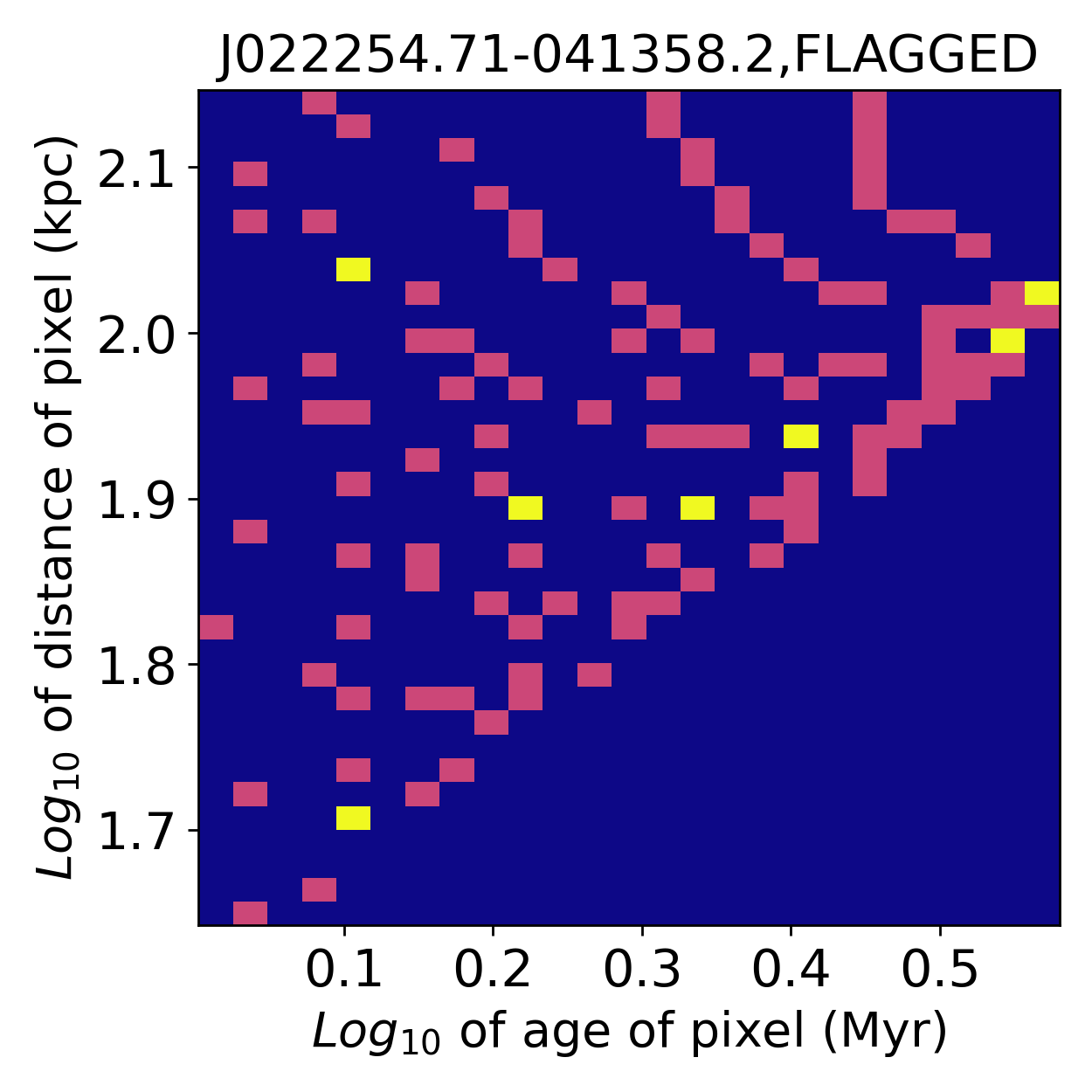}
\includegraphics[width=1.3in,height=1.3in]{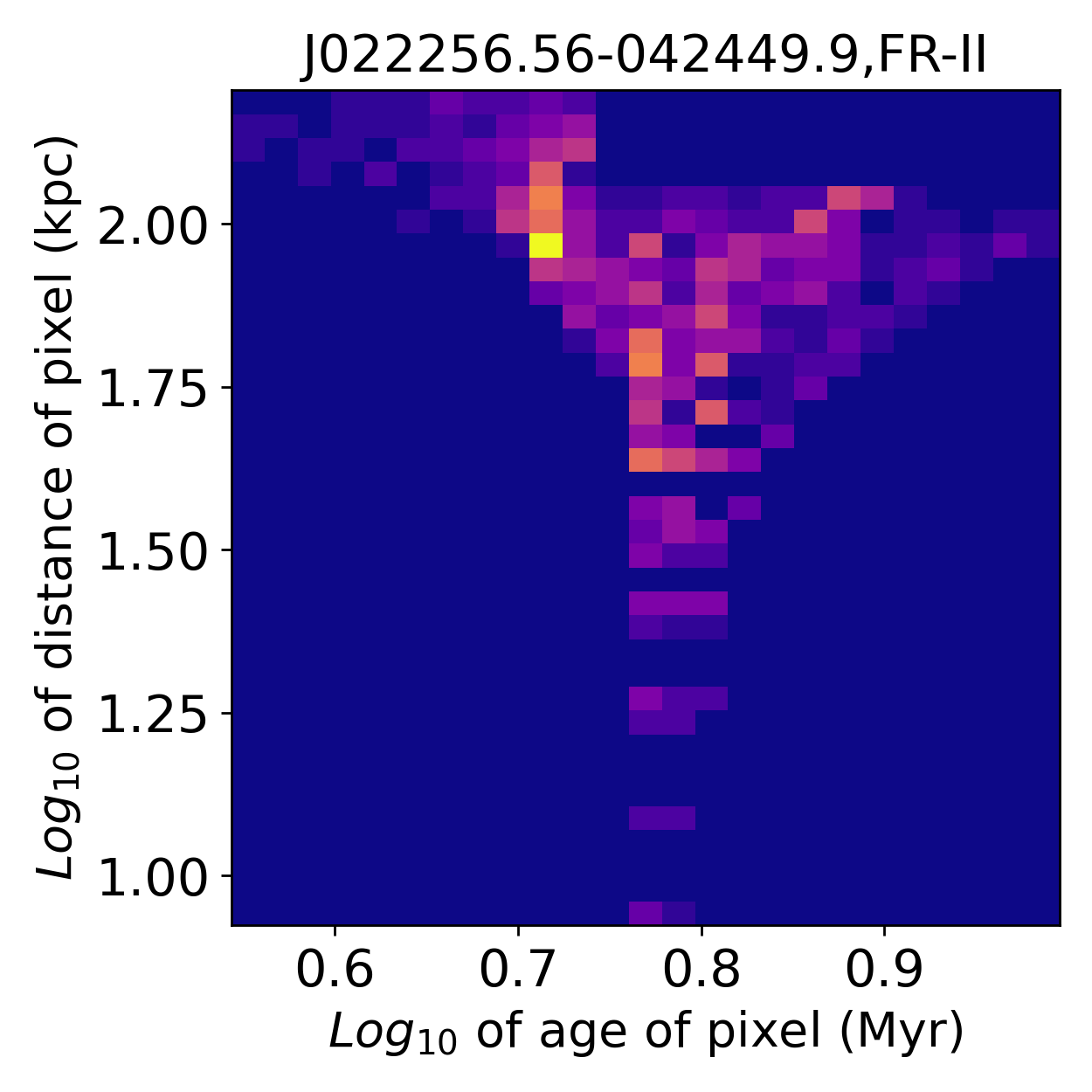}\includegraphics[width=1.3in,height=1.3in]{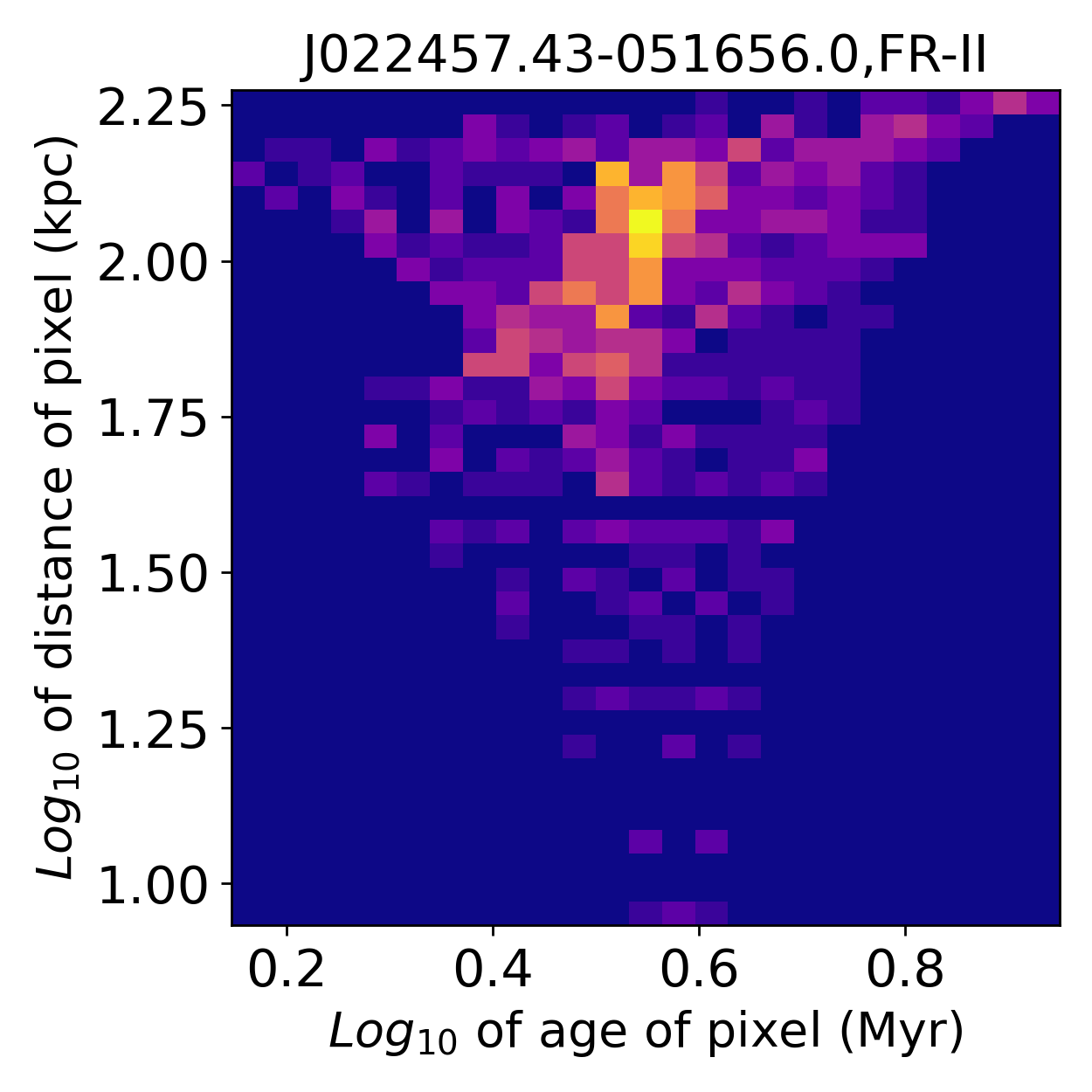}\includegraphics[width=1.3in,height=1.3in]{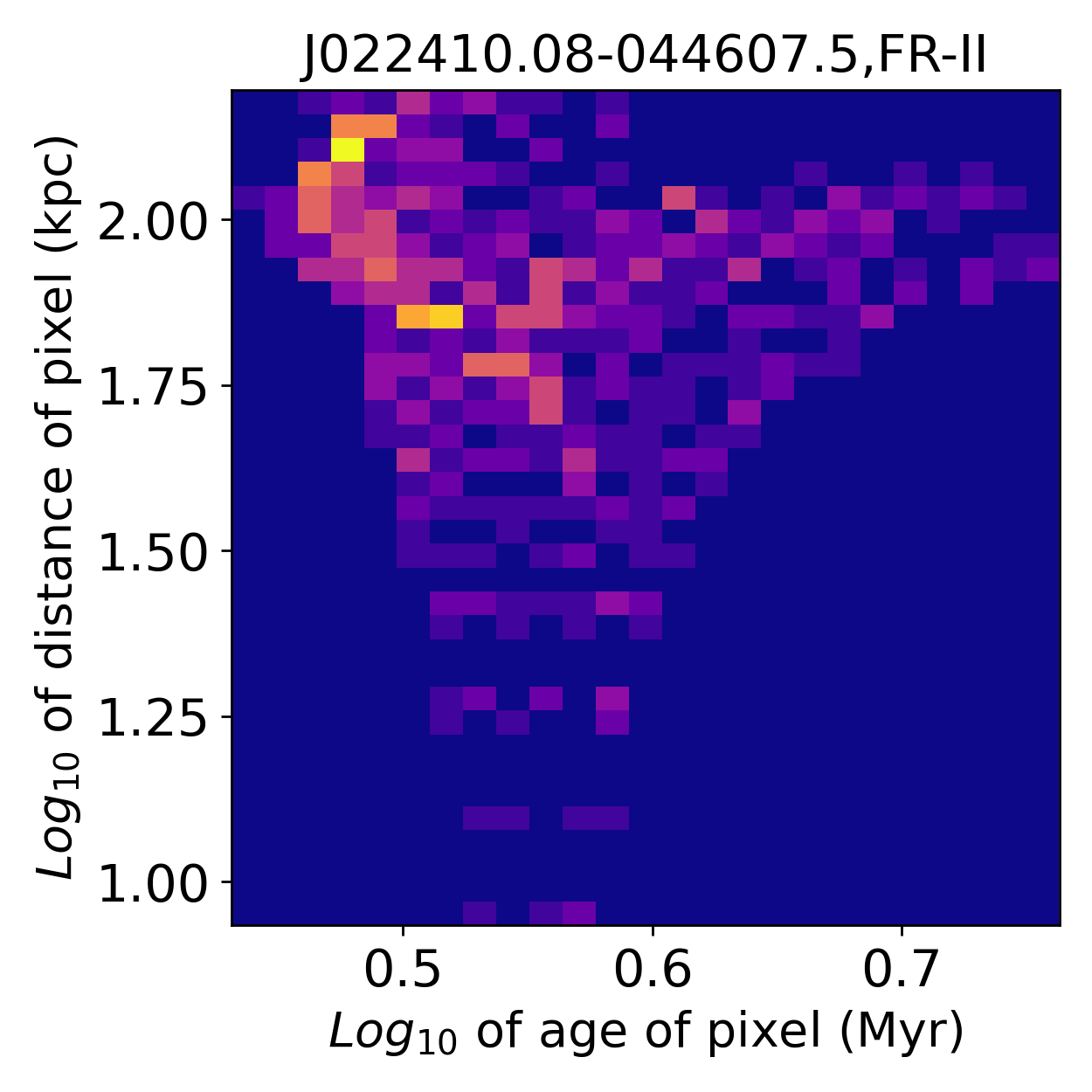}\includegraphics[width=1.3in,height=1.3in]{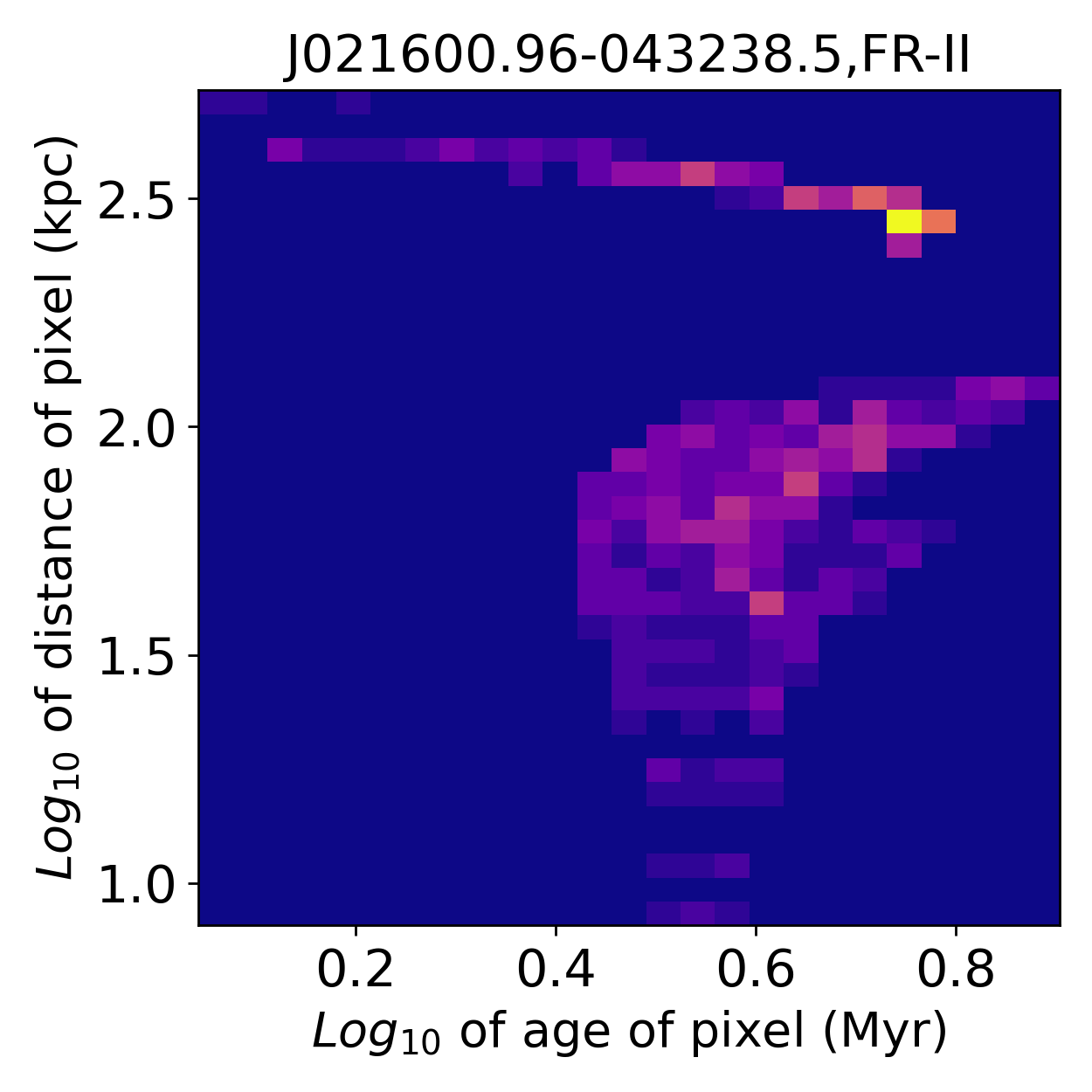}
\includegraphics[width=1.3in,height=1.3in]{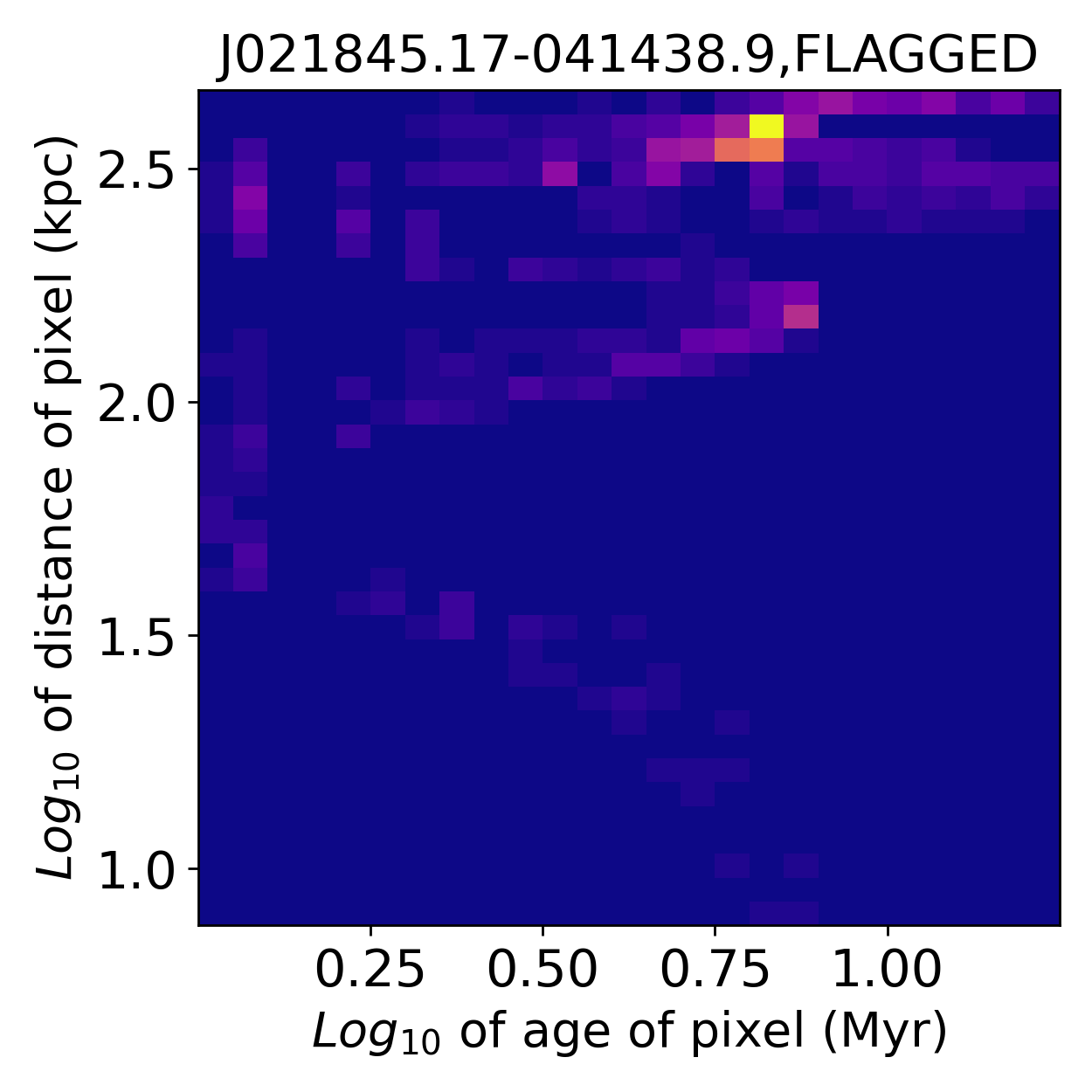}\includegraphics[width=1.3in,height=1.3in]{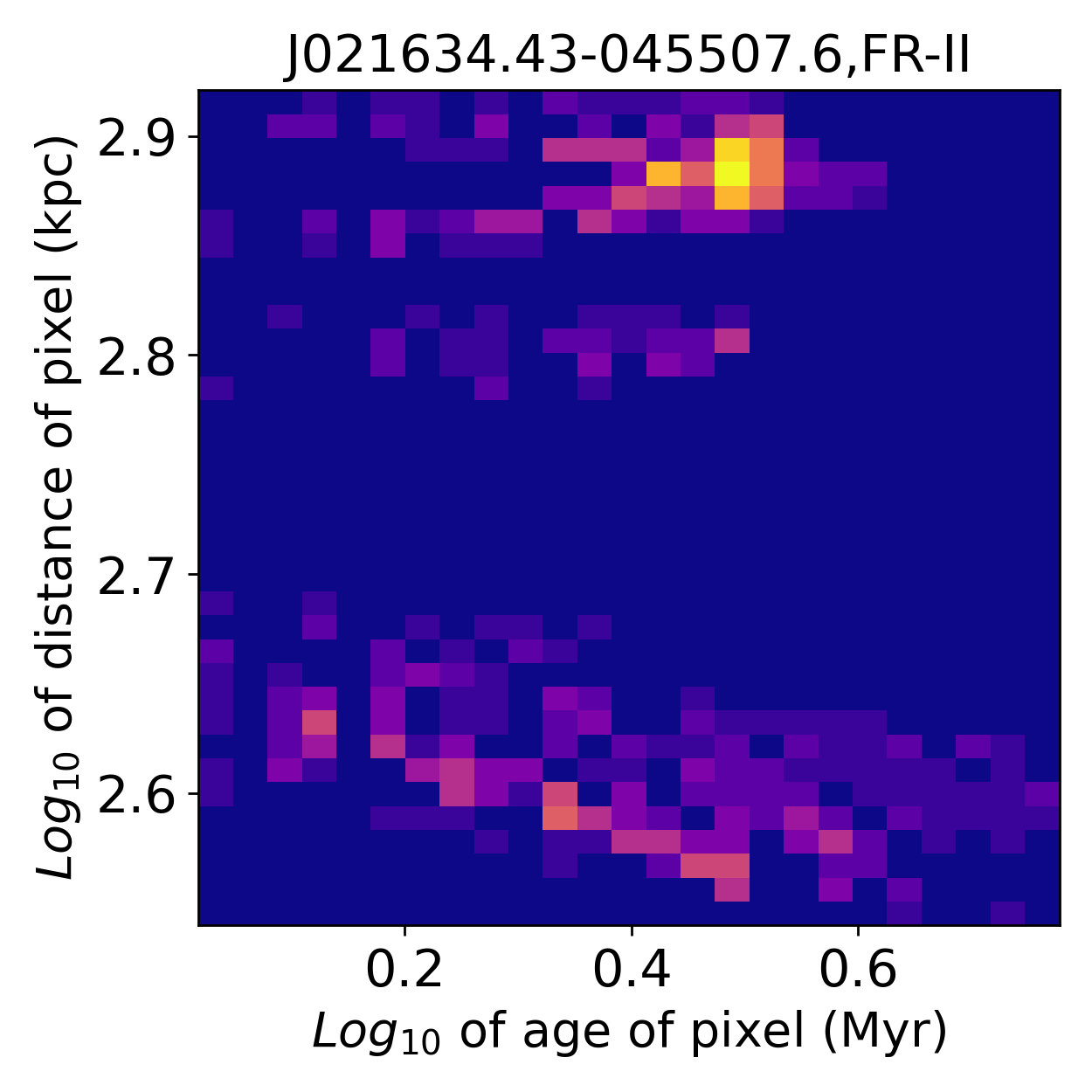}\includegraphics[width=1.3in,height=1.3in]{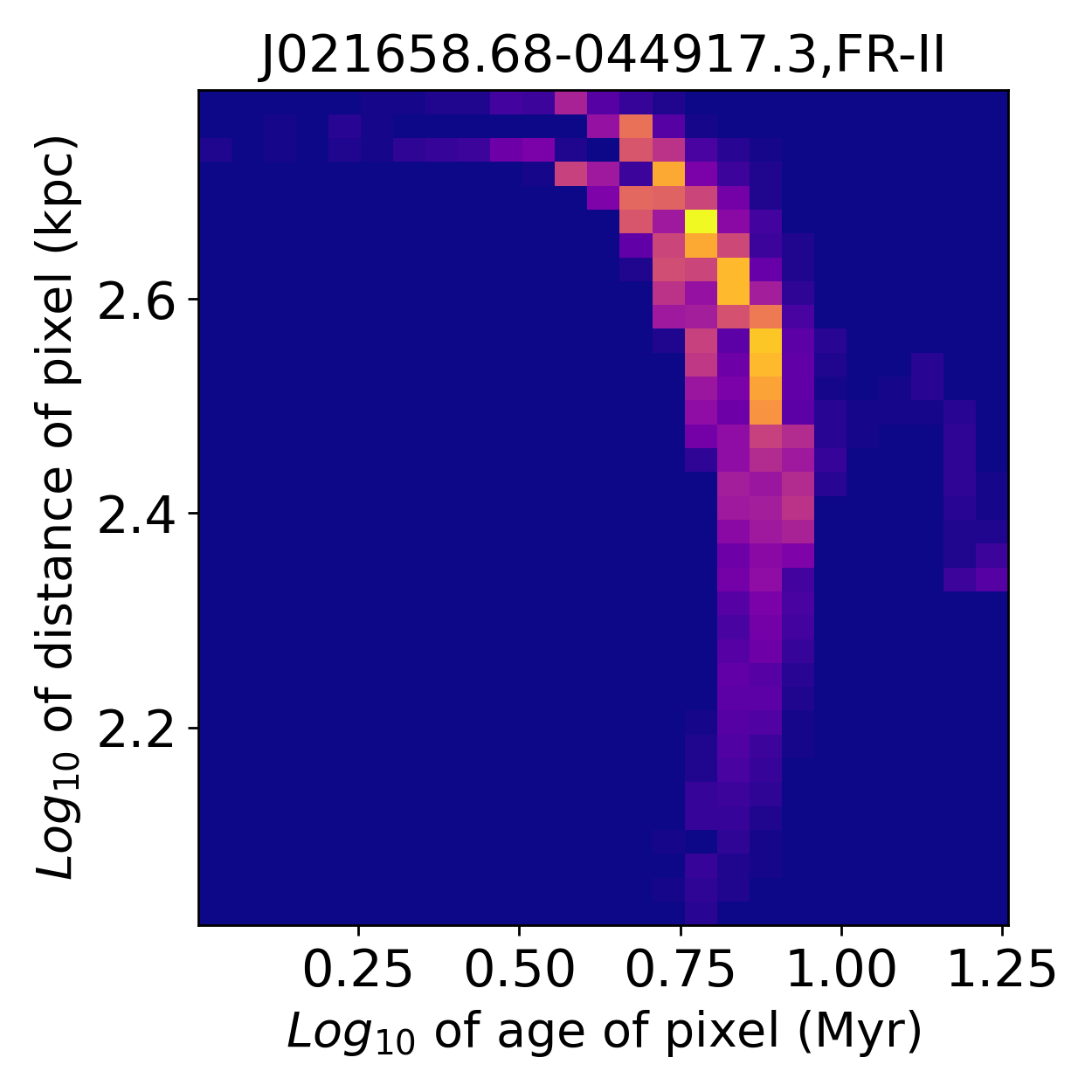}\includegraphics[width=1.3in,height=1.3in]{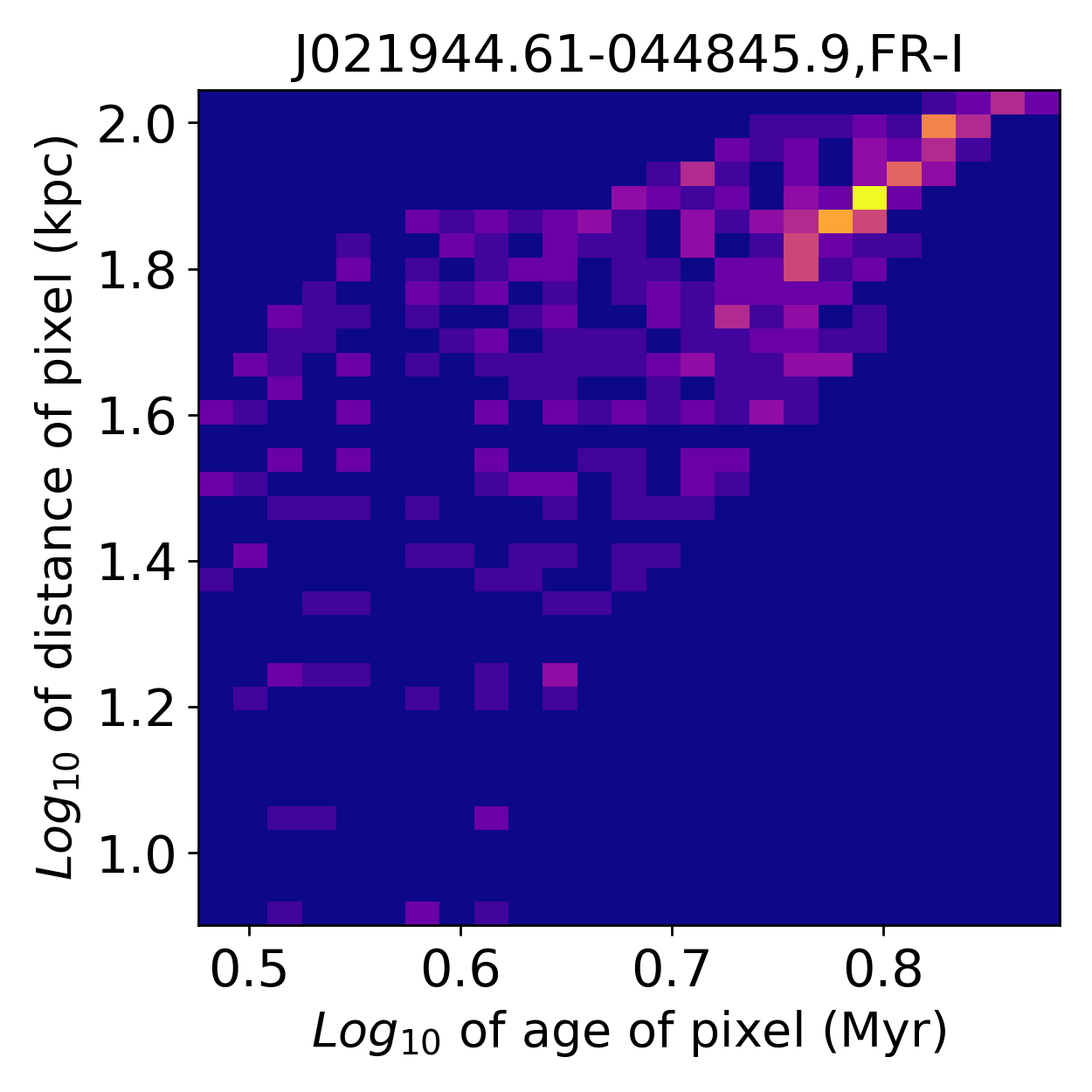}
\caption{Distribution of age pixel by pixel with respect to its distance from the core. The x-axis is age of the source in logarithmic scale and y-axis is the distance of the corresponding pixel from the core of the source in logarithmic scale. The title in each plot gives source IAU name and morphology.}
\label{a1}
\end{figure*}

This section shows error maps and age distribution plot for each source that is visualized in a different way. Figure \ref{ap} shows error distribution for each source. Figure \ref{a1} shows the correlation between the age distribution to the size distribution of pixels for each source. For each pixel in the source that reports an age value, we calculate the distance of that pixel from the core of the source and hence obtain an array of distances that correlate to their respective age. The aim is to understand the pattern we obtain for a well sampled source in the pixel space for different FR types. For example, for target 1 (fig \ref{t1a}), we see that pixel number density increases with increasing age and size. Hence, we expect to see linear relationship for Head-Tail sources. As for well resolved and sampled FR-type II sources we can see clear double arcs along the y axis, for example target 10, target 27 (Fig. \ref{t1c} and \ref{t7c}). These sources show increase in age as we reach close to core, which is represented by a downward curve of the pixel density in the plot. With sources like target 14, target 17 (fig \ref{t4b} and \ref{t5a}); which show FR-type I morphology, we see an arc that starts from low age and ends at higher age with increasing distance from the core. We can observe an upward curve of the pixel density in the plots. The Figure also shows some plots with 'FLAGGED' title which is due to anomalous behavior seen in the age distribution plots. These distinct patterns can be used as training datasets to obtain filters for autonomous selection and classification of the sources. Although this is just one of the thought applications, more exploration can be done into the method for future use.

\bsp	
\label{lastpage}
\end{document}